\documentclass[12pt,preprint]{aastex}
\usepackage{psfig,epsf,multicol,natbib,ccaption}

\newcommand{\uv}{\mbox{$u$-$v$}}

\newcommand{\etall}{\mbox{et~al.}}
\newcommand{\no}{\mbox{$n_{e_o}$}}

\newcommand{\rcI}{\mbox{$\theta_{c_1}$}}
\newcommand{\rcII}{\mbox{$\theta_{c_2}$}}

\newcommand{\dTo}{\mbox{$\Delta T_0$}}
\newcommand{\Xo}{\mbox{$S_{x  0}$}}

\newcommand{\LameH}{\mbox{$\Lambda_{e \mbox{\tiny H}}$}}

\newcommand{\Om}{\mbox{$\Omega_{\mbox{\scriptsize $M$}}$}}
\newcommand{\Ob}{\mbox{$\Omega_{\mbox{\scriptsize $B$}}$}}
\newcommand{\Ol}{\mbox{$\Omega_{\mbox{\scriptsize $\Lambda$}}$}}

\newcommand{\Msun}{\mbox{M$_\odot$}}
\newcommand{\kB}{\mbox{$k_{\mbox{\tiny B}}$}}

\newcommand{\sigT}{\mbox{$\sigma_{\mbox{\tiny T}}$}}
\newcommand{\Tcmb}{\mbox{$T_{\mbox{\tiny CMB}}$}}

\newcommand{\mutot}{\mbox{$\mu_{\mbox{\scriptsize tot}}$}}
\newcommand{\nH}{\mbox{$n_{\mbox{\tiny H}}$}}

\newcommand{\Da}{\mbox{$D_{\!\mbox{\tiny A}}$}}

\newcommand{\chandra}{{\it Chandra}}
\newcommand{\rosat}{{\it ROSAT}}

\newcommand{\asca}{{\it ASCA}}

\newcommand{\wmap}{{\it WMAP}}

\newcommand{\Mgas}{\mbox{$M_{\mbox{\scriptsize gas}}$}}
\newcommand{\Mtot}{\mbox{$M_{\mbox{\scriptsize total}}$}}

\newcommand{\tc}{\mbox{$t_{\mbox{\scriptsize cool}}$}}
\newcommand{\tH}{\mbox{$t_{\mbox{\tiny Hubble}}$}}
\newcommand{\rII}{\mbox{$r_{\mbox{\scriptsize 200}}$}}

\newcommand{\likel}{\mbox{$\mathcal{L}$}}

\newcommand{\fg}{\mbox{$f_{\mbox{\scriptsize gas}}$}}

\newcommand{\rhocrit}{\mbox{$\rho_{\mbox{\scriptsize crit}}$}}

\newcommand{\rtfh}{\mbox{$r_{\mbox{\scriptsize 2500}}$}}
\newcommand{\rfh}{\mbox{$r_{\mbox{\scriptsize 500}}$}}
\newcommand{\rth}{\mbox{$r_{\mbox{\scriptsize 200}}$}}
\newcommand{\nodatah}{--}
\newcommand{\nodatae}{ }

\begin{document}

\title{X-ray and Sunyaev-Zel'dovich Effect Measurements\\
         of the Gas Mass Fraction in Galaxy Clusters}

\author{
Samuel J. LaRoque\altaffilmark{1},
Massimiliano Bonamente\altaffilmark{2,3},
John E. Carlstrom\altaffilmark{1,4},
Marshall K. Joy\altaffilmark{3},
Daisuke Nagai\altaffilmark{1,5},
Erik D. Reese\altaffilmark{6},
Kyle S. Dawson\altaffilmark{7}
}

\altaffiltext{1}{Kavli Institute for Cosmological Physics, Department
  of Astronomy and Astrophysics, University of Chicago, Chicago, IL 60637}

\altaffiltext{2}{Department of Physics, University of Alabama,
  Huntsville, AL 35812}

\altaffiltext{3}{NASA Marshall Space Flight Center, Huntsville, AL 35812}

\altaffiltext{4}{Department of Physics, Enrico Fermi Institute,
  University of Chicago, Chicago, IL 60637}

\altaffiltext{5}{Theoretical Astrophysics, California Institute of Technology, Mail Code 130-33, Pasadena, CA 91125}

\altaffiltext{6}{Department of Physics, University of California,
  Davis, CA 95616}

\altaffiltext{7}{Department of Physics, University of California, Berkeley, CA 94720 (now at LBNL)}

\begin{abstract}
We present gas mass fractions of 38 massive galaxy clusters spanning
redshifts from 0.14 to 0.89, derived from {\it Chandra} X-ray data and
OVRO/BIMA interferometric Sunyaev-Zel'dovich Effect (SZE)
measurements. We use three models for the gas distribution: (1) an
isothermal $\beta$-model fit jointly to the X-ray data at radii beyond
100~kpc and to all of the SZE data, (2) a non-isothermal double
$\beta$-model fit jointly to all of the X-ray and SZE data, and (3) an
isothermal $\beta$-model fit only to the SZE spatial data.  We show
that the simple isothermal model well characterizes the intracluster
medium (ICM) outside of the cluster core, and provides consistently
good fits to clusters spanning a wide range of morphological
properties.  The X-ray and SZE determinations of mean gas mass
fractions for the 100~kpc-cut isothermal $\beta$-model are
\fg(X-ray)$=0.110^{+0.003}_{-0.003}\,^{+0.006}_{-0.018}$ and
\fg(SZE)$=0.116^{+0.005}_{-0.005}\,^{+0.009}_{-0.026}$, where
uncertainties are statistical followed by systematic at 68\%
confidence.  For the non-isothermal double $\beta$-model,
$\fg$(X-ray)$=0.119^{+0.003}_{-0.003}\,^{+0.007}_{-0.014}$ and
$\fg$(SZE)$=0.121^{+0.005}_{-0.005}\,^{+0.009}_{-0.016}$.  For the
SZE-only model,
$\fg$(SZE)$=0.120^{+0.009}_{-0.009}\,^{+0.009}_{-0.027}$.  The
agreement in the results shows that the core can be satisfactorily
accounted for by either excluding the core in fits to the X-ray data
(the 100~kpc-cut model) or modeling the intracluster gas with a
non-isothermal double-$\beta$ model.  We find that the SZE is largely
insensitive to structure in the core.  Our results indicate that the
ratio of the gas mass fraction within \rtfh\ to the cosmic baryon
fraction, $\frac{\fg }{\Ob/\Om }$, is $0.68^{+0.10}_{-0.16}$ where the
range includes statistical and systematic uncertainties at 68\%
confidence.  Finally, by assuming that cluster gas mass fractions are
independent of redshift, we find that the results are in agreement
with standard $\Lambda$CDM cosmology and are inconsistent with a flat
matter dominated ($\Om=1$) universe.

\end{abstract}


%
%
\section{Introduction}
\label{sec:intro}

Massive galaxy clusters ($M \sim 10^{15}\,\Msun$) are thought to be
the relatively recent descendants of rare high density fluctuations in
the primordial universe.  The evolution of massive clusters is
critically dependent on cosmology, in particular on the matter density
\Om\ and the normalization of the power spectrum
$\sigma_8$. Furthermore, as massive galaxy clusters have collapsed
from volumes of order 1000 Mpc$^3$, their composition should reflect
that of the non-relativistic components of the universe, i.e., the
baryon budget of a cluster should reflect $\Ob/\Om$.

For these reasons there is considerable interest in using cluster
observations to constrain cosmology, and there has been significant
success. For example, early X-ray observations were used to constrain
the gas mass fraction of clusters and thereby set a lower limit to
\Ob/\Om.  Using constraints on \Ob\ from Big Bang Nucleosynthesis,
this led to upper limits on \Om\ that strongly ruled out a flat matter
dominated universe \citep{white1993a}; a precise measurement of \Om\
was not possible due to insufficient understanding of the cluster
baryon budget.  A low \Om\ was also indicated by the discovery of high
redshift ($z \sim 1$) massive clusters, which should be extremely rare
in a matter dominated, flat universe \citep{bahcall1998,donahue1998}.

Cosmological constraints can be obtained by exploiting the expected
redshift independence of cluster gas mass fractions; only if the
correct cosmology is used in the derivation of the gas mass fractions
for a sample of clusters spanning a reasonable redshift range would
the resulting gas mass fractions be constant with redshift
\citep{sasaki1996,pen1997}. This technique is independent of the
uncertainty in the cluster baryon budget as well as the value of the
Hubble constant.

Subsequently there have been many more cosmological studies using
cluster X-ray and Sunyaev-Zel'dovich Effect (SZE) measurements
\citep[e.g., see ][]{white1993a,david1995,evrard1997,
myers1997,mohr1999a,ettori1999,grego2001,allen2002, sanderson2003a,
allen2004}. In general the results are in good agreement, along with
cosmic microwave background (CMB) \citep[i.e., ][]{spergel2006} and
Type Ia supernova measurements \citep{perlmutter1999, riess1998}, with
the now standard $\Lambda$CDM cosmology.

Future large scale cluster surveys are being pursued which exploit the
critical cosmological dependence of the evolution of cluster abundance
to provide independent, precise determinations of \Om, \Ol, $\sigma_8$
and the equation of state of dark energy, $w(z)$ \citep[e.g.,
][]{kneissl2001,romer2001, kosowsky2003,ruhl2004}.  The ability to
extract cosmology from large surveys using SZE, X-ray, or any other
cluster observable, will depend on the ability to link the
measurements to cluster mass.  An improved understanding of cluster
structure and evolution is of interest in itself and would clearly
also be helpful for improving cosmological measurements with clusters.

In this paper we investigate the gas mass fractions for a sample of 38
clusters spanning redshifts from 0.14 to 0.89, using {\it Chandra}
X-ray data and BIMA/OVRO interferometric radio SZE data. This SZE
cluster sample is the largest yet compiled by over a factor of two.
Combining the SZE data with the high resolution {\it Chandra} X-ray
data we are able to compare the SZE and X-ray gas mass fractions using
two different models for the ICM.  Comparison of the SZE and X-ray
results allows us to place constraints on the possible systematic
effects associated with our incomplete knowledge of cluster
properties.  The results also provide constraints on the cluster
baryon budget, and thereby provide an important observational
benchmark for cluster simulations. The baryon budget is sensitive to
the details of cluster formation, gas cooling and star formation
history of the ICM \citep{ettori2006}.  Lastly, we consider the
cosmological constraints derived by assuming the cluster gas mass
fractions of our sample do not evolve with redshift.

The paper is organized as follows: we review the theory underlying the
X-ray emission and the SZE in clusters in \S\ref{sec:xray_sze} and
describe the X-ray and SZE data in \S\ref{sec:data}.  In
\S\ref{sec:methods} we describe the cluster models and analysis
methods used to determine the gas mass fractions, and we present tests
of these models.  In \S\ref{subsec:systematics} we discuss additional
sources of uncertainty in the gas mass fraction measurements, both
statistical and systematic.  The results are presented in
\S\ref{sec:results}, and constraints on the cluster baryon budget and
cosmological parameters are detailed in
\S\ref{sec:physics_and_cosmology}.  Finally, we summarize our
conclusions in \S\ref{sec:conclusions}.  All uncertainties are at the
68.3\% confidence level, and a $\Lambda$CDM cosmology with \Om=0.3,
\Ol=0.7, $h$=0.7 is assumed unless otherwise stated.

\section{X-ray emission and Sunyaev Zel'dovich Effect}
\label{sec:xray_sze}

X-ray emission in clusters arises predominantly from thermal
bremsstrahlung for gas electron temperatures $T_e \gtrsim 3$~keV.  The
X-ray emissivity is not a directly observable quantity; instead X-ray
proportional counters and CCDs measure the X-ray surface brightness
over some frequency band \citep[e.g.,][]{sarazin1988},
\begin{equation}
\label{eq:xray_sb}
S_x = \frac{1}{4\pi (1+z)^4} \! \int \!\!  \, n_e \nH \LameH \,d\ell
\end{equation}
where the integral is along the line of sight and $\Lambda$ is called
the X-ray cooling function and is proportional to $T_e^{1/2}$.  The
observed X-ray emission can also be used to measure the gas
temperature using instruments with spectral capability.

The thermal SZE is a small ($\lesssim 1$~mK) distortion in the CMB
spectrum caused by inverse Compton scattering of the CMB photons off
of energetic electrons in the hot intracluster gas
\citep{sunyaev1970,sunyaev1972}.  The spectral distortion can be
expressed for dimensionless frequency $x \equiv h\nu/\kB \Tcmb$ as a
temperature change $\Delta T$ relative to the CMB temperature \Tcmb:
\begin{equation}
\label{eq:thermal_sz}
\frac{\Delta T}{\Tcmb} = f(x) y = f(x) \int \!\! \, \sigT n_e
\frac{\kB T_e}{m_e c^2}\,d\ell,
\end{equation}
where $y$ is the Compton-$y$ parameter, \sigT\ is the Thomson
scattering cross-section of the electron, $f(x)$ contains the
frequency dependence of the SZE, and the integral is along the line of
sight.  The frequency dependence can be expressed as
\begin{equation}
\label{eq:sze_freq_dependence}
f(x)=\left( x \frac{e^x + 1}{e^x - 1} -4 \right)
(1+\delta_{\mbox{\tiny SZE}}(x,T_e))
\end{equation}
where $\delta_{\mbox{\tiny SZE}}(x,T_e)$ is a relativistic correction.
We adopt the analytic corrections of \citet{itoh1998} which are good
to fifth order in $\kB T_e/m_e c^2$.  The linear density and
temperature dependences of the SZE make it a complementary probe to
the X-ray emission, which varies as $n_e^2 T_e^{1/2}$.  The SZE is a
decrement at low frequencies ($\la 218$ GHz) and an increment at high
frequencies due to the up-scattering of photons by the hot electrons.
There is also a kinetic SZE that arises from scattering of CMB photons
in a cluster with line of sight motion relative to the CMB rest frame.
Discussion of the SZE in this work refers to the thermal SZE unless
otherwise specified.  We address the possible effects of the kinetic
SZE on our results in \S\ref{subsec:systematics}.

With knowledge of the gas temperature, one can use either $S_x$ or
$\Delta T$ to determine the gas distribution and obtain \Mgas.  The
total mass \Mtot\ can be determined by assuming that the gas is in
hydrostatic equilibrium (HSE) with the cluster potential.  The gas
mass fraction \fg\ is then \Mgas/\Mtot.

%
%

\section{Data}
\label{sec:data}

The cluster sample consists of clusters that have both X-ray data from
the {\it Chandra X-ray Observatory} and SZE data from the BIMA/OVRO
SZE imaging project, which uses the Berkeley-Illinois-Maryland
Association (BIMA) and Owens Valley Radio Observatory (OVRO)
interferometers to image the SZE.  The BIMA/OVRO SZE imaging project
has generally pursued hot clusters (published $T_X \gtrsim 5$~keV)
which by inference should be massive and have a strong SZE.
Properties of the cluster sample, including full cluster names,
redshifts, positions, and \chandra\ and BIMA/OVRO observing
information, are listed in Table~\ref{tab:data}.

The X-ray data were reduced using the Chandra Interactive Analysis of
Observations (CIAO) 3.2 software package with CALDB 3.1.  Details of
the data reduction procedure are given in \citet{bonamente2004} and
\citet{bonamente2006}; briefly, CIAO and the CALDB are used to apply
corrections for charge transfer inefficiency (CTI) and buildup of a
contaminant on the optical blocking filter.  The observations are
filtered to contain only photons with energies between 0.7 and
7.0~keV, and to remove contamination from solar flares.  The data are
then binned into images for each chip relevant to the observation, and
X-ray backgrounds are extracted from appropriate cluster-free regions
of the chips \citep{bonamente2004}.  Exposure maps are also created
with CIAO, and all spatial analysis is performed on the binned images
and exposure maps.  Cluster spectra are extracted from a circular
region centered on the cluster containing roughly 95\% of the cluster
counts, again restricted to the energy range 0.7---7.0~keV.  Spectra
are grouped so that Gaussian statistics can be used in the spectral
fitting, and a background spectrum is also extracted using the same
background regions used in the image analysis.  The software package
XSPEC \citep{arnaud1996} is used to fit a model to the spectrum and
determine the X-ray spectroscopic temperature and metallicity of the
cluster.

Interferometric radio observations of the cluster SZE were performed
at the Berkeley-Illinois-Maryland Association observatory (BIMA) and
at the Owens Valley Radio Observatory (OVRO).  The millimeter-wave
arrays were equipped with 26-36 GHz receivers for the SZE observations
\citep{carlstrom1996}.  Most of the OVRO and BIMA telescopes were
placed in a compact configuration to provide sensitivity on angular
scales subtended by distant clusters (typically $\sim 1'$) and a few
telescopes were placed at longer baselines for simultaneous point
source imaging \citep{reese2002}.

The SZE data consist of the position in the Fourier domain ($u$-$v$
plane) and the visibilities --- the real and imaginary Fourier
component pairs as functions of $u$ and $v$, which are the Fourier
conjugate variables to right ascension and declination.  The effective
resolution of the interferometer, the synthesized beam, depends on the
$u$-$v$ coverage and is therefore a function of the array
configuration and source position. A typical size for the synthesized
beam for the short baseline data is $\sim 1^\prime$.  Details of the
SZE data reduction can be found in \citet{grego2001} and
\citet{reese2002}.  Briefly, the SZE data were reduced using the
MIRIAD \citep{sault1995} and MMA \citep{scoville1993} software
packages. Absolute flux calibration was performed using Mars
observations adopting the brightness temperature from the
\citet{rudy1987} Mars model. The gain was monitored with observations
of bright radio point sources, and remained stable at the 1\% level
over a period of months.  Data were excised when one telescope was
shadowed by another, when cluster observations were not bracketed by
two phase calibrators, when there were anomalous changes in the
instrumental response between calibrator observations, or when there
was spurious correlation.

Images were made with the DIFMAP software package \citep{pearson1994}
to inspect the data quality and, using only long baseline data, to
identify and fix the positions of radio point sources.  The point
source fluxes are included as free parameters in the model fitting,
using the same methodology as \citet{reese2002}.

\section{Analysis Methods and Modeling}
\label{sec:methods}

\subsection{Cluster Density Models}
\label{subsec:density_models}

The isothermal $\beta$-model has frequently been used in the analysis of 
X-ray and SZE cluster images \citep{cavaliere1976,
cavaliere1978,jones1984,elbaz1995,grego2001,reese2002,ettori2004}. 
The 3-dimensional electron number density is given by
\begin{equation}
n_e({\mathbf{r}}) = \no \left ( 1 + \frac{r^2}{r_c^2} \right )^{-3\beta/2},
\label{eq:single_beta}
\end{equation}
where $n_e$ is the electron number density, $r$ is the radius from the
center of the cluster, $r_c$ is the core radius of the intracluster
medium (ICM), and $\beta$ is a power law index.  A convenient feature of
the isothermal $\beta$-model is that the X-ray surface brightness and
SZE decrement profiles take simple analytic forms
\begin{eqnarray}
S_x & = & \Xo \left ( 1 + \frac{\theta^2}{\theta_c^2} \right
        )^{(1-6\beta)/2}, \label{eq:easy_x_signal} \\
\Delta T & = & \dTo \left ( 1 + \frac{\theta^2}{\theta_c^2} \right)
^{(1-3\beta)/2}, \label{eq:easy_sz_signal}
\end{eqnarray}
where \Xo\ is the central X-ray surface brightness, \dTo\ is the
central thermodynamic SZE temperature decrement/increment, and
$\theta_c$ is the angular core radius of the cluster.  

However, recent deep X-ray Chandra observations and numerical
simulations indicate that the $\beta$-model is not a good description
in the outskirts ($r>1$---$1.5\rtfh$) of clusters
\citep[][]{borgani2004,vikhlinin2005b}; to avoid biases associated
with this effect, we compute masses enclosed within \rtfh, the radius
at which the mean enclosed mass density is equal to $2500 \rhocrit$.
Results are not extrapolated beyond this radius. Incidentally, \rtfh\
is also the outer limiting radius at which both the \chandra\
\citep[e.g.,][]{allen2004} and BIMA/OVRO data \citep{grego2001}
provide strong constraints on the ICM model.

In some clusters the isothermal $\beta$ model fails to provide a good
description of the X-ray surface brightness observed in the cluster
core.  This is the case, for instance, in highly relaxed clusters with
sharply peaked central X-ray emission.  We have therefore developed two
extensions of the isothermal $\beta$-model to overcome this
limitation; we describe these new models and their application to the
X-ray and SZE data below.

\subsubsection{The 100~kpc-Cut Model}
\label{subsubsec:100kpc_cut_mod}

First we consider a single isothermal $\beta$-model fit to the X-ray
data with the central 100~kpc excised.  The 100~kpc radius is a good
compromise, as it is large enough to exclude the cooling region in
cool-core clusters while keeping a sufficient number of photons to
enable the mass modeling.  The X-ray spectroscopic temperature is also
determined using photons extracted from a radial shell between 100~kpc
and \rtfh. This is referred to as the 100~kpc-cut model in the
remainder of this work.

There is no simple way to excise the central 100~kpc from the
interferometric SZE data, because these data are fit in the Fourier
plane.  We therefore fit the entire SZE dataset, while using the X-ray
spectroscopic temperature from the 100~kpc-cut model. The inclusion of
the dense core in the SZE data should have little effect on the
derived cluster parameters, because the SZE as a probe of pressure is
less sensitive to behavior in the dense core than are the X-ray data.
It should also have little effect on the best-fit shape parameters,
$\theta_c$ and $\beta$, because these fits are driven mainly by the
X-ray data.  The SZE data therefore mainly constrain the overall
normalization of the SZE signal, which is insensitive to the details
of the cluster core \citep[e.g.,][]{nagai2005}.

\subsubsection{The Non-isothermal Double $\beta$-Model}
\label{subsubsec:non_isoth_mod}

We also develop a more sophisticated cluster plasma model that takes
into account temperature profiles.  A motivation for considering this
model is to assess the biases arising from the isothermal assumption
and the effects of the core exclusion in the previous model.

The model uses a second $\beta$-model component to describe the
sharply peaked X-ray emission present in the cores of some clusters
\citep{mohr1999a}.  The 3D density profile of the double-$\beta$ model
is expressed by,
\begin{equation}
n_e({\mathbf{r}}) = \no \left[ f\: \!\left ( 1 +
\frac{r^2}{\,r_{c_1}\!\!\!\!^2}\right) ^{-3\beta/2} + (1-f)\:\left (1 +
\frac{r^2}{\,r_{c_2}\!\!\!\!^2}\right)^{-3\beta/2}\right]
\label{eq:double_beta}
\end{equation}
where the two core radii, $r_{c_1}$ and $r_{c_2}$, describe the
narrow, peaked central density component and the broad, shallow outer
density profile, respectively, and $f$ represents the fractional
contribution of the narrow, peaked component to the central density
\no\ ($0\leq f \leq 1$).  This model has enough freedom to
simultaneously fit the X-ray surface brightness in the outer regions
and the central emission excess seen in some clusters.  We set $f=0$
(equivalent of the single $\beta$-model) if the reduced $\chi^2$ for
such a fit is less than 1.5.

The 3D temperature profile is modeled assuming that the ICM is in
hydrostatic equilibrium with a NFW dark matter density distribution
\citep{navarro1996, navarro1997},
\begin{equation}
\frac{\rho_{\mbox{\tiny
DM}}(r)}{\rhocrit(z)}=\frac{\delta_c}{(r/r_s)(1+r/r_s)^2}
\label{eq:nfw}
\end{equation}
where $\rhocrit(z)$ is the critical density of the universe at
redshift $z$, $r_s=\rII/c\;$ is the scale radius, $c$ is the
concentration parameter of the dark matter, and $\delta_c$ is the
characteristic over-density of the halo and is related to the
concentration parameter $c$.  Using the best-fit 3D gas density model,
we solve the hydrostatic equation iteratively for the 3D model
temperature profile.  This 3D temperature profile is then weighted by
the X-ray cooling function and the square of the cluster density,
convolved with the appropriate instrumental response, and integrated
along the line of sight, yielding a projected temperature profile
model.  Finally, the projected temperature profile model is compared
to the observed temperature profile extracted from the \chandra\ data.

Altogether, the model is parametrized by the combination of the ICM
density model, \no, \rcI, \rcII, $\beta$, and $f$; the concentration
parameter of the dark matter, $c$; and the outer radius, \rth.  This
model is referred to as the non-isothermal $\beta$-model in the
remainder of this work.

\subsubsection{The SZE-only Model}
\label{subsubsec:sze_only_mod}

We also adopt a third model, used to fit the SZE spatial data
independently of the X-ray spatial data.  We use the full SZE dataset
with no 100~kpc exclusion, but use the 100~kpc-cut X-ray spectral data
to determine the gas mass fractions.  The SZE spatial data alone do
not have sufficient resolution to provide strong individual
constraints on $\theta_c$ and $\beta$ \citep{grego2001}, so we fix
$\beta$ at the median value of $0.7$ for our sample.  Values of
$\beta$ in our sample are almost entirely in the range 0.5 to 1.0,
with most clustered between 0.6 and 0.8; this range is also consistent
with many independent studies using other cluster samples
\citep[e.g.,][]{jones1999,mohr1999a,ettori2004}.  To test the effect
of changing $\beta$, we have repeated the analysis with $\beta$ fixed
at 0.6 and 0.8, finding that all changes in the parameters are small
relative to the 68\% statistical uncertainties.  We refer to this
model as the SZE-only model in the remainder of this work.

\subsection{Model Fitting and Likelihood Analysis}
\label{subsec:MCMC}
We determine the best-fit values and confidence intervals of the model
parameters using likelihood analysis based on a Markov chain Monte
Carlo (MCMC) method.  We work with log likelihoods for both the
spatial and spectral data.  For spatial fitting to X-ray and SZE
images, the log likelihoods can be written \citep{cash1979,reese2002}
\begin{eqnarray}
\ln(\mathcal{L}_{X-ray})=\sum_i \left[ D_i\ln(M_i) - M_i - \ln(D_i!) \right],
        & & \mbox{for X-ray data (Poisson)}, \label{eq:likex} \\
\ln(\mathcal{L}_{SZE})=\sum_i -\frac{1}{2} \left ( \Delta R^2_i + \Delta I^2_i \right )
        W_i & & \mbox{for SZE data (Gaussian)}, \label{eq:likesz}
\end{eqnarray}
where $M_i$ and $D_i$ are the model prediction and data in pixel $i$
in the X-ray data, while $\Delta R^2_i$ and $\Delta I^2_i$ are the
difference between the real and imaginary components of the SZE data
and model at each point $i$ in the Fourier plane and
$W_i=1/\sigma_i^2$ is a measure of the Gaussian noise. Calculation of
the spectral log likelihood is done with XSPEC.  The spatial and
spectral log likelihoods are then added together to compute the joint
log likelihood.  \citet{bonamente2006} describe the likelihood
analysis for the non-isothermal double $\beta$-model; in that case the
X-ray and SZE central densities were linked and \Da\ was a free
parameter in the fit.  Our approach differs only in that \Da\ is fixed
for an assumed cosmology and the X-ray and SZE central densities are
fit separately so that the individual mass results can be compared.
We therefore describe the likelihood calculation here only in the
context of the 100~kpc-cut and SZE-only models.

In the 100~kpc-cut case, the X-ray model is given by
Eq.~[\ref{eq:easy_x_signal}] plus a constant X-ray background,
$B_x$. The model has five free parameters, including the
two-dimensional cluster position ($x_c$, $y_c$), central surface
brightness \Xo, and the shape parameters $\theta_c$ and $\beta$.  The
background, $B_x$, is determined from the data using a method
described in \citet{bonamente2004} and held constant during the model
fitting.  The model is calculated at each pixel position and then
multiplied by the exposure map before being compared to the
image. X-ray point sources and any obvious substructure are masked
out, and the masked pixels are ignored in the likelihood calculation.
Any pixels within 100~kpc of the cluster center are also ignored.  The
Poisson log likelihood is then computed using Eq.~[\ref{eq:likex}].

The interferometric SZE data are analyzed directly in the Fourier
plane, where the noise characteristics and spatial filtering of the
interferometer are well understood.  In both the 100~kpc-cut and
SZE-only cases, the composite SZE model consists of the isothermal
$\beta$-model and any point sources detected in the field.  The
parameters of interest are the two-dimensional cluster position,
central decrement \dTo, $\theta_c$, and $\beta$ and the positions and
fluxes of the point sources.  The SZE $\beta$-model plus point source
model are constructed in the image plane, multiplied by the primary
beam, and fast Fourier transformed to the \uv\ plane for comparison
with the data.  The Gaussian log likelihood is then calculated using
Eq.~[\ref{eq:likesz}].

X-ray spectral likelihoods are calculated using $\chi^2$ information
from XSPEC.  A photo-absorbed Raymond-Smith thermal plasma model
\citep{raymond1977} is fit to the 0.7---7.0~keV X-ray spectral data,
with HI column density fixed at the value from \citet{dickey1990} and
solar abundances from \citet{feldman1992}.  XSPEC is used to create a
table of $T_e$ and metallicity relative to solar, $Z$, values versus
log likelihood for the spectral model using the relation $\ln \likel =
-\frac{1}{2} \chi^2$.

Best fit values and confidence intervals for model parameters are
determined using a MCMC method which efficiently handles the large
number of parameters involved in the model fitting.  Implementation of
a MCMC method for determining the angular diameter distance to the
cluster A611 from SZE and X-ray data is described in detail in
\citet{bonamente2004}, including choice of parameter support and
convergence and mixing tests.

For the 100~kpc-cut model, we use the likelihood analysis based on the
MCMC method and fit the SZE and X-ray data jointly; $\theta_c$ and
$\beta$ are linked between X-ray and SZE datasets \citep{reese2000},
and the X-ray spectral model is also included.  Since the SZE data do
not constrain $\theta_c$ and $\beta$ well individually, the X-ray
dataset drives the fit to these parameters while \dTo\ is allowed to
find its best-fit value.  The central 100~kpc are excluded from the
X-ray data but not the SZE data.  Since the datasets are independent,
the X-ray and SZE spatial log likelihoods and spectral log likelihood
are added to determine the joint likelihood for this model.

For the SZE-only model, the MCMC method is used to fit the SZE data to
an isothermal $\beta$-model; $\beta$ is fixed at 0.7 while $\theta_c$
and \dTo\ are allowed to find their best fit values.  An X-ray
spectral model with central 100~kpc excluded from the X-ray data is
also included.  Since the SZE and X-ray datasets are independent, the
SZE spatial log likelihood and X-ray spectral log likelihood are added
to determine the joint likelihood for this model.

\subsubsection{Gas Mass, Total Mass and Gas Mass Fraction}
\label{subsubsec:props_from_model}

With the best-fit ICM model and X-ray temperature in hand, it is
straightforward to compute the gas mass and total mass of the cluster.
For the $\beta$-model, the enclosed gas mass is obtained by
integrating the best-fit 3D gas density profile:
\begin{equation}
\Mgas(r) = A \int_{0}^{r/\Da} \left (1+\frac{\theta^2}{\theta_c^2}
\right)^{-3\beta/2}\: \theta^2 d\theta,
\label{eq:mgas_single}
\end{equation}
where $A=4\pi \mu_e \no m_p\, \Da^3$, and $\mu_e$, the mean molecular
weight of the electrons, is determined from the X-ray spectral data.
Assuming the isothermal gas temperature, we can compute the gas mass
independently from both the X-ray and SZE datasets.  For the X-ray
data, the model central electron density \no\ can be expressed
analytically as \citep{birkinshaw1991}:
\begin{equation}
\no = \left( \frac{\Xo \:4\pi (1+z)^4 \:\frac{\mu_H}{\mu_e}\:
  \Gamma(3\beta)}{\LameH \Da \pi^{1/2}\: \Gamma(3\beta-\frac{1}{2})\,
  \theta_c} \right)^{1/2}.
\label{eq:xray_ne0}
\end{equation}
For the SZE, the model central electron density can be expressed as
\citep[e.g.,][]{grego2001}:
\begin{equation}
\no = \left( \frac{\dTo \,m_e c^2 \:\Gamma(\frac{3}{2}\beta)}{f_{(x, T_e)}
  \Tcmb \sigT\, \kB T_e \Da \pi^{1/2} \:\Gamma(\frac{3}{2}\beta -
  \frac{1}{2})\, \theta_c} \right)
\label{eq:sz_ne0}
\end{equation}
For the 100~kpc-cut model, we compute the gas mass using
Eq.~[\ref{eq:mgas_single}] by extrapolating the model into the cluster
centers.  For the non-isothermal $\beta$-model, the gas mass is
obtained by integrating the best fit central density over a
distribution similar to Eq.~[\ref{eq:mgas_single}], but which accounts
for the additional density component.  In this case $\mu_e$ is treated
as a constant, as its value changes by only $0.3\%$ over the radial
range considered.  In the SZE-only analysis, the gas mass is computed
using Eq.~[\ref{eq:mgas_single}] with model central density from
Eq.~[\ref{eq:sz_ne0}].

The total mass, \Mtot, can be obtained by solving the hydrostatic
equilibrium equation as:
\begin{equation}
\Mtot(r)=-\frac{kr^2}{G\mutot m_p} \left( T_e(r)\frac{d n_e(r)}{dr} +
n_e(r) \frac{dT_e(r)}{dr} \right).
\label{eq:mtot_hse_full}
\end{equation}
Under the isothermal assumption, this reduces to the simple analytic
form \citep[e.g.,][]{grego2001}:
\begin{equation}
\Mtot(r)=\frac{3\beta kT_e}{G \mu m_p} \frac{r^3}{r_c^2 + r^2},
\label{eq:mtot_hse_iso}
\end{equation}
which can be used to calculate total masses for both the 100~kpc-cut
and SZE-only models.  For the non-isothermal $\beta$-model, the
temperature derivative of Eq.~[\ref{eq:mtot_hse_full}] is simple to
compute numerically.  We then compute X-ray and SZE gas mass fractions
as $\fg =\Mgas/\Mtot$ for the sample of 38 clusters.

\subsection{Comparison of the Density Models Fit to X-ray Data}
\label{subsec:example_joint}

We now compare the results of the X-ray surface brightness modeling
and the cluster parameters extracted using different ICM models. The
primary goal of this comparison is to assess the effects of the
isothermal assumption and different treatments of the cluster core.

\begin{figure}[tp]
\centerline{
  \includegraphics[scale=0.35]{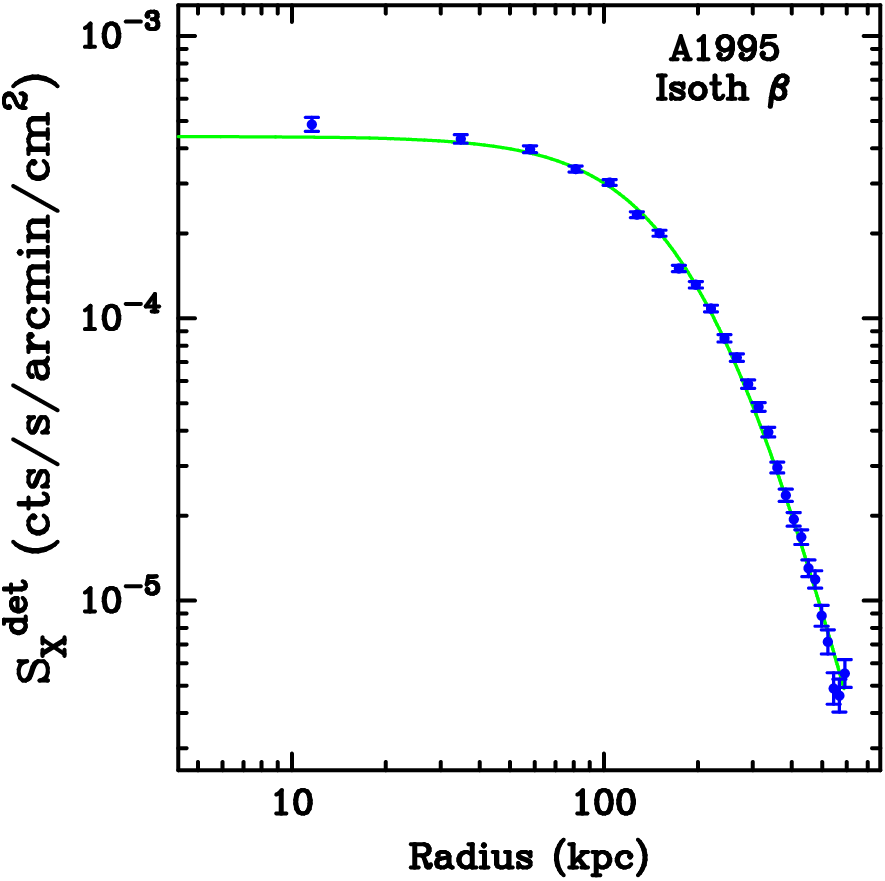}
  \includegraphics[scale=0.35]{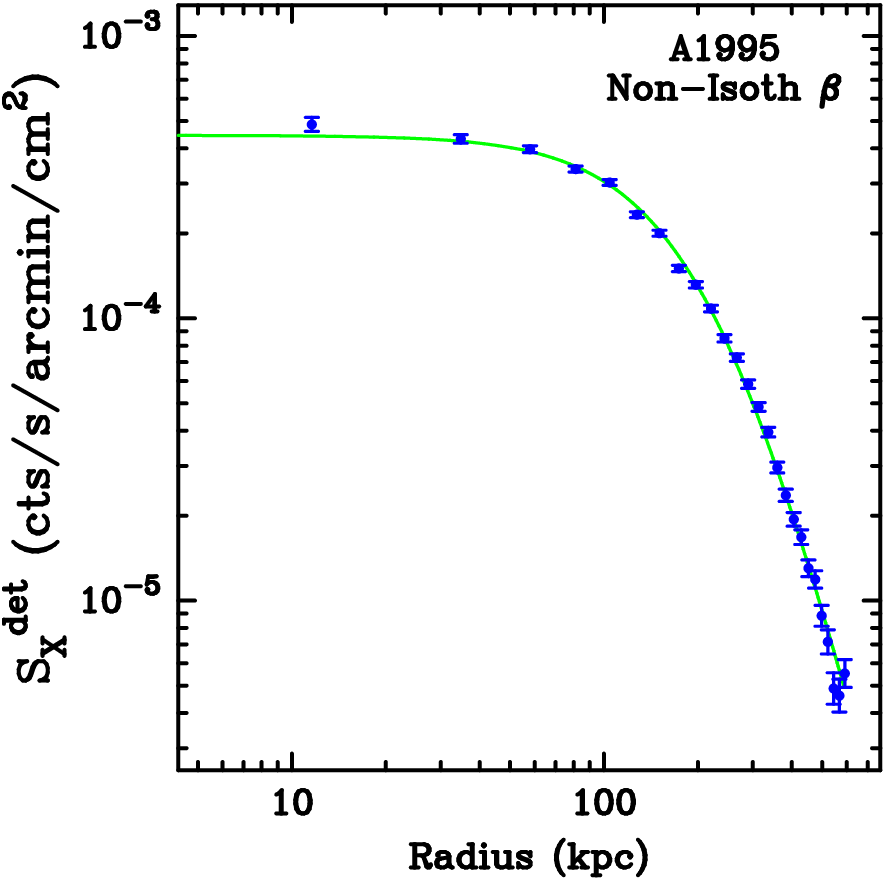}
  \includegraphics[scale=0.35]{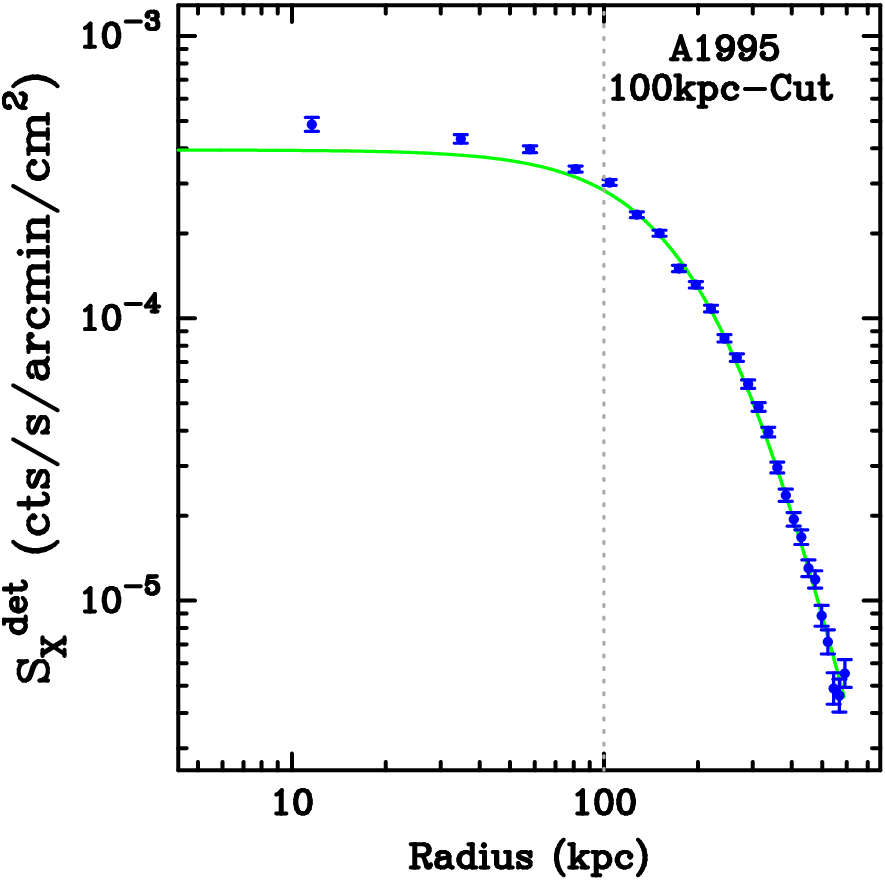}
}
\vspace{0.1in}
\centerline{
  \includegraphics[scale=0.35]{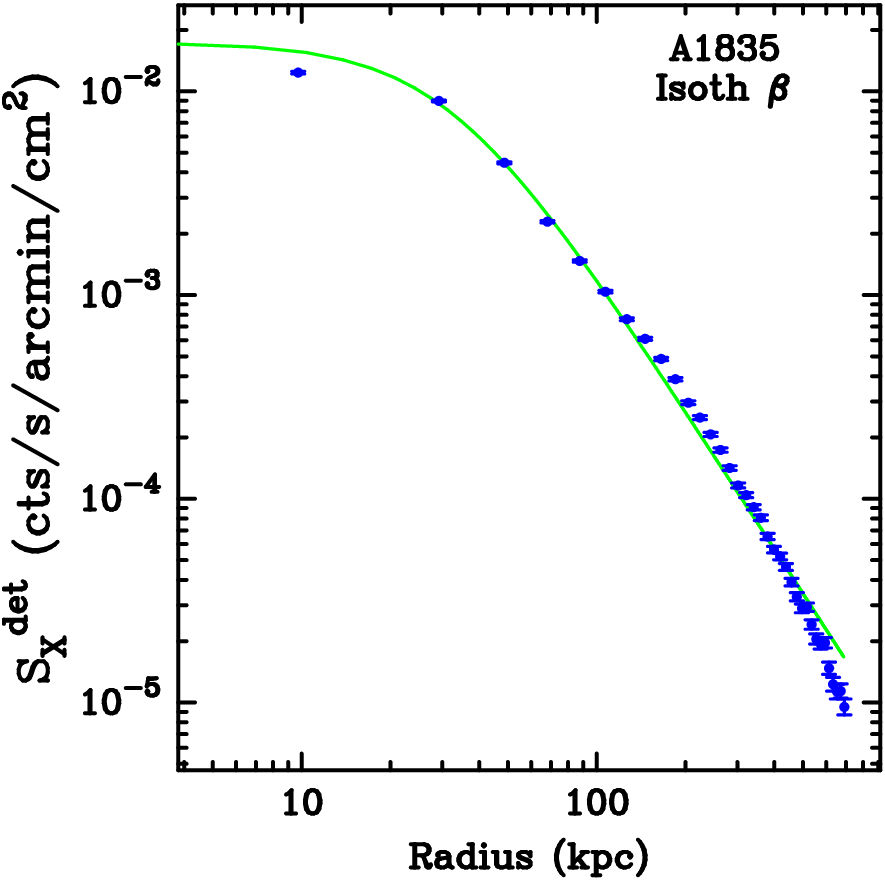}
  \includegraphics[scale=0.35]{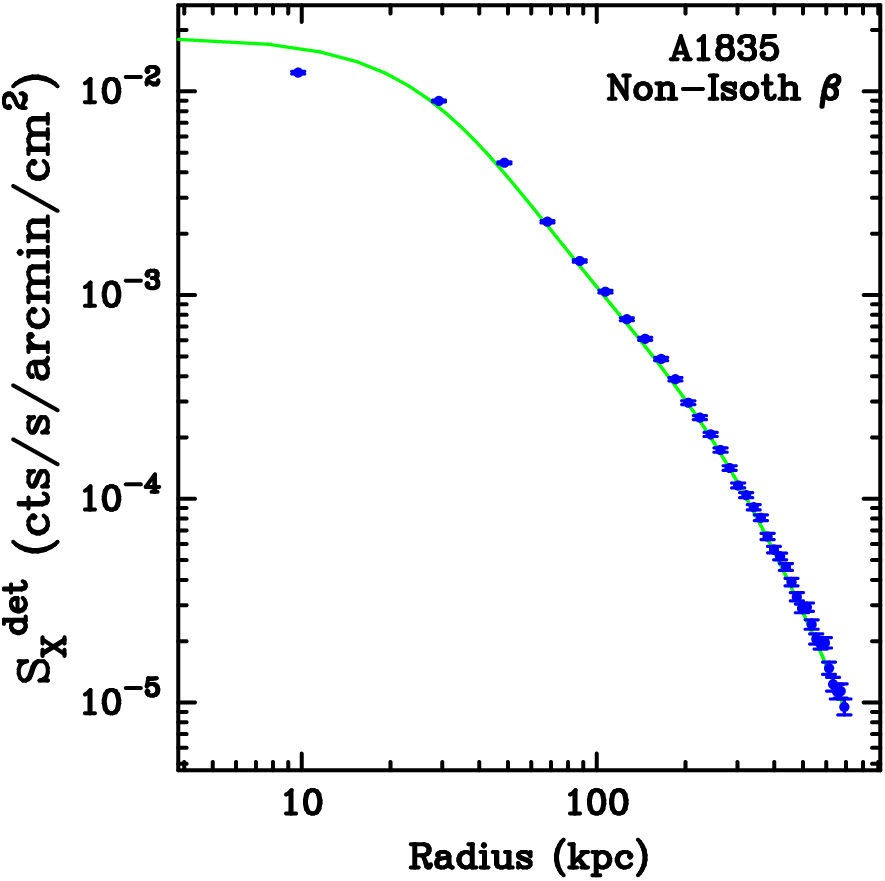}
  \includegraphics[scale=0.35]{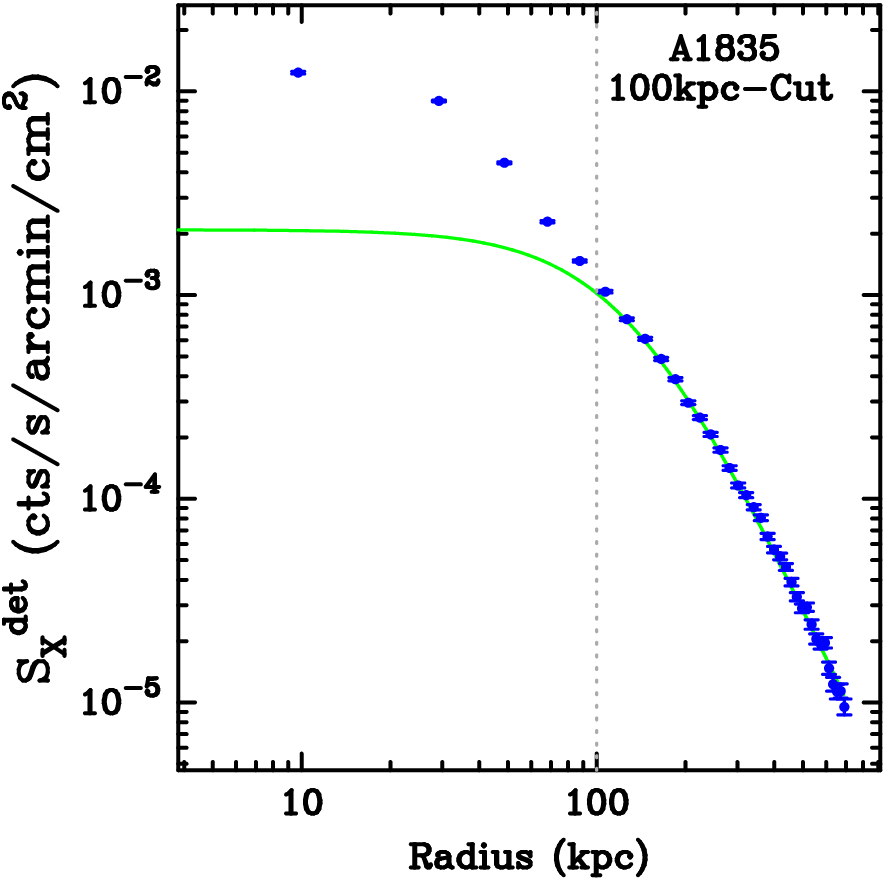}
}
\caption[]{Comparison of X-ray surface brightness profiles for the
  isothermal $\beta$-model, non-isothermal $\beta$-model, and
  100~kpc-cut model as applied to both the non cool-core cluster A1995
  (upper panels) and the cool-core cluster A1835 (lower panels).  The
  dotted vertical line in the 100~kpc-cut panels denotes the 100~kpc
  radius.}
\label{fig:model_cmp}
\end{figure}

Figure~\ref{fig:model_cmp} shows the isothermal $\beta$-model,
non-isothermal $\beta$-model, and 100~kpc-cut model as fit to both the
non cool-core cluster A1995 and the cool-core cluster A1835.  The
X-ray surface brightness profiles are background subtracted and
constructed using concentric annuli centered on the cluster
\citep[e.g.,][]{bonamente2006}.  The model fitting is done to the
entire two-dimensional cluster image.  Table~\ref{tab:fit_param_cmp}
lists the spatial ($\theta_c$ and $\beta$), spectral (isothermal
spectroscopic gas temperature $T_X$), and inferred (\Mgas, \Mtot, \fg)
cluster quantities obtained using these three ICM models.  The radius
\rtfh\ is computed using parameters from the 100~kpc-cut models.
Temperature profiles and data points for the non-isothermal models can
be found in \citet{bonamente2006}, who demonstrate that the
spectroscopic data are well fit by the model temperature profiles.

In the case of A1995 (a non cool-core cluster), we find that the
results are largely insensitive to the chosen model.  The surface
brightness profiles appear well fit by all three models, and the
derived gas masses, total masses, and gas mass fractions are in good
agreement.  Although $\theta_c$ and $\beta$ differ slightly, these
parameters are sufficiently degenerate that the difference has a very
small effect on the derived masses.  These results illustrate that the
simple 100~kpc-cut model works as well as the more sophisticated
non-isothermal double $\beta$-model for the non cool-core clusters.

Analysis of the cool-core cluster A1835, on the other hand, highlights
the importance of a proper treatment of the cluster core.  The most
striking differences are the masses derived from the isothermal
$\beta$-model versus those from the other two models.  This
illustrates the shortcomings of a brute force application of the
isothermal $\beta$-model to cool-core clusters.  The mass
discrepancies can be attributed to a poorer fit to the surface
brightness at $r>100$~kpc; this arises because the fit is driven
mainly by the extremely high signal to noise data in the cluster core.
A1835 and other cool-core clusters tend to have extremely small core
radii and $\beta$ parameters when fit by a single isothermal
$\beta$-model.  We also find that the X-ray spectroscopic temperature,
$T_{\rm X}$, is biased low when the core is included in the
determination of $T_{\rm X}$, which has an additional impact on the
total mass estimate of the cluster.

The models can be quantitatively compared using a goodness of fit
analysis.  Goodness of fit for the X-ray data is assessed using
Monte-Carlo simulations, following \citet{winkler1995}.  For a given
cluster, we construct the best-fit model and compare with the data to
determine the fit statistic from Eq.~[\ref{eq:likex}].  Poisson noise
is then randomly added to the best-fit model, creating $10^4$ new
realizations, and the fit statistic is calculated for each by
comparing them with the original best-fit model.  The fraction of
simulations that give a lower fit statistic than that of the best-fit
model compared to the data is called the ``goodness''
\citep[e.g.,][]{jonker2005}, with values near 0.5 indicating a good
fit to the data and values near zero or one indicating a poor fit.

In the case of A1995, all three models (standard isothermal $\beta$,
100~kpc-cut, and non-isothermal) provide acceptable descriptions of
the data, with respective goodness values of 0.416, 0.427, and 0.496.
Acceptable goodness values of 0.547, 0.548, and 0.603, for the same
three models, are also found for A1835.  However, the goodness value
for the standard isothermal $\beta$-model may be biased by the
extremely high count rate in the core of this cluster.  To investigate
this, we use the best-fit parameters from the standard $\beta$-model,
but test them with the 100~kpc-cut dataset.  The result is a goodness
of zero; the standard $\beta$-model is a poor description of the
cluster gas distribution between 100~kpc and \rtfh.  Tests on clusters
in our sample indicate that 90---95\% of the cluster gas mass within
\rtfh\ lies beyond a 100~kpc radius \citep{laroque2005}.  It is
therefore critical that the model give a good description of the data
between 100~kpc and \rtfh, which the standard isothermal $\beta$-model
fails to do.  By contrast, when the best-fit non-isothermal
$\beta$-model parameters are fit to the data between 100~kpc and
\rtfh, the goodness is 0.776, implying a much better description of
the data in this region.

\subsection{Testing the Model with Simulations}
\label{subsec:example_simulations}
 
The ability of the 100~kpc-cut model to recover the gas mass can be
further assessed by fitting the model to projected data from simulated
clusters with precisely known gas masses.  This test is carried out
using cosmological simulations of galaxy clusters generated with the
Adaptive Refinement Tree (ART) N-body+gasdynamics code
\citep{kravtsov1999,kravtsov2002}. These simulations include radiative
cooling and UV heating, star formation, and stellar feedback processes
in addition to the standard gas dynamics.  Mock \chandra\ observations
of these simulated clusters are performed by using the true gas
density, temperature and metallicity in each gas cell along with the
MEKAL spectral emissivity code \citep{mewe1985} to generate an
energy-dependent X-ray flux between 0.1 and 10~keV.  This flux is then
projected along the line of sight, converted to a photon count rate
appropriate for $z=0.01$, and convolved with an instrumental response
simulating that of ACIS-I between 0.7 and 7.0~keV.  This generates a
counts image similar to the \chandra\ level 2 event files described in
\S\ref{sec:data}.  The spectroscopic temperature over a region of
interest is determined using an absorbed MEKAL model in XSPEC
(Nagai~\etall, in preparation).  Temperatures derived using the MEKAL
model differ by at most a few percent from those derived from the
relativistic Raymond-Smith model used in our analysis.  These
temperatures should therefore well approximate the measured \chandra\
temperatures for the clusters in our sample.

We have fit $\beta$-models to the counts images described above for
four different simulated clusters, spanning a range in mass from
$8.7\times10^{13}$ to $6.0\times10^{14}\,\Msun$ at \rtfh\, and
including both relaxed and unrelaxed clusters.  Three fits are
performed for each cluster; one each to projections along three
perpendicular axes.  The likelihood analysis follows the methods of
\S\ref{subsec:MCMC}, and the central 100~kpc of each simulated cluster
are excluded from the fit.  Any obvious gas clumps are also removed
from the mock images prior to fitting.  A spectroscopic temperature is
measured between 100~kpc and \rtfh, and the masses are determined at
\rtfh.  Goodness values for 100~kpc-cut fits to the simulated data are
consistently in the range 0.35 to 0.65, implying an acceptable
description of the data.

\begin{figure}[tp]
\centerline{
  \includegraphics[scale=0.7]{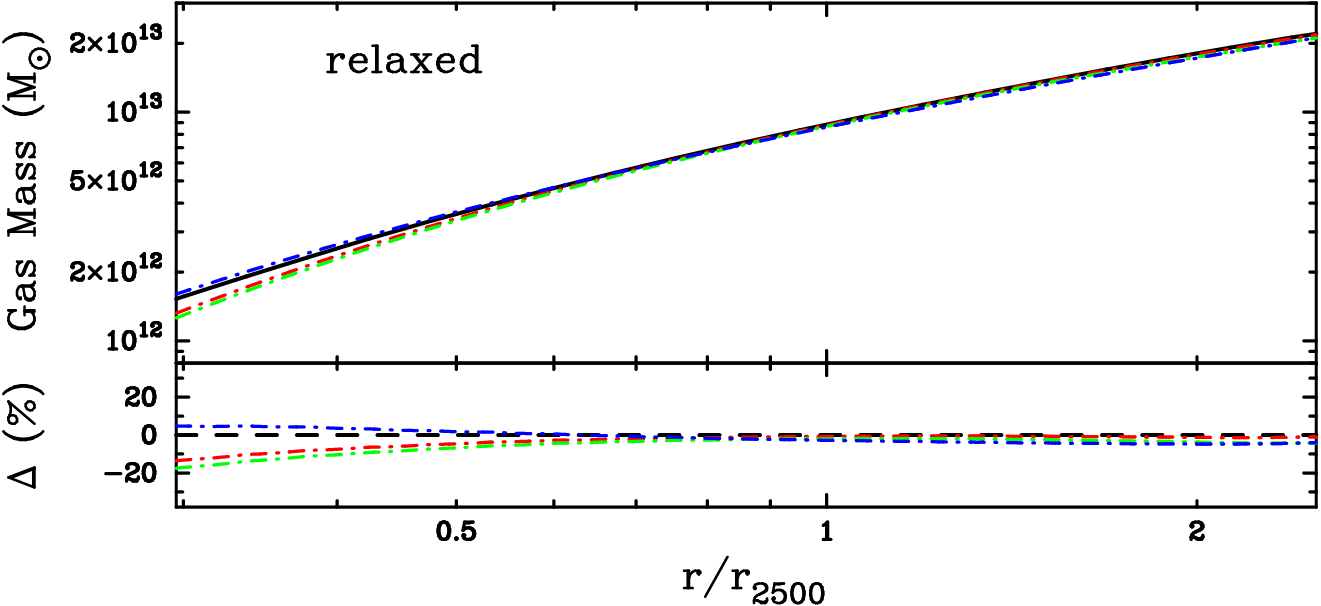} }
\vspace{0.1in}
\centerline{
  \includegraphics[scale=0.7]{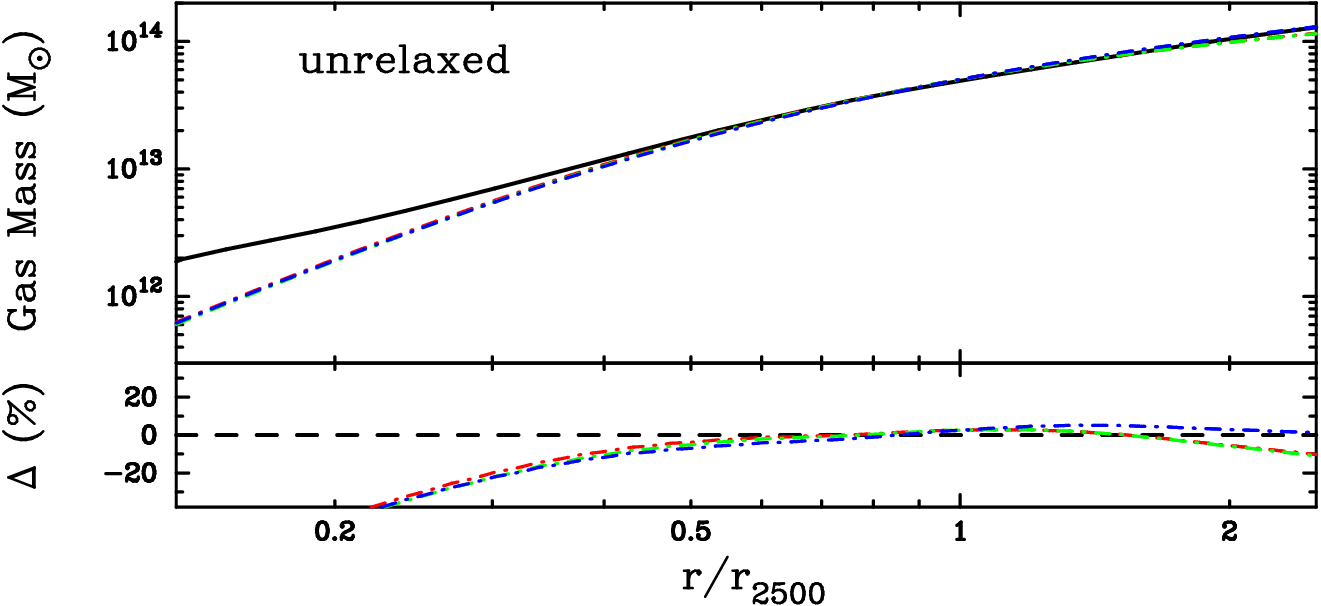}
}
\caption{Ability of the 100~kpc-cut model to recover the gas mass at
  \rtfh\ in two simulated clusters, one highly relaxed (top panel) and
  one highly unrelaxed (bottom panel).  The upper part of each panel
  shows the true gas mass profile (dark solid line) with dash-dotted
  lines showing profiles recovered from fitting the model to $x$, $y$,
  and $z$ projections.  The lower part of each panel shows the
  fractional deviation of recovered mass from true mass as a function
  of radius.}
\label{fig:simul_masses}
\end{figure}

Figure~\ref{fig:simul_masses} shows the mass results for both a highly
relaxed and a highly unrelaxed cluster.  
In both relaxed and unrelaxed clusters, the gas mass is recovered to
better than 10\% between 0.5\rtfh\ and 2\rtfh, and to better than 5\%
within \rtfh\ for the twelve different fits.  This shows that the
simple isothermal $\beta$-model with central 100~kpc excluded does an
excellent job of recovering the true three-dimensional gas mass
distribution from the two-dimensional projected data at radii
encompassing \rtfh\ in simulated clusters.

\section{Additional Sources of Uncertainty}
\label{subsec:systematics}

Before presenting the results, we discuss possible sources of both
statistical and systematic uncertainty in our measurements and
estimate their effects on the X-ray and SZE derived gas mass fraction
in clusters.  \citet{bonamente2006} address sources of statistical and
systematic uncertainty in the X-ray and SZE measurements as they apply
to calculation of the angular diameter distance \Da.  Some are found
to have a negligible effect on our measurements, such as uncertainties
in the assumed Galactic $N_H$ used to calculate spectroscopic
temperatures, and interference associated with other anisotropies in
the CMB.  However, we do review non-negligible effects discussed in
\citet{bonamente2006} in the context of the \fg\ calculations, and we
address additional uncertainties unique to the \fg\ analysis.  The
effect of these uncertainties on the gas mass fractions is summarized
in Table~\ref{tab:systematics}.  The effects are quantified below in
terms of their impact on the gas mass fraction for individual
clusters.  Note, however, that in mean results for the sample
presented in \S\ref{subsec:results}, statistical uncertainties average
down by a factor of one over square root of the sample size, while
systematic uncertainties do {\it not} average down.

\subsection{Selection Effects}
\label{subsubsec:systematics_selection}

Clusters chosen for SZE observation are generally selected on the
basis of X-ray luminosity, $L_x$.  Selecting by X-ray brightness could
bias the SZE sample toward elliptical clusters elongated along the
line of sight; selecting instead by luminosity alleviates this
problem.  However, selecting by $L_x$ may introduce other biases
because of the preferential selection of cool-core clusters or
clusters undergoing major merging events.  Another SZE selection bias
involves point sources; preferentially selecting clusters without
bright ($\gtrsim 20$~mJy) point sources in the field may bias our
sample toward non cool-core clusters, or toward clusters at higher
redshifts whose properties might differ from nearby clusters
\citep[e.g.,][]{grego2001,reese2002}.  X-ray selection biases could
also be a factor, as we only include clusters which can be fit with
the 100~kpc-cut and non-isothermal double $\beta$-models.  Three
clusters, CL\,J0152.7$-$1357, Abell~520, and MACS\,J0717.5+3745, are
not included in our sample due to their highly irregular morphologies.

We assess the bias associated with sample selection by attempting to
construct a subsample largely free of these effects.  The subsample
consists of clusters from the BCS catalog \citep{ebeling1998}, which
is compiled with data from a single instrument (\rosat) and has a
well-defined flux limit of $4.4\times10^{-12}$~erg~cm$^{-2}$~s$^{-1}$
in the \rosat\ 0.1---2.4~keV band.  The subsample contains only
clusters with redshifts $z\leq0.3$ to minimize biases associated with
cluster redshift evolution.  To reduce selection effects associated
with line-of-sight elongation, we impose a flux cut of
$6.6\times10^{-12}$~erg~cm$^{-2}$~s$^{-1}$ on the subsample, which is
50\% higher than the BCS flux limit.  We further select clusters with
intrinsic luminosities in the range $0.35 \leq L_x \leq
1.5\times10^{45}$~erg~s$^{-1}$ in the 2---10~keV \chandra\ band.  This
eliminates clusters in the extreme high and low luminosity ranges of
our sample, which may have unusual dynamical properties.  The final
subsample consists of the ten clusters A267, A665, A773, Zw3146,
A1413, A1914, A2204, A2218, A2261, and R2129 (A1413 and R2129 are not
included when the SZE-only model is considered, for reasons discussed
in \S\ref{subsec:results}).

To quantify the effects of selection biases, we calculate the weighted
mean \fg\ values using the 100~kpc-cut, non-isothermal, and SZE-only
model results (see \S\ref{subsec:results}) for both the full cluster
sample and the unbiased subsample.  We find that the mean \fg\ changes
by less than 3\% between the full and unbiased samples, for all three
of these models. We therefore conclude that selection effects do not
appear to have a substantial impact on the gas mass fraction or
cosmological results.

\subsection{Kinetic SZE} 
\label{subsubsec:systematics_cmb_ksz}

The kinetic SZE is a distortion in the CMB spectrum caused by the
peculiar velocity of a cluster along the line of sight.
\citet{reese2002} report that, for a cluster with $T_x=8.0$~keV and
with typical line of sight peculiar velocity of 300~km~s$^{-1}$
\citep{watkins1997, colberg2000}, the kinetic SZE would be 4\% of the
thermal SZE for our 30 GHz observations.  This effect can be positive
or negative depending on whether the cluster is moving toward us or
away from us.  Since the SZE-derived \fg\ is linearly proportional to
the SZE decrement, the kinetic SZE introduces a $\pm 4\%$ additional
statistical uncertainty into the \fg\ measurement of each cluster.

\subsection{Radio Point Source Contamination}
\label{subsubsec:systematics_ptsrc}

Point sources unaccounted for in the SZE data can cause the decrement
to be ``filled in'' by the point source flux, diminishing the
magnitude of the SZE signal.  It is also possible to overestimate the
magnitude of the decrement if the point source is in a negative beam
sidelobe.  Point sources are identified in the data using images from
DIFMAP.  All cluster fields are also cross-referenced with 1.4~GHz
maps from the NRAO VLA Sky Survey (NVSS) \citep{condon1998}, which
occasionally allow us to distinguish faint point sources with flat or
gently declining spectra from the noise in the 30~GHz maps.  For cases
in which NVSS shows a bright source that does not show up at 30~GHz,
we place a point source model at the NVSS position and marginalize
over its flux.  Strongly inverted sources with 30~GHz fluxes near the
noise limit would not show up in NVSS, but should be rare.
\citet{cooray1998a} report spectral indices for 55 point sources with
fluxes measured between 1.4~GHz and 30~GHz, finding only three with mildly
inverted spectra.

NVSS contains only point sources with fluxes greater than 2~mJy at
1.4~GHz.  To test the effect of point sources too faint to be
cross-correlated with NVSS, \citet{laroque2002} randomly distribute
point sources in the field of the cluster A2163 according to a number
count versus flux density relationship calibrated from 41 clusters
observed with BIMA.  Considering undetected point sources with
intrinsic fluxes between 10~$\mu$Jy and 300~$\mu$Jy at 30~GHz, they
find that the best-fit SZE decrement can change by $\pm 4\%$.

Most clusters have massive central cD galaxies, and many of them are
radio bright.  \citet{laroque2002} include a point source model at the
optical position of the cD galaxy in A2163 and marginalize over its
flux.  They find an increase in magnitude of the best-fit central
decrement of 2\%, and the best-fit point source flux is consistent
with zero.  \citet{reese2002} uses a similar treatment, placing point
source models at the centers of the clusters A697, A2261, and A1413.
The best-fit fluxes are all consistent with zero, and again increase
the magnitude of \dTo\ by only $\sim 2\%$.  We conclude that the
effect of undetected point sources is likely dominated by the
off-center sources, and should add an additional $\sim \pm 4\%$
statistical uncertainty to the SZE gas mass fractions for each
cluster.

\subsection{Instrument Calibration}
\label{subsubsec:systematics_calibration}

The absolute calibration of the ACIS response is uncertain at the 5\%
level in the 0.7---7.0~keV band (http://cxc.harvard.edu/cal), after
CTI corrections and correction for contamination on the optical
blocking filters are applied.  Measurements of $T_X$ are also subject
to $\sim 5\%$ systematic uncertainty, owing mainly to calibration
errors between energy bins.  The \chandra\ calibration uncertainty
therefore translates to $\sim3\%$ uncertainty in the X-ray gas mass
($M_{\mbox{\scriptsize gas}}^{^{\mbox{\scriptsize X-ray}}} \propto
\Xo^{1/2}$) and 5\% in the HSE total mass ($\Mtot \propto T_X$) for a
6\% uncertainty in the X-ray \fg.  Absolute calibration of the
BIMA/OVRO interferometric imaging system is known conservatively to
4\% \citep{reese2002}, which when combined with the uncertainty in
$T_X$ translates to a 6\% systematic uncertainty in the SZE derived
gas mass ($M_{\mbox{\scriptsize gas}}^{^{\mbox{\scriptsize sze}}}
\propto \dTo/T_X$) and 8\% systematic uncertainty in the SZE-derived
\fg.

\subsection{X-ray Background}
\label{subsubsec:systematics_bkg}

X-ray background is usually estimated from regions as far from a
cluster as possible; however, due to space constraints on the chip,
the background regions might contain some amount of cluster flux,
particularly for low-$z$ clusters which subtend a larger solid angle
on the sky.  We therefore test the effect of overestimating the X-ray
background due to cluster contamination on a relatively nearby cool
core cluster, Zw3146.  This cluster is on chip I3, so the original
background is extracted from circular regions on chips I0, I1, and I2
at appropriate distances from the readouts
\citep[e.g.,][]{bonamente2004}.  To assess the effect of
overestimating the background, we move the extraction regions closer
to the cluster until the best-fit background increases by $\sim 25\%$
due to cluster contamination.  This should be seen as an extreme case,
only applicable to a small number of bright, nearby clusters in our
sample.  The spectral fits are redone with these new background
regions, and the resulting temperature and higher background level are
applied in the X-ray/SZE joint fit.

We find that the higher X-ray background modifies the best-fit
parameters, but the magnitude of the effect on individual parameters
is small ($\lesssim 5\%$) with a combined effect on \fg\ of $\sim
-7\%$.  We repeat this same analysis on one of the non cool-core
clusters, A2259, which has a much smaller central surface brightness
to background ratio, and find that the results are much the
same. Given that typical statistical uncertainty in $B_x$ is at most
$\sim 5\%$, much less than the systematic uncertainty of 25\%
considered here, fixing the background in our analysis should have a
negligible effect on the \fg\ results presented in this paper.
Accounting for the fact that some clusters are more susceptible than
others to an overestimated X-ray background, we add a one-sided $+2\%$
systematic uncertainty to the X-ray derived \fg\ results.

\subsection{Effects of Asphericity}
\label{subsubsec:example_ellipticity}

All cluster analysis presented in this paper uses a spherical model.
Here we assess the effects of this assumption on the results presented
in this paper.  To assess this effect, we refit the X-ray data of all
38 clusters using a 100~kpc-cut isothermal $\beta$-model and
accounting for projected ellipticity $\eta$ and position angle of the
clusters on the sky.  We find a mean projected ellipticity $\left <
\eta \right >=0.79$ with rms scatter 0.09, where $\eta$ is the ratio
of a cluster's minor axis to major axis.  These results are consistent
with the previous studies of \citet{mohr1995}.  Accounting for the
inclination angle \citep[e.g.,][]{cooray1998b}, an ellipsoidal cluster
with $\eta = 0.8$ and axis of symmetry inclined at $45\degr$ to the
line of sight can have a derived total mass in error by 12\% and
derived gas mass in error by 6\% if modeled as a sphere.  Extending
this to the more complex triaxial distributions, we conservatively
estimate a statistical uncertainty of 10\% in the X-ray derived \Mgas\
and 15\% in \Mtot.  This translates to a $\sim 20\%$ statistical
uncertainty in the X-ray derived gas mass fractions, given that the
changes in gas mass and total mass can either add or partially cancel
depending on inclination angle and geometry \citep{piffaretti2003}.
The effect on the SZE \fg\ results should be smaller since the SZE
depends linearly on density.  \citet{grego2000} calculate the gas mass
fraction of A370 assuming both prolate and oblate geometries and find
a resulting uncertainty in \fg\ of 20\%.  This cluster has
$\eta=0.64$, however, so we assume a milder 10\% statistical
uncertainty in the SZE \fg\ results for each cluster.

\subsection{Hydrostatic Equilibrium}
\label{subsubsec:hydrostatic}

The HSE condition is a key assumption that enables measurements of the
gravitationally bound mass of clusters from the X-ray and SZE
data. The assumption is that the ICM is supported against
gravitational collapse only by thermal pressure of the hot
intracluster gas.  The hydrostatic mass would therefore underestimate
the true gravitational bound mass of the cluster if non-thermal
pressure support is present.  One form of non-thermal pressure comes
from turbulent gas motions in clusters.  Recent numerical simulations
suggest that this provides about 10\% of the total pressure support in
clusters \citep{faltenbacher2005,rasia2006,lau2006} and similar
results are obtained using {\it XMM-Newton} observations of the Coma
cluster \citep{schuecker2004}.  More direct measurements of the
turbulent gas motions in clusters, however, await measurements of
Doppler broadening of the iron lines with high-resolution X-ray
spectroscopy experiments \citep{inogamov2003}.  Cosmic rays and
magnetic fields may also provide non-thermal pressure support.  While
direct measurements of the non-thermal pressure support may be
difficult, detailed comparisons of the hydrostatic mass with other
mass estimates for the same cluster (e.g., gravitational lensing and
the velocity dispersion of galaxies in clusters) will likely provide
an important check on the hydrostatic mass estimate and potentially
interesting constraints on the non-thermal pressure support in
clusters.  For now we assign a one-sided systematic uncertainty of
$-10\%$ to the total masses which accounts for the assumed
contribution from non-thermal pressure.  We caution, however, that
this estimate is somewhat uncertain and this effect will need to be
investigated in greater detail as more data become available.

\subsection{Isothermal Assumption}
\label{subsubsec:systematics_temperature}

The isothermal assumption can potentially affect the gas mass fraction
measurements and the cosmological constraints through its effects on
both gas mass and total mass.  The X-ray derived gas mass is affected
little by the isothermal assumption, because the X-ray emission has a
very weak dependence on the assumed temperature ($\Lambda \propto
T^{1/2}$).  The SZE derived gas mass is more sensitive to this
assumption since it depends linearly on the cluster gas temperature.
Our choice of a radial range between 100~kpc and $\rtfh$ should
minimize the bias due to the temperature structure for two reasons:
(1) use of the X-ray temperature with the 100~kpc core exclusion
should avoid the most significant bias associated with cool-core
clusters, and (2) focusing on the inner $\rtfh$ should minimize the
bias due to the temperature decline observed in the outskirts of
clusters \citep{vikhlinin2005}.

The total mass derived from the X-ray and SZE data is also affected by
the isothermal assumption.  If the temperature profile deviates from
the isothermal gas distribution, the average temperature depends on
the weighting scheme. For example, the spectroscopic temperature,
$T_{\rm spec}$, may differ by about 10\% from the gas mass-weighted
temperature $T_{\rm mg}$ in the radial range between 70~kpc and
$r_{\mbox{\scriptsize 500}}$ \citep{vikhlinin2005}.  But the
difference is likely smaller in the radial range (100~kpc $<r<\rtfh$)
considered in this study.  We therefore expect that the bias on the
cluster masses and hence our gas mass fraction measurements is not
significant.

The effect of the isothermal assumption can be quantified using the
results of Table~\ref{tab:fgas}, presented in \S\ref{subsec:results}
below.  There we find that the X-ray and SZE weighted mean gas mass
fractions are nearly identical for the non-isothermal model, while the
SZE gas mass fraction is $\sim 5\%$ larger than the X-ray value for
the 100~kpc-cut model.  This hints at a possible systematic bias of
about $-5\%$ in the spectral temperature under the isothermal
assumption, since the SZE \fg\ has an additional $T_X^{-1}$ dependence
relative to the X-ray.  We therefore assign an additional $-5\%$
systematic uncertainty to the gas mass fractions derived from the
X-ray 100~kpc-cut fits, and a $-10\%$ systematic uncertainty to the
gas mass fractions from the SZE 100~kpc-cut fits.

\subsection{Cluster Substructure}
\label{subsubsec:systematics_substructure}

The X-ray derived gas mass fractions are particularly sensitive to any
non-negligible small scale structure in the intracluster gas.  When
gas clumps, it has density higher than the local average, $\left < n_e
\right >$, and the X-ray surface brightness is increased by a factor
of $C\equiv \left< n_e^2 \right> / \left< n_e \right >^2$ relative to
the SZE decrement \citep{nagai2000,mathiesen2001}. Since $n_e\propto
S_x^{1/2}$ for the X-ray data and $n_e\propto \Delta T$ for the SZE
data, X-ray gas mass fractions would be systematically overestimated
by a factor of $C^{1/2}$ with respect to SZE gas mass fractions in the
presence of significant clumping.  Cluster substructure has two
important effects on the X-ray observables.  First, gas clumps can
bias the measured X-ray gas mass high by $C^{1/2}$.  Second, cluster
substructure can bias the average spectroscopic temperature of the
cluster as a whole \citep[e.g.,][]{mathiesen2001,mazzotta2004}.

The \chandra\ observations provide important handles on both of these
effects, because their superb resolution and sensitivity enable
detection of the prominent clumps that contribute significantly to the
X-ray surface brightness and the determination of the average
temperature of the cluster.  We test this using the mock {\it Chandra}
observations of the simulated clusters introduced in
\S\ref{subsec:example_simulations}.  The test shows that the gas mass
is recovered to better than 5\% using the simple 100~kpc-cut model,
indicating that the clumping factor should be small if the most
prominent clump is removed from the image.  These results should also
hold for the high $z$ clusters, as a significant fraction of the low
energy photons from the cool clumps are redshifted out of the
0.7---7~keV band, making the instrument less sensitive to clumping
effects (Nagai~\etall, in preparation).

For the same reasons, the bias on the Chandra spectroscopic
temperature is also small.  Tests on several clusters in our sample
show that not removing bright clumps may bias the temperatures low by
10---15\%, in reasonable agreement with the previous findings
\citep{mathiesen2001,mazzotta2004}.  Further tests, leaving in only
fainter clumps, show a decrease in spectroscopic temperature of only
2---3\%, smaller than the assumed bias due to the isothermal
assumption.

We therefore conclude that there is no evidence that cluster
substructure significantly affects our results when bright spots are
removed from the X-ray images prior to fitting. The agreement in the
mean X-ray and SZE gas mass fractions, particularly for the
non-isothermal model which avoids the bias due to the isothermal
assumption, is consistent with this conclusion (cf.\
\S\ref{subsubsec:clumping}).

\section{Results and Discussion}
\label{sec:results}

\subsection{Gas Mass Fraction Results}
\label{subsec:results}

In Table~\ref{tab:joint_icm_params} we present the X-ray and SZE
measurements of cluster gas mass fractions as determined from the
100~kpc-cut model (\S\ref{subsubsec:100kpc_cut_mod}).  The X-ray and
SZE data are fit jointly, with the cluster core radius and $\beta$
parameter linked between datasets while the central surface brightness
\Xo\ and central SZE decrement \dTo\ are allowed to vary.  The gas
mass fractions are plotted as a function of redshift in
Figure~\ref{fig:fg_vs_z}.  X-ray surface brightness profiles and the
best-fit 100~kpc-cut models are shown in Appendix A for the 38
clusters in our sample.  The X-ray surface brightness profiles are
well described by the isothermal $\beta$-model beyond 100~kpc from the
cluster center.


In Table~\ref{tab:non_isoth_mod_params} we present X-ray and SZE gas
mass fractions determined from the non-isothermal model
(\S\ref{subsubsec:non_isoth_mod}).  The X-ray and SZE data are again
fit jointly with shape parameters linked, while the X-ray and SZE
central density parameters are allowed to vary separately.  The gas
mass fractions are plotted as a function of redshift in
Figure~\ref{fig:fg_vs_z}.  Surface brightness profiles for the
non-isothermal model are shown in \citet{bonamente2006}, and show
consistently good fits to the cluster radial profiles out to \rtfh.

In Table~\ref{tab:sze_masses} we present SZE gas mass fractions
determined from the SZE-only model (\S\ref{subsubsec:sze_only_mod}).
We find that this model constrains \dTo\ and $\theta_c$ individually
in 23 of our 38 clusters, and those 23 are shown in the table.  Gas
mass fractions derived with this method are plotted versus redshift in
Figure~\ref{fig:fg_vs_z}.

The mean X-ray and SZE gas mass fractions for all three models are
summarized in Table \ref{tab:fgas}, where the uncertainties are
statistical followed by systematic at 68\% confidence.  The
statistical uncertainties include the effects of cluster asphericity,
radio point sources, and the kinetic SZE, combined in quadrature;
systematic uncertainties include errors associated with instrument
calibration, X-ray background, and the assumptions of HSE and
isothermality (see \S\ref{subsec:systematics} and
Table~\ref{tab:systematics}).  The results show that the X-ray and SZE
measurements of the mean gas mass fraction in clusters agree to better
than 10\% for the three ICM models considered.

\subsection{Effects of Cluster Cores on Derived Gas Mass Fractions} 
\label{subsubsec:cores_nocores}

The results summarized in Table~\ref{tab:fgas} show that the X-ray and
SZE derived gas mass fractions agree remarkably well as long as care
is taken to account for the effect of the cluster core on the X-ray
emission. The results show that the core can be satisfactorily
accounted for by either ignoring the core in fits to the X-ray data
(the 100~kpc-cut model) or modeling the ICM with the non-isothermal
double-$\beta$ model.

It is also noteworthy that the gas mass fraction results derived from
spatial fits to only the SZE data including the core also agree
remarkably well with the above models. This indicates that the SZE is
not very sensitive to the cluster core, and is in line with
expectations since the SZE is a probe of the ICM pressure.

The effect of cluster cores on the gas mass fractions is further
investigated by comparing results for cool-core and non cool-core
subsamples.  There is a population of clusters in our sample with
bright and sharply peaked cores (cf.\ Appendix A), including A586,
MC0744, Zw3146, A1413, MC1311, A1689, R1347, MS1358, A1835, MC1423,
A2204, A2261, and R2129.  These clusters have central cooling times
(calculated {\it with} the cores included in the X-ray data)
$t_{\mbox{\scriptsize cool}} \approx \frac{3k_B T_X \mu_H}{2\LameH n_e
\mu_{tot}}$ less than 0.5\,\tH, and we refer to this subsample as the
cool-core sample because the sharply peaked X-ray emission is
indicative of strong radiative cooling in the cluster core.  The
results of a comparison of cool-core and non cool-core weighted mean
gas mass fractions are shown in Table~\ref{tab:cores_nocores}, where
in addition to the three main ICM models considered above, we include
results from two additional naive models which make no effort to
compensate for the cluster cores: (1) a no-cut isothermal
$\beta$-model fit jointly to the full X-ray and SZE datasets, and (2)
a spatial model fit to only the SZE data as before, but using the
X-ray temperature derived from all of the X-ray data instead of the
data with central 100~kpc excluded.

Apart from a small $< 10$\% offset in the X-ray results, the derived
gas mass fractions for the 100~kpc-cut and non-isothermal models for
the cool-core and non cool-core clusters are in excellent agreement.
Figure~\ref{fig:fg_vs_z} also demonstrates this agreement, as there is
no systematic offset seen between the cool-core (triangles) and non
cool-core (squares) clusters.  The results for the naive models
clearly show the importance of accounting for the core in the
cool-core subsample; large offsets are found for the derived gas mass
fraction of the cool-core subsample, while the same naive models
provide perfectly consistent results for the non cool-core
subsample. The primary cause of the biased results of the naive models
for the cool-core clusters is the poor spatial fits and the low
$T_X$. A depressed $T_X$ results in a lower derived total cluster mass
as well as a higher SZE gas mass.

\subsection{Constraints on Clumping in the Intracluster Gas}
\label{subsubsec:clumping}

The comparison of the X-ray and SZE gas mass fraction measurements
shows that the results are in good agreement.  These results suggest a
clumping factor of the ICM consistent with zero.  The high resolution
\chandra\ data allow one to remove any visible clumps before the data
are fit, and the agreement between X-ray and SZE gas mass fraction
measurements indicates that any clumps below the noise level of the
data have a negligible effect on the mass measurements.  Further
indications in favor of a negligible clumping factor are discussed in
\S\ref{subsubsec:systematics_substructure}.

\subsection{Comparison with previous studies}
\label{subsec:previous_studies}

Finally, we compare our X-ray and SZE gas mass measurements to the
results from previous studies in the literature. We first compare our
X-ray results with \chandra\ results obtained using a different
analysis technique \citep{allen2002,allen2004}\footnote{Although mean
gas mass fraction results for other cluster samples are available in
the literature
\citep[e.g.,][]{evrard1997,mohr1999a,ettori2003,sanderson2003}, the
detailed comparisons with our results are non-trivial as they are
often calculated at different radii and assume different
cosmologies.}, in which the cluster mass is modeled assuming that the
cluster gas is in hydrostatic equilibrium with a NFW potential and
using a deprojected 3D gas temperature profile. These authors report
mean gas mass fractions of $\fg=0.113\pm0.005$ for a sample of six
highly relaxed clusters \citep{allen2002} and $\fg=0.117\pm0.002$ for
a larger sample of 26 relaxed clusters \citep{allen2004}; these
results are in good agreement with our own.  We also compare the
results of eight clusters we have in common with the \citet{allen2004}
dataset and find that the measurements of individual clusters are
consistent with no systematic offset.  

For the SZE gas mass fractions, we compare results of our SZE-only
model fitting directly with the results of \citet{grego2001}.  There
are many similarities between our cluster sample and analysis method
and those of \citeauthor{grego2001}; for instance, they use BIMA and
OVRO data, and 13 of our 23 clusters have data in common with their
sample.  They also use an isothermal $\beta$-model to analyze the SZE
data.  Their approach also differs somewhat from ours, in that they
(1) use spectroscopic temperatures from the literature (generally from
\asca), and (2) measure the gas mass fraction at $65\arcsec$ from the
cluster center, then scale the results to \rfh\ using relations
calibrated from numerical simulations \citep{evrard1996,evrard1997}.
Despite these differences, \citeauthor{grego2001}\ measure a mean gas
mass fraction of $\fg=0.081^{+0.009}_{-0.011}\,h^{-1}$, which is
equivalent to $0.116^{+0.013}_{-0.016}$ for $h=0.7$, in good agreement
with our SZE derived gas mass fraction (cf.\ Table~\ref{tab:fgas}).
Similarly, if we calculate gas mass fractions from our SZE-only model
at 65\arcsec\ and scale to \rfh\ using the same relations, we obtain a
weighted mean \fg\ of $0.086\pm0.007\,h^{-1}$, in good agreement with
the \citeauthor{grego2001} results. The cluster gas mass fraction
results derived from the implied SZE for nearby clusters obtained from
correlating known clusters with \wmap\ data is also consistent with
our results \citep{afshordi2005}.

\section{Constraints on Cluster Physics and Cosmology}
\label{sec:physics_and_cosmology}

\subsection{The cluster baryon budget}
\label{subsec:baryonbudget}
              
Currently, the relation between \fg\ and the cosmological baryon
fraction \Ob/\Om\ is not accurately known \citep{ettori2006}.
However, since recent CMB studies provide precise constraints on
cosmological parameters \citep[e.g.,][]{spergel2006}, we can use our
\fg\ measurements to obtain constraints on the ratio
\begin{equation}
\eta_{\mbox{\scriptsize gas}} \equiv \frac{\fg}{(\Ob/\Om)}.
\end{equation}
Combining our \fg\ results with the \wmap\ values of \Ob\ and \Om\
($\Ob h^2 = 0.0223\pm^{0.0007}_{0.0009}$, $\Om
h^2=0.127^{+0.007}_{-0.013}$) and the {\it Hubble Space Telescope}
measurement of $h=0.72\pm0.08$ \citep{freedman2001}, we obtain the
constraints on $\eta_{\mbox{\scriptsize gas}}$ given in
Table~\ref{tab:fg_bc_results}, where uncertainties are again
statistical followed by systematic at 68\% confidence.  Since the \fg\
data are calculated assuming \Om=0.3, \Ol=0.7, $h=0.7$, very similar
to the \wmap\ constraints, no significant bias is expected in the
results of Table~\ref{tab:fg_bc_results}.  These results indicate that
the ratio of the gas mass fraction in clusters within \rtfh\ to the
cosmic baryon fraction, $\frac{\fg }{\Ob/\Om }$, is
$0.68^{+0.10}_{-0.16}$ where the range includes statistical and
systematic uncertainties, in accord with the conclusions reached by
\citet{afshordi2005}. This ratio provides a benchmark for simulations
of galaxy clusters. We note that recent numerical simulations that
include radiative cooling and star formation have baryon budgets in
line with our results, although they find that the gas tends to
overcool in the cluster cores \citep{kravtsov2005}.

\subsection{Cosmological Parameters}
\label{subsec:fg_cosmo_2}

Cluster gas mass fractions can provide strong constraints on
cosmological parameters. The normalization of the cluster gas mass
fraction depends sensitively on the ratio of \Ob/\Om\
\citep[e.g.,][]{white1993a,david1995,myers1997,mohr1999a,ettori1999,grego2001,allen2002},
while the redshift evolution of the gas mass fraction depends
sensitively on \Om, \Ol\ and even the equation of dark energy, $w$
\citep{sasaki1996,pen1997,allen2004}. Methods involving the
normalization, however, depend sensitively on {\it a priori} knowledge
of the cluster baryon budget which, as discussed in
\S~\ref{subsec:baryonbudget}, is poorly known. By contrast, methods
using the redshift evolution are powerful in that they are independent
of the cluster baryon budget and even prior knowledge on other
cosmological parameters such as \Ob\ and $h$.

We therefore consider constraints on cosmological parameters from the
redshift evolution of the cluster gas mass fraction. For parameter
estimation, we follow the procedure described in \citet{allen2004},
and fit the data to the model,
\begin{equation}
\label{eq:fg_shape}
f_{\mbox{\scriptsize gas}}^{\mbox{\scriptsize mod}}(z)=
N
\left(\frac{\Da(z,\Omega_{\mbox{\scriptsize $M$}}
^{\mbox{\scriptsize fid}},
\Omega_{\mbox{\scriptsize $\Lambda$}}
^{\mbox{\scriptsize fid}})}
{\Da(z,\Omega_{\mbox{\scriptsize $M$}}
^{\mbox{\scriptsize mod}},
\Omega_{\mbox{\scriptsize $\Lambda$}}
^{\mbox{\scriptsize mod}})}
\right) ^{(3/2 \,[\mbox{\scriptsize{X-ray}}], \,\,1
\,[\mbox{\scriptsize{SZE}}])}
\end{equation}
where we assume a $\Lambda$CDM cosmology with \Om=0.3, \Ol=0.7, as a
fiducial cosmology, indicated by ``fid''.  Using this model, we fit an
arbitrary normalization $N$, and grid the variables \Om\ and \Ol\
between 0 and 1.3 with the {\it WMAP} constraint $\Om +
\Ol=1.003\pm^{0.013}_{0.017}$ \citep{spergel2006}.  We cannot place
useful constraints on \Om\ and \Ol\ from the X-ray or SZE data without
a prior on the total density parameter.

\begin{figure}
\centerline{
  \includegraphics[scale=0.45]{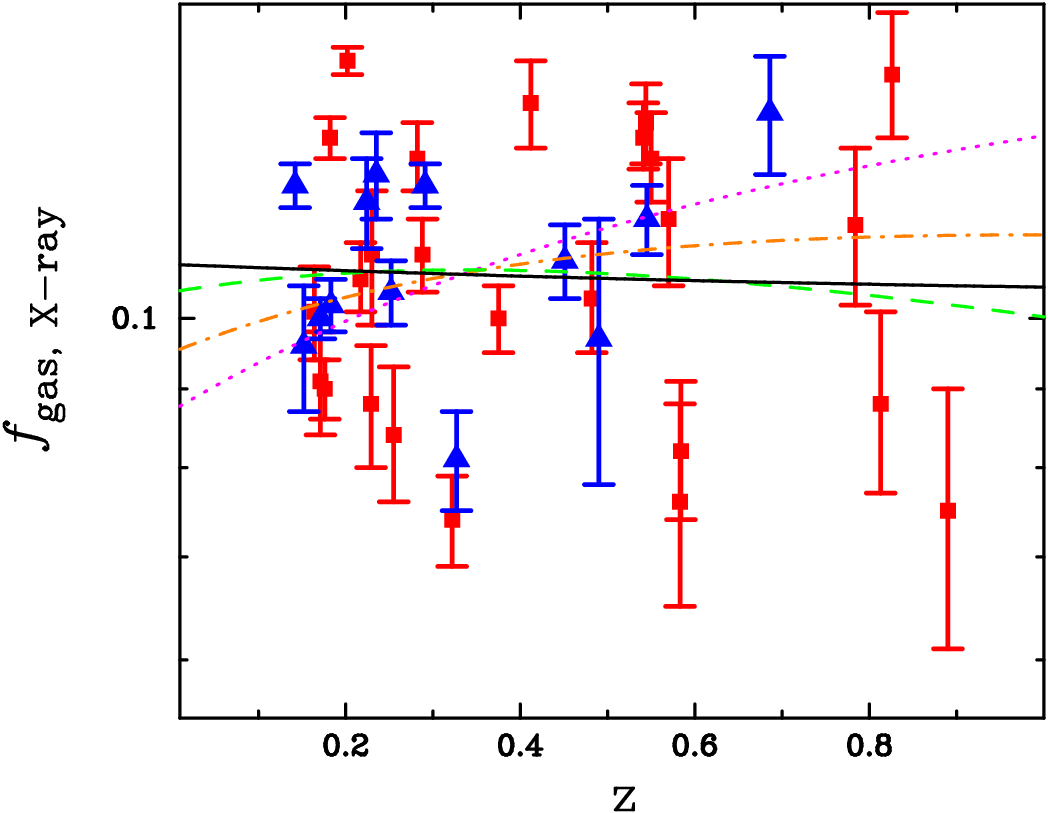}
  \includegraphics[scale=0.45]{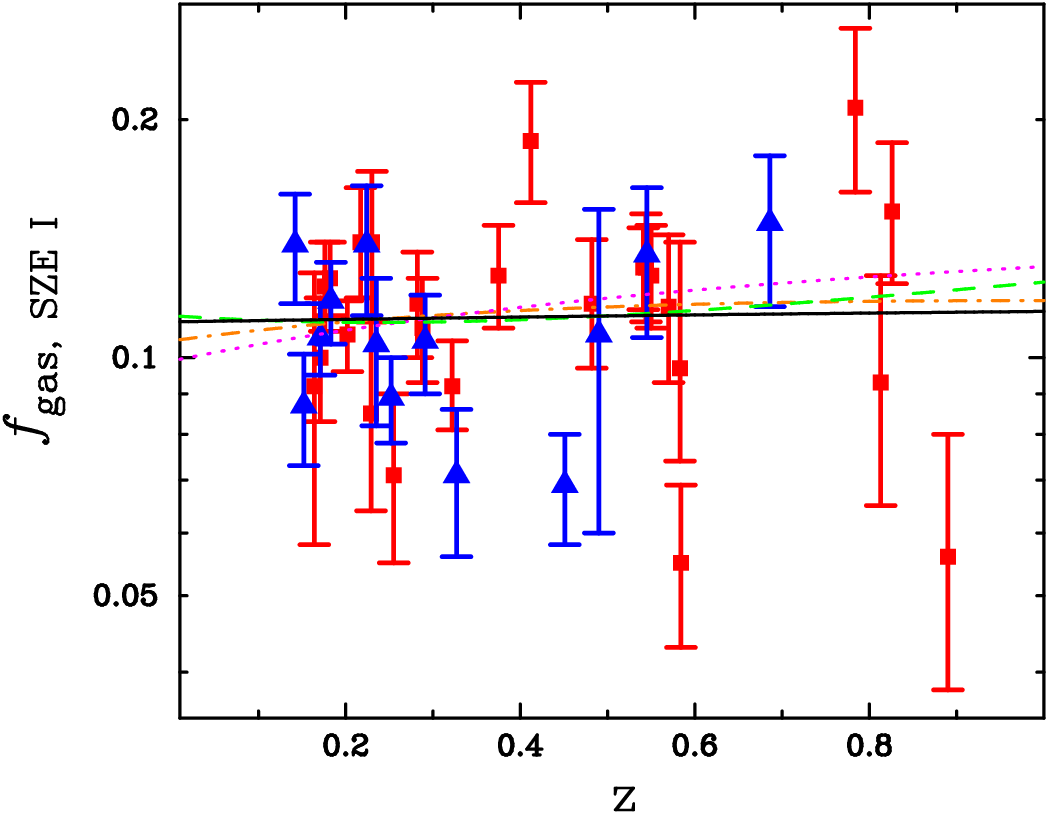}}
\caption[Constraints on Cosmology from Shape of \fg-$z$
  Distribution]{100~kpc-cut derived gas mass fractions versus redshift
  for X-ray (left) and SZE~I (right) data, showing best fit models for
  different cosmologies.  The solid lines (black) show the best fit
  cosmologies when the prior $\Om+\Ol=1.00$ is assumed.  The dashed
  (green) lines show the best fit cosmologies when no prior is
  assumed.  The dash-dotted (orange) lines show the best fit
  normalization when an \Om=0.3, \Ol=0 cosmology is assumed, and the
  dotted (magenta) line shows the best fit normalization for a
  cosmology with \Om=1.0, \Ol=0.}
\label{fig:fg_vs_z_cosmocmp}
\end{figure}

Marginalizing over the \wmap\ total density prior, we find best-fit
parameters of $\Om=0.26^{+0.24}_{-0.15}$, $\Ol=0.74^{+0.15}_{-0.24}$
from the 100~kpc-cut X-ray data and $\Om=0.40^{+0.51}_{-0.26}$ and
$\Ol=0.60^{+0.26}_{-0.51}$ from the isothermal $\beta$ model analysis.
Similarly, the non-isothermal model returns
$\Om=0.40^{+0.28}_{-0.20}$, $\Ol=0.60^{+0.20}_{-0.28}$ from X-ray and
$\Om=0.18^{+0.27}_{-0.17}$, $\Ol=0.82^{+0.17}_{-0.27}$ from SZE.  Our
results are consistent with independent tests, such as those of \wmap,
and inconsistent with a matter dominated universe (\Om=1, \Ol=0)
model. In Figure \ref{fig:fg_vs_z_cosmocmp} we show the \fg\ data, the
best-fit model using the {\it WMAP} prior (solid line), the best-fit
model without the {\it WMAP} prior (unconstrained; dashed line), and
the model corresponding to two other fiducial cosmologies (\Om=1,
\Ol=0: dotted line; \Om=0.3, \Ol=0: dash-dotted line).  Only the
100~kpc-cut results are plotted for simplicity.  It is evident that
the X-ray data disfavors the flat, matter-dominated model.
Interestingly, systematic uncertainty should be negligible compared to
statistical uncertainty with this method, since most of the systematic
uncertainties affect the normalization and do not introduce trends
with redshift.  Cosmological constraints from this method are
therefore expected to improve as the cluster sample size is increased.

\section{Conclusions}
\label{sec:conclusions}

We have presented X-ray and SZE measurements of gas mass fractions for
a sample of 38 massive galaxy clusters using {\it Chandra} X-ray
observations and BIMA/OVRO SZE radio interferometric observations.
Three models for the cluster gas distributions are used in the gas
mass fraction derivations: (1) an isothermal $\beta$-model fit to the
X-ray data at radii beyond 100~kpc and to all of the SZE data, (2) a
non-isothermal double $\beta$-model
fit to all of the X-ray and SZE data, and (3) an isothermal
$\beta$-model fit to the SZE spatial data with $\beta$ fixed at 0.7
and a spectral temperature from the 100~kpc-cut X-ray data.  

The cluster gas distributions are well fit by the non-isothermal model
out to \rtfh.  We have also demonstrated that the simple 100~kpc-cut
isothermal model provides a good description of the intracluster
medium (ICM) outside of the cluster core for objects with a wide range
of morphological properties.

For all three models, the mean gas mass fractions determined from the
X-ray and SZE data are in excellent agreement.  Specifically, with the
isothermal $\beta$-model we find a mean X-ray gas mass fraction
$\fg$(X-ray)$=0.110^{+0.003}_{-0.003}\,^{+0.006}_{-0.018}$ and a mean
SZE gas mass fraction
$\fg$(SZE)$=0.116^{+0.005}_{-0.005}\,^{+0.009}_{-0.026}$, where
uncertainties are statistical followed by systematic at 68\%
confidence.  For the non-isothermal double $\beta$-model, the mean gas
mass fractions are
$\fg$(X-ray)$=0.119^{+0.003}_{-0.003}\,^{+0.007}_{-0.014}$ and
$\fg$(SZE)$=0.121^{+0.005}_{-0.005}\,^{+0.009}_{-0.016}$.  For the
SZE-only model,
$\fg$(SZE)$=0.120^{+0.009}_{-0.009}\,^{+0.009}_{-0.027}$.

It is noteworthy that the gas mass fractions derived from spatial fits
to only the SZE data including the core also agree remarkably well
with the other models. This indicates that the SZE is largely
insensitive to structure in the cluster core. The lack of sensitivity
to the core structure is expected since the SZE is a probe of the gas
pressure; this is promising for extracting cosmology from upcoming
large scale SZE cluster surveys \citep[e.g., see ][]{carlstrom2002}.

To investigate the effect of the cluster cores on the results, we
divided the cluster sample into two subsamples depending on whether or
not the clusters exhibit cool cores. The three ICM models above lead
to consistent gas mass fraction results for each subsample. However,
when the subsamples are fitted with naive ICM models that make no
effort to account for the cluster cores, the resulting gas mass
fractions for the cool-core subsample show large offsets from the
above results, while the results for the non cool-core clusters are
consistent with the above results.  This clearly indicates the
importance of accounting for the cores in clusters that exhibit cool
cores.

Understanding the baryon budget in clusters is one of the most
important aspects of cluster physics.  Our X-ray and SZE measurements
indicate that the ratio of the gas mass fraction within \rtfh\ to the
cosmic baryon fraction, $\frac{\fg }{\Ob/\Om }$, is
$0.68^{+0.10}_{-0.16}$ where the range includes statistical and
systematic uncertainties at 68\% confidence. This result is in accord
with the conclusions reached by \citet{afshordi2005}. The measurement
provides a benchmark for simulations of galaxy clusters.  We note that
recent numerical simulations that include radiative cooling and star
formation have baryon budgets in line with our measurements, although
they find that the gas tends to overcool in the cluster cores
\citep{kravtsov2005}.

Our results also have implications for cosmological constraints based
on measurements of cluster gas mass fractions.  Constraints from the
normalization of the gas mass fraction depend sensitively on {\it a
priori} knowledge of the cluster baryon budget $\eta_{\rm gas}$,
which, as discussed above, is still somewhat uncertain.  However, a
cosmological test based on the redshift evolution of the gas mass
fraction is independent of the cluster baryon budget and even prior
knowledge of cosmological parameters such as \Ob\ and $h$.
Marginalizing over the \wmap\ total density prior, we use this method
and find best-fit parameters of $\Om=0.26^{+0.24}_{-0.15}$,
$\Ol=0.74^{+0.15}_{-0.24}$ and $\Om=0.40^{+0.51}_{-0.26}$,
$\Ol=0.60^{+0.26}_{-0.51}$ from the 100~kpc-cut X-ray and SZE samples,
and $\Om=0.40^{+0.28}_{-0.20}$, $\Ol=0.60^{+0.20}_{-0.28}$ and
$\Om=0.18^{+0.27}_{-0.17}$, $\Ol=0.82^{+0.17}_{-0.27}$ from the
non-isothermal X-ray and SZE samples, respectively.  Our results are
consistent with standard $\Lambda$CDM cosmology, and inconsistent
with \Om=1, \Ol=0.  Interestingly, the systematic uncertainty for this
method is negligible compared to statistical uncertainty, as most
sources of systematic uncertainty are not expected to introduce trends
with redshift.  Cosmological constraints using this method are
therefore expected to improve rapidly as the cluster sample size
increases with upcoming X-ray and SZE cluster surveys.

\acknowledgements 
We are grateful to the late Leon van Speybroeck for compiling
\chandra\ data for a large ensemble of clusters, and to his colleagues
on the \chandra\ project as well, especially C.~Jones and H.~Tananbaum
who generously shared proprietary data with us.  The support of the
BIMA and OVRO staff over many years is also gratefully acknowledged,
including J.R.\ Forster, C.\ Giovanine, R.\ Lawrence, S.\ Padin, R.\
Plambeck, S.\ Scott and D.\ Woody.  We thank C.\ Alexander, K.\ Coble,
A.\ Cooray, L.\ Grego, G.\ Holder, J.\ Hughes, W.\ Holzapfel, A.\
Miller, J.\ Mohr, S.\ Patel and P.\ Whitehouse for their contributions
to the SZE instrumentation, observations and analysis.  Finally, we
thank A.~Kravtsov for many useful discussions.

This work was supported by NASA LTSA grant NAG5-7985 and also in part
by NSF grants PHY-0114422 and AST-0096913, the David and Lucile
Packard Foundation, the McDonnell Foundation, and a MSFC director's
discretionary award.  Research at the Owens Valley Radio Observatory
and the Berkeley-Illinois-Maryland Array was supported by National
Science Foundation grants AST 99-81546 and 02-28963.  SL and DN
acknowledge support from the NASA Graduate Student Researchers
Program.  DN is also supported by the Sherman Fairchild Postdoctoral
Fellowship at Caltech.

\footnotesize
\bibliography{clusters}
\bibliographystyle{apj}

\clearpage

\newpage

\begin{figure}[ht]
\centerline{
\psfig{figure=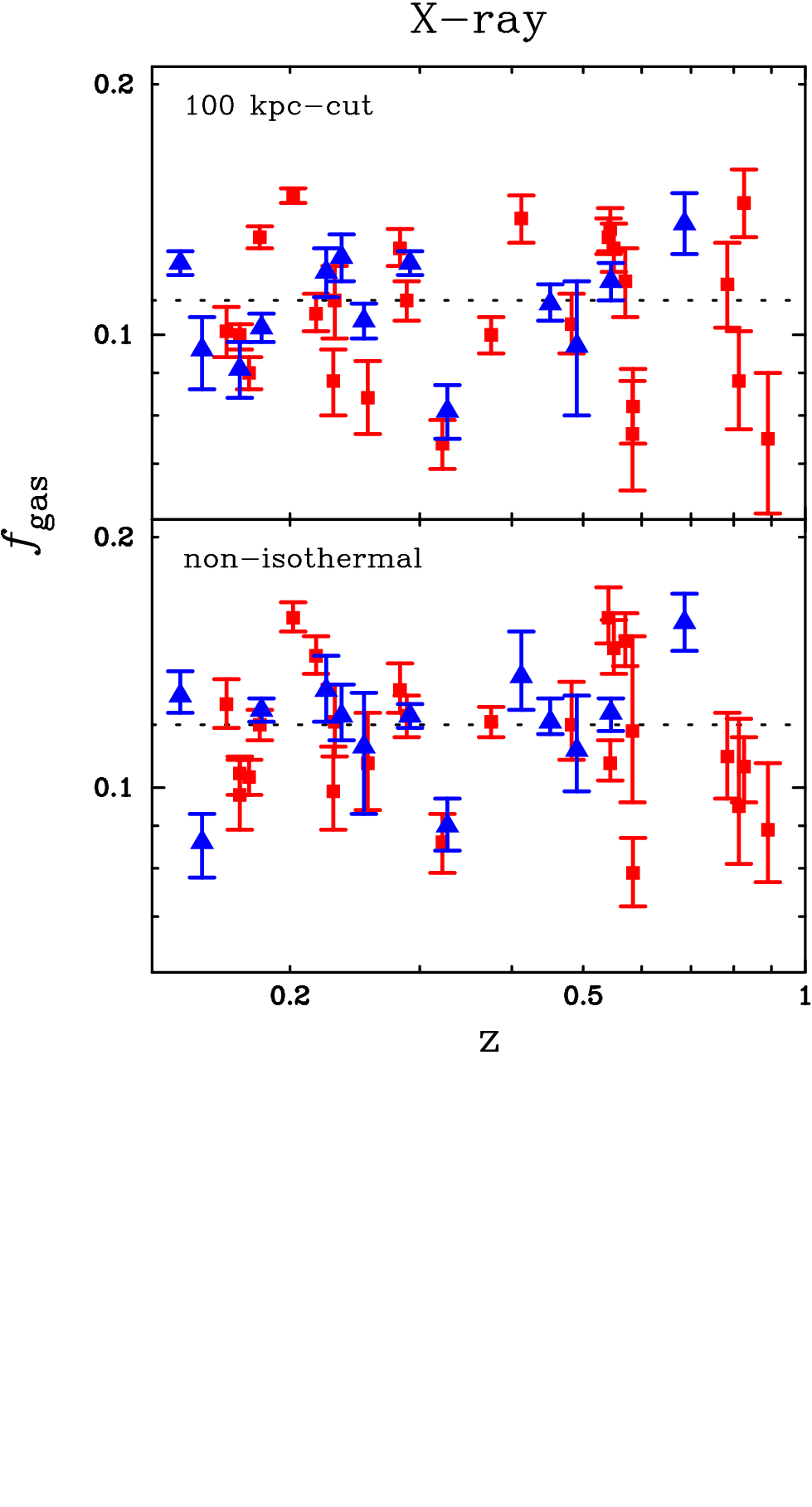,angle=0,width=3.2in}
\psfig{figure=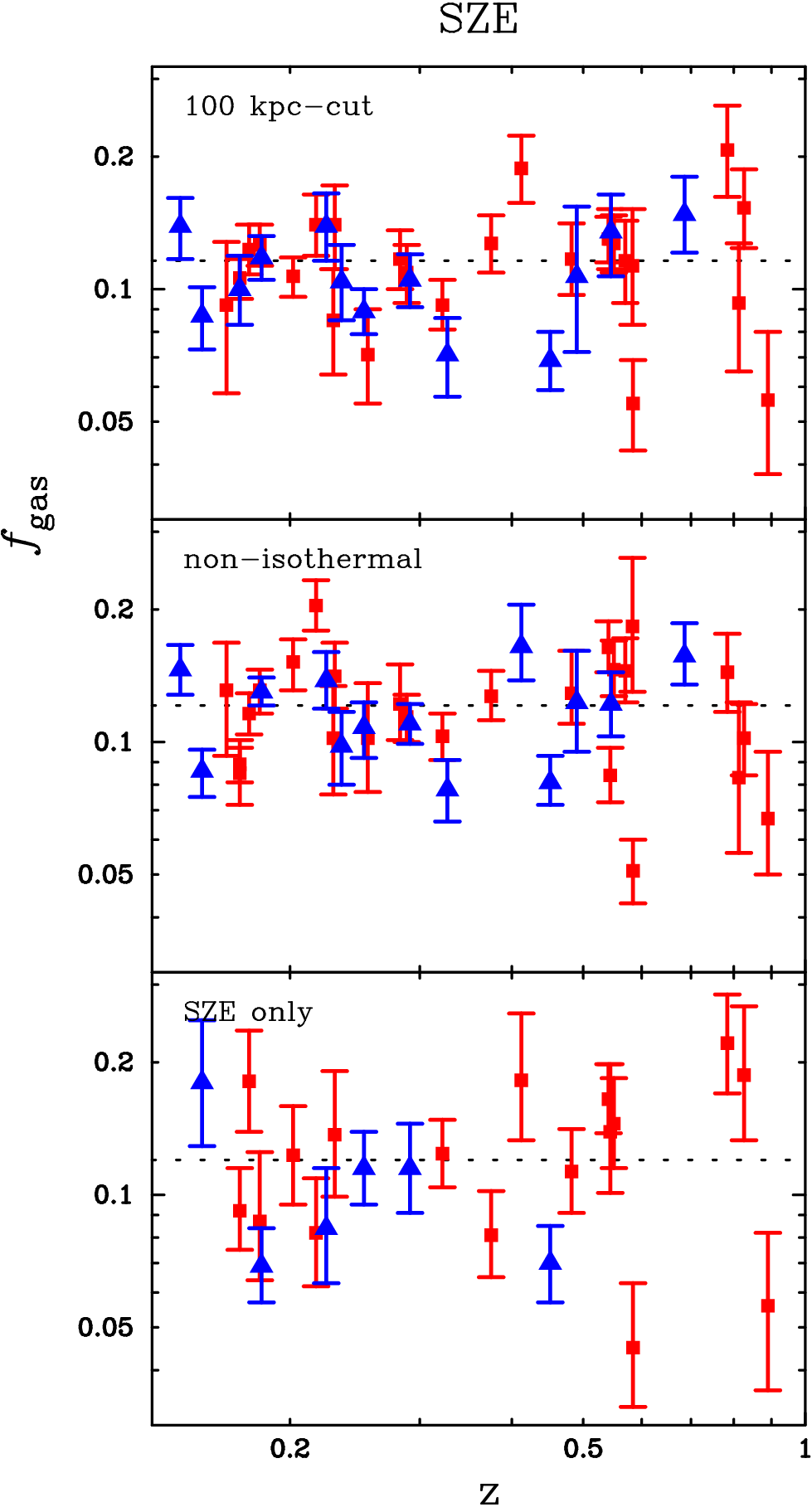,angle=0,width=3.2in}
}
\caption{Gas mass fraction versus redshift derived from the X-ray data
(left panels) and from the SZE data (right panels).  Triangles (blue)
denote cool-core clusters and squares (red) denote non cool-core
clusters.  Dashed lines show the weighted mean \fg\ for each sample.
Error bars are statistical from Tables~\ref{tab:joint_icm_params},
\ref{tab:non_isoth_mod_params}, and \ref{tab:sze_masses}.}
\label{fig:fg_vs_z}
\end{figure}

\begin{deluxetable}{lrcccccccccl}
\rotate
\tabletypesize{\scriptsize}
\tablewidth{0pt}
\tablecaption{Cluster Data\label{tab:data}}
\tablehead{
\colhead{} &
\colhead{} &
\multicolumn{5}{c}{\chandra\ X-ray Data} &
\multicolumn{4}{c}{Interferometric SZE Data}&
\colhead{}
\\
\colhead{} &
\colhead{} &
\multicolumn{5}{c}{\hrulefill} &
\multicolumn{4}{c}{\hrulefill} &
\colhead{}
\\
\colhead{Cluster} &
\colhead{z} &
\colhead{ObsID} &
\colhead{Chip} &
\colhead{(ks)} &
\colhead{(hh mm ss)} &
\colhead{($\circ\;\;$ $\prime\;\;$ $\prime\prime$ )} &
\colhead{BIMA (hr)} &
\colhead{OVRO (hr)} &
\colhead{(hh mm ss)} &
\colhead{($\circ\;\;$ $\prime\;\;$ $\prime\prime$ )} &
\colhead{$z$ reference}
}
\startdata
CL~0016+1609 & 0.541 
        & \phn520 & I3 & \phn67.4 & 00 18 33.5 & +16 26 12.5 & 43 & 100 & 00 18 33.3 & +16 26 04.0
&  \citet{stocke1991}
\\
ABELL~68 & 0.255 
	& 3250 & I3 & \phn10.0 & 00 37 06.2 & +09 09 33.2 & 54 & \nodatah & 00 37 04.0 & +09 10 02.5
& \citet{struble1999}
\\
ABELL~267 & 0.230 
	& 1448 & I3 & \phn\phn7.4 & 01 52 42.1 & +01 00 35.7 & 50 & \nodatah & 01 52 42.3 & +01 00 26.0
 & \citet{struble1999}
\\
ABELL~370  & 0.375 
	& \phn515 & S3 & \phn65.3 & 02 39 53.2 & -01 34 35.0 &26 & 33 & 02 39 52.4 & -01 34 43.8
& \citet{struble1999}
\\
MS~0451.6-0305 & 0.550 
	& \phn902 & S3 & \phn42.2 & 04 54 11.4 & -03 00 52.7 & \nodatah & 30 & 04 54 11.6 & -03 01 01.3 
& \citet{gioia1994} \\
	& \nodatae
	& \phn529 & I3 & \phn13.9 & \nodatae& \nodatae& \nodatae& \nodatae& \nodatae& \nodatae
& \nodatae
\\
MACS~J0647.7+7015 & 0.584
	& 3196 & I3 & \phn19.3 & 06 47 50.2 & +70 14 54.6 & \nodatah & 23 & 06 47 50.2 & +70 14 56.1
& \citet{laroque2003}\\
	& \nodatae
	& 3584 & I3 & \phn20.0 & \nodatae& \nodatae & \nodatae& \nodatae& \nodatae& \nodatae
& \nodatae
\\
ABELL~586 & 0.171
	& \phn530 & I3 & \phn10.0 & 07 32 20.2 & +31 37 55.6& 45 & \nodatah & 07 32 19.6 & +31 37 55.3
& \citet{struble1999}
\\
MACS~J0744.8+3927 & 0.686
	& 3197 & I3 & \phn20.2 & 07 44 52.8 & +39 27 26.7 & \phn8 & 17 & 07 44 52.4 & +39 27 33.2
& \citet{laroque2003}\\
	& \nodatae
	& 3585 & I3 & \phn19.4 & \nodatae& \nodatae& \nodatae& \nodatae& \nodatae& \nodatae
& \nodatae
\\
ABELL~611 & 0.288
	& 3194 & S3 & \phn36.1 & 08 00 56.6 & +36 03 24.1 & \nodatah & 57 & 08 00 56.5 & +36 03 22.9
& \citet{struble1999}
\\
ABELL~665 & 0.182
	& 3586 & I3 & \phn29.7 & 08 30 58.1 & +65 50 51.6 & 52 & 16  & 08 30 58.6 & +65 50 49.8 
 & \citet{struble1999}\\
	& \nodatae
	& \phn531 & I3 & \phn9.0 & \nodatae& \nodatae& \nodatae& \nodatae& \nodatae& \nodatae
& \nodatae
\\
ABELL~697& 0.282 
	& 4217 & I3 & \phn19.5 & 08 42 57.5 & +36 21 56.2 & \nodatah & 47 & 08 42 57.8 & +36 21 54.5
& \citet{struble1999}
\\
ABELL~773 & 0.217
	& \phn533 & I3 & \phn11.3 & 09 17 52.8 & +51 43 38.9 & 26 & 66 & 09 17 53.5 & +51 43 49.8 
& \citet{struble1999}\\
	& \nodatae
	& 3588 & I3 & \phn\phn9.4 & \nodatae& \nodatae& \nodatae& \nodatae& \nodatae& \nodatae
& \nodatae
\\
ZW~3146& 0.291
	& \phn909 & I3 & \phn46.0 & 10 23 39.7 & +04 11 09.5 & 25 & 15 & 10 23 37.8 & +04 11 17.8
& \citet{allen1992}
\\
MS~1054.5-0321& 0.826
        & 512 & S3  & \phn89.1 & 10 56 59.4 & -03 37 34.2 & \nodatah  & 43  & 10 56 59.1 & -03 37 34.0  
& \citet{luppino1995}
\\
MS~1137.5+6625 & 0.784
	& \phn536 & I3 & \phn77.0 & 11 40 22.3 & +66 08 16.0 & 88 & \nodatah & 11 40 23.1  & +66 08 05.3
& \citet{donahue1999}
\\
MACS~J1149.5+2223 & 0.544 
	& 1656 & I3 & \phn18.5 & 11 49 35.5 & +22 24 02.3 & 39 & \nodatah & 11 49 34.9 & +22 23 54.8 
& \citet{laroque2003}\\
	& \nodatae
	& 3589 & I3 & \phn20.0 & \nodatae& \nodatae& \nodatae& \nodatae& \nodatae& \nodatae
& \nodatae
\\
ABELL~1413& 0.142
	& 1661 & I3 & \phn\phn9.7 & 11 55 18.0 & +23 24 17.0 & 28 & \nodatah &  11 55 17.7 & +23 24 39.5
& \citet{struble1999}\\
	& \nodatae
	& \phn537 & I3 & \phn\phn9.6 & \nodatae& \nodatae& \nodatae& \nodatae& \nodatae& \nodatae
& \nodatae
\\
CL~J1226.9+3332& 0.890 
	& 3180 & I3 & \phn31.7 & 12 26 57.9 & +33 32 47.4 & 33 & \nodatah & 12 26 58.0 & +33.32 57.9 
& \citet{ebeling2001a}\\
	& \nodatae
	& 932  & S3 & \phn9.9  &  \nodatae& \nodatae& \nodatae& \nodatae& \nodatae& \nodatae
& \nodatae
\\
MACS~J1311.0-0310 & 0.490
	&3258 & I3 & 14.9 & 13 11 01.7 & -03 10 38.5 & 39 & \nodatah &  13 11 02.2 &  -03 10 45.6 
& \citet{allen2004}
\\

ABELL~1689& 0.183
	& 1663 & I3 & \phn10.7 & 13 11 29.5 & -01 20 28.2 & 16 & 26 & 13 11 29.1 & -01 20 29.7
& \citet{struble1999}\\
	& \nodatae
	& \phn540 & I3 & \phn10.3 &  \nodatae& \nodatae& \nodatae& \nodatae& \nodatae& \nodatae
& \nodatae
\\
RX~J1347.5-1145& 0.451
	& 3592 & I3 & \phn57.7 & 13 47 30.6 & -11 45 08.6 & 22 & \phn3 &13 47 30.6 & -11 45 12.3
& \citet{schindler1995}
\\

MS~1358.4+6245& 0.327
	& \phn516 & S3 & \phn48.1 & 35 59 50.6 & +62 31 04.1& 70 & \nodatah & 13 59 50.2 & +62 31 07.0
& \citet{gioia1994}
\\
ABELL~1835& 0.252
	& \phn495 & S3 & \phn19.5 & 14 01 02.0 & +02 52 41.7 & 27 & 23 & 14 01 01.8 & +02 52 45.6
& \citet{struble1999}\\
	& \nodatae
	& \phn496 & S3 & \phn10.7 & \nodatae& \nodatae& \nodatae& \nodatae& \nodatae& \nodatae
& \nodatae
\\
MACS~J1423.8+2404& 0.545
	& 4195 & S3 & 115.6 & 14 23 47.9 & +24 04 42.6  & 35 & \nodatah & 14 23 47.7 & +24 04 37.3
& \citet{laroque2003}
\\
ABELL~1914 & 0.171 
	& 3593 & I3 & \phn18.9 & 14 26 00.8 & +37 49 35.7 & 24 & \nodatah & 14 26 01.3 & 37 49 38.6 
& \citet{struble1999}\\
	& \nodatae
	& \phn542 & I3 & \phn\phn8.1 & \nodatae& \nodatae& \nodatae& \nodatae& \nodatae& \nodatae
& \nodatae
\\
ABELL~1995& 0.322
	& \phn906 & S3 & \phn56.7 & 14 52 57.9 & +58 02 55.8 & 50 & 58 & 14 52 58.1 & +58 02 57.0
& \citet{patel2000}
\\
ABELL~2111& 0.229
	& \phn544 & I3 & \phn10.3 &15 39 41.0 & +34 25 08.8 & 36 & \nodatah & 15 39 40.2  & +34 25 00.4
& \citet{struble1999}
\\
ABELL~2163& 0.202
	& 1653 & I1 & \phn71.1 & 16 15 46.2 & -06 08 51.3 & 23 & 37 & 16 15 43.6 & -06 08 46.6
& \citet{struble1999}
\\
ABELL~2204 & 0.152
       & 499 & S3   & \phn8.6 &  16 32 46.9 & +05 34 31.9  & 30 & \nodatah & 16 32 46.6 & +05 34 20.6
& \citet{struble1999}\\
	& \nodatae
       &  6104 & I3  & \phn9.6 &  \nodatae& \nodatae& \nodatae& \nodatae& \nodatae& \nodatae
& \nodatae
\\	
ABELL~2218& 0.176 
	& 1666 & I0 & \phn41.7 & 16 35 51.9 & +66 12 34.5 & 32 & 70 & 16 35 48.7 & +66 12 28.1 
& \citet{struble1999}
\\
RX~J1716.4+6708& 0.813 
	& \phn548 & I3 & \phn51.7 & 17 16 48.8 & +67 08 25.3 & 37 & \nodatah & 17 16 51.2 & +67 07 49.6
& \citet{henry1997}
\\
ABELL~2259& 0.164 
	& 3245 & I3 & \phn10.0 & 17 20 08.5 & +27 40 11.0 & 25 & \nodatah & 17 20 09.0 & +27 40 09.4 
& \citet{struble1999}
\\
ABELL~2261& 0.224
	& \phn550 & I3 & \phn\phn9.1 & 17 22 27.1 & +32 07 57.8 & 23 & 40 & 17 22 26.9 & +32 07 59.9 
& \citet{struble1999}
\\
MS~2053.7-0449& 0.583
	& \phn551 & I3 & \phn44.3 & 20 56 21.2 & -04 37 47.8 & \nodatah & 154 & 20 56 21.0 & -04 37 47.2
& \citet{stocke1991}\\
	& \nodatae
        & 1667 & I3 & \phn44.5 & \nodatae& \nodatae& \nodatae& \nodatae& \nodatae& \nodatae
& \nodatae
\\
MACS~J2129.4-0741& 0.570
	& 3199 & I3 & \phn8.5 & 21 29 26.0 & -07 41 28.7 & \nodatah & 24 & 21 29 24.9 & -07 41 43.9
& \citet{laroque2003}\\
	& \nodatae
        & 3595 & I3 & \phn18.4 & \nodatae& \nodatae& \nodatae& \nodatae& \nodatae& \nodatae
& \nodatae
\\
RX~J2129.7+0005& 0.235
	& \phn552 & I3 & \phn10.0 & 21 29 39.9 &+00 05 19.8 & 47 & \nodatah & 21 29 38.1 & +00 05 12.4
& \citet{ebeling1998}
\\
MACS~J2214.9-1359& 0.483
	& 3259 & I3 & \phn19.5 & 22 14 57.3 & -14 00 12.3 &41 & 11 & 22 14 58.4 & -14 00 10.9
& Note (\tablenotemark{a}\ ) \\
	& \nodatae       
& 5011 & I3 & \phn16.1 & \nodatae& \nodatae & \nodatae& \nodatae& \nodatae& \nodatae
& \nodatae
\\
MACS~J2228.5+2036& 0.412
	& 3285 & I3 & \phn19.9 & 22 28 33.0 & +20 37 14.4 & 39 & \nodatah & 22 28 33.1 & +20 37 14.2
& \citet{bohringer2000}
\\
\enddata
\tablenotetext{a}{Redshift derived from the Fe lines in the Chandra x-ray spectrum, this paper.}
\end{deluxetable}

\begin{deluxetable}{cccccccc}
\tablewidth{0pt}
\tablecolumns{7}
\tablecaption{Comparison of Parameters for Different ICM Models \label{tab:fit_param_cmp}}
\tablehead{
\colhead{Cluster} &
\colhead{Model} &
\colhead{$\theta_c$} &
\colhead{$\beta$} &
\colhead{$T_X$} &
\colhead{\Mgas(\rtfh)} &
\colhead{\Mtot(\rtfh)} &
\colhead{\fg}
\\
\colhead{} &
\colhead{} &
\colhead{($\arcsec$)} &
\colhead{} &
\colhead{(keV)} &
\colhead{($10^{13}\Msun$)} &
\colhead{($10^{14}\Msun$)} &
\colhead{}
}
\startdata

A1995 & Isoth. $\beta$ & $50.3^{+1.5}_{-1.5}$ &
$0.921^{+0.024}_{-0.023}$ & $8.63^{+0.39}_{-0.40}$ &
$3.52^{+0.12}_{-0.12}$ & $4.69^{+0.39}_{-0.39}$ & 
$0.075^{+0.004}_{-0.004}$\\

\nodata & Non-Isoth. $\beta$ & $50.8^{+1.8}_{-1.5}$ &
$0.925^{+0.029}_{-0.024}$ & \nodata &
$3.77^{+0.21}_{-0.23}$ & $4.37^{+0.75}_{-0.71}$ & 
$0.086^{+0.011}_{-0.009}$\\

\nodata & 100~kpc-cut & $57.5^{+2.7}_{-2.6}$ &
$1.01^{+0.04}_{-0.04}$ & $8.22^{+0.44}_{-0.45}$ &
$3.51^{+0.14}_{-0.14}$ & $4.74^{+0.50}_{-0.48}$ & 
$0.074^{+0.005}_{-0.005}$\\

A1835 & Isoth. $\beta$ & $8.13^{+0.09}_{-0.09}$ &
$0.543^{+0.001}_{-0.001}$ & $8.37^{+0.23}_{-0.23}$ &
$4.04^{+0.08}_{-0.08}$ & $2.55^{+0.11}_{-0.11}$ & 
$0.158^{+0.004}_{-0.003}$\\

\nodata & Non-Isoth. $\beta$ & $9.08^{+0.25}_{-0.21}$\,
\tablenotemark{a} &
$0.784^{+0.016}_{-0.014}$ & \nodata &
$6.49^{+0.30}_{-0.33}$ & $5.77^{+0.70}_{-0.71}$ & 
$0.112^{+0.009}_{-0.008}$\\

\nodata & \nodata & $61.6^{+1.9}_{-1.8}$\,\,\tablenotemark{b} & 
\nodata & \nodata & \nodata & \nodata\\

\nodata & 100~kpc-cut & $33.6^{+1.0}_{-1.0}$ &
$0.690^{+0.007}_{-0.008}$ & $11.4^{+0.7}_{-0.6}$ &
$5.78^{+0.23}_{-0.20}$ & $5.56^{+0.50}_{-0.42}$ & 
$0.104^{+0.005}_{-0.005}$\\

\enddata
\tablenotetext{a}{Inner $\theta_c$.}
\tablenotetext{a}{Outer $\theta_c$.}
\end{deluxetable}

\begin{deluxetable}{lcc}
\tablecaption{Sources of Uncertainty in \fg\ Measurements
\label{tab:systematics}} 
\tablewidth{260pt} 
\tablehead{
Source & 
X-ray \fg & 
SZE \fg
} 
\startdata 
\multicolumn{3}{c}{Statistical Effects\tablenotemark{1}} \\
 & & \\
Kinetic SZE & \nodata & $\pm4\%$ \\
Radio Point Sources & \nodata & $\pm4\%$ \\ 
Asphericity & $\pm20\%$ & $\pm10\%$ \\ 
\hline
 & & \\
\multicolumn{3}{c}{Systematic Effects\tablenotemark{2}} \\
 & & \\
Instrument Calibration & $\pm6\%$ & $\pm 8\%$ \\ 
X-ray Background & $+2\%$ & \nodata \\ 
HSE & $-10\%$ & $-10\%$ \\
Isothermality\tablenotemark{3} & $-5\%$ & $-10\%$ \\
\enddata
\tablenotetext{1}{Statistical uncertainties average down by a factor
  of $\sqrt{38}$ for the sample.}
\tablenotetext{2}{Systematic uncertainties do not average down.}
\tablenotetext{3}{Applies to isothermal 100~kpc-cut model only.}
\end{deluxetable}

\begin{deluxetable}{lccccccccccc}
\tabletypesize{\tiny}
\rotate
\renewcommand{\arraystretch}{0.90}
\tablewidth{0pt}
\tablecolumns{12}
\tablecaption{Model Parameters, Cluster Masses, and Gas Mass Fractions:
Isothermal $\beta$ model with 100 kpc cut\label{tab:joint_icm_params}}
\tablehead{
\colhead{Cluster} & 
\colhead{\Xo\tablenotemark{a}} & 
\colhead{$\theta_c$} & 
\colhead{$\beta$} & 
\colhead{\dTo} & 
\colhead{$T_X$} &
\colhead{$M_{\mbox{\scriptsize gas}}^{^{\mbox{\scriptsize xray}}}$} &
\colhead{$M_{\mbox{\scriptsize gas}}^{^{\mbox{\scriptsize sze}}}$} &
\colhead{\Mtot} &
\colhead{$f_{\mbox{\scriptsize gas}}^{^{\mbox{\scriptsize xray}}}$} &
\colhead{$f_{\mbox{\scriptsize gas}}^{^{\mbox{\scriptsize sze}}}$} &
\colhead{\tc/\tH} 
 \\
\colhead{} &
\colhead{} & 
\colhead{($\arcsec$)} & 
\colhead{} & 
\colhead{(mK)} & 
\colhead{(keV)} &
\colhead{($10^{13}\,\Msun$)} &
\colhead{($10^{13}\,\Msun$)} &
\colhead{($10^{14}\,\Msun$)} &
\colhead{} &
\colhead{} &
\colhead{}
}
\startdata
CL0016 & 
$1.27^{+0.04}_{-0.04}$ &
$42.8^{+2.1}_{-2.0}$ &
$0.742^{+0.025}_{-0.022}$ &
$-1.361^{+0.083}_{-0.083}$ &
$10.46^{+0.61}_{-0.60}$ &
$4.38^{+0.29}_{-0.29}$ & $4.34^{+0.26}_{-0.26}$ &
$3.33^{+0.39}_{-0.37}$ &
$0.131^{+0.007}_{-0.006}$ & $0.130^{+0.016}_{-0.015}$ &
$1.5$ 
\\

A68 & 
$1.60^{+0.13}_{-0.11}$ &
$55.1^{+5.3}_{-5.0}$ &
$0.764^{+0.046}_{-0.042}$ &
$-0.712^{+0.099}_{-0.103}$ &
$\phn9.55^{+1.09}_{-0.96}$ &
$3.65^{+0.34}_{-0.31}$ & $3.07^{+0.45}_{-0.44}$ &
$4.32^{+0.91}_{-0.74}$ &
$0.084^{+0.009}_{-0.008}$ & $0.071^{+0.019}_{-0.016}$ &
$1.4$ 
\\

A267 & 
$2.04^{+0.28}_{-0.23}$ &
$43.1^{+5.8}_{-5.1}$ &
$0.720^{+0.049}_{-0.040}$ &
$-0.692^{+0.077}_{-0.080}$ &
$\phn5.89^{+0.66}_{-0.54}$ &
$2.24^{+0.20}_{-0.18}$ & $2.83^{+0.32}_{-0.31}$ &
$2.03^{+0.42}_{-0.33}$ &
$0.110^{+0.011}_{-0.011}$ & $0.140^{+0.032}_{-0.029}$ &
$0.8$ 
\\

A370 & 
$0.73^{+0.02}_{-0.02}$ &
$61.8^{+3.6}_{-3.5}$ &
$0.811^{+0.039}_{-0.037}$ &
$-0.866^{+0.090}_{-0.088}$ &
$\phn8.67^{+0.51}_{-0.49}$ &
$2.78^{+0.21}_{-0.20}$ & $3.51^{+0.37}_{-0.36}$ &
$2.77^{+0.37}_{-0.33}$ &
$0.100^{+0.005}_{-0.005}$ & $0.127^{+0.020}_{-0.018}$ &
$2.1$
\\

MS0451 & 
$2.35^{+0.11}_{-0.10}$ &
$40.0^{+1.9}_{-1.7}$ &
$0.773^{+0.026}_{-0.023}$ &
$-1.476^{+0.086}_{-0.086}$ &
$\phn9.95^{+0.76}_{-0.69}$ &
$4.76^{+0.31}_{-0.30}$ & $4.79^{+0.28}_{-0.28}$ &
$3.76^{+0.53}_{-0.46}$ &
$0.127^{+0.009}_{-0.008}$ & $0.127^{+0.020}_{-0.018}$ &
$1.0$ 
\\

MC0647 & 
$2.19^{+0.20}_{-0.18}$ &
$24.1^{+2.0}_{-1.8}$ &
$0.687^{+0.023}_{-0.020}$ &
$-1.312^{+0.116}_{-0.123}$ &
$14.06^{+1.78}_{-1.59}$ &
$4.90^{+0.47}_{-0.42}$ & $3.26^{+0.30}_{-0.29}$ &
$5.98^{+1.30}_{-1.05}$ &
$0.082^{+0.009}_{-0.008}$ & $0.055^{+0.014}_{-0.012}$ &
$1.0$
\\

A586 & 
$3.36^{+0.43}_{-0.36}$ &
$45.3^{+5.1}_{-4.6}$ &
$0.723^{+0.039}_{-0.033}$ &
$-0.651^{+0.081}_{-0.084}$ &
$\phn6.35^{+0.46}_{-0.39}$ &
$2.26^{+0.13}_{-0.11}$ & $2.49^{+0.32}_{-0.32}$ &
$2.48^{+0.34}_{-0.28}$ &
$0.091^{+0.007}_{-0.007}$ & $0.100^{+0.019}_{-0.017}$ &
$0.5$
\\

MC0744 & 
$1.40^{+0.12}_{-0.10}$ &
$25.8^{+2.0}_{-2.0}$ &
$0.723^{+0.028}_{-0.027}$ &
$-1.299^{+0.134}_{-0.138}$ &
$\phn8.14^{+0.80}_{-0.72}$ &
$3.07^{+0.27}_{-0.25}$ & $3.36^{+0.34}_{-0.35}$ &
$2.26^{+0.42}_{-0.35}$ &
$0.136^{+0.012}_{-0.011}$ & $0.148^{+0.032}_{-0.027}$ &
$0.3$
\\

A611 & 
$3.21^{+0.37}_{-0.29}$ &
$25.7^{+2.1}_{-2.2}$ &
$0.632^{+0.017}_{-0.017}$ &
$-0.761^{+0.084}_{-0.082}$ &
$\phn6.79^{+0.41}_{-0.38}$ &
$2.36^{+0.11}_{-0.11}$ & $2.33^{+0.26}_{-0.26}$ &
$2.14^{+0.22}_{-0.20}$ &
$0.110^{+0.006}_{-0.006}$ & $0.109^{+0.017}_{-0.016}$ &
$0.6$
\\

A665 & 
$6.51^{+1.46}_{-1.11}$ &
$17.7^{+2.9}_{-2.5}$ &
$0.454^{+0.006}_{-0.005}$ &
$-1.330^{+0.117}_{-0.119}$ &
$\phn8.37^{+0.37}_{-0.33}$ &
$2.62^{+0.10}_{-0.09}$ & $2.53^{+0.22}_{-0.23}$ &
$2.00^{+0.14}_{-0.12}$ &
$0.131^{+0.004}_{-0.004}$ & $0.126^{+0.014}_{-0.013}$ &
$1.1$ 
\\

A697 & 
$2.46^{+0.06}_{-0.06}$ &
$42.7^{+1.6}_{-1.5}$ &
$0.605^{+0.011}_{-0.011}$ &
$-1.223^{+0.125}_{-0.121}$ &
$10.21^{+0.70}_{-0.65}$ &
$4.39^{+0.28}_{-0.26}$ & $4.09^{+0.41}_{-0.42}$ &
$3.47^{+0.40}_{-0.37}$ &
$0.127^{+0.007}_{-0.006}$ & $0.117^{+0.019}_{-0.017}$ &
$1.2$
\\

A773 & 
$2.31^{+0.17}_{-0.15}$ &
$38.7^{+2.5}_{-2.5}$ &
$0.594^{+0.013}_{-0.012}$ &
$-1.101^{+0.118}_{-0.116}$ &
$\phn8.16^{+0.56}_{-0.52}$ &
$2.75^{+0.17}_{-0.15}$ & $3.64^{+0.39}_{-0.39}$ &
$2.59^{+0.30}_{-0.27}$ & $0.106^{+0.006}_{-0.005}$ &
$0.140^{+0.024}_{-0.021}$ &
$1.0$
\\

Zw3146 & 
$7.56^{+0.32}_{-0.28}$ &
$30.4^{+0.9}_{-0.9}$ &
$0.736^{+0.009}_{-0.009}$ &
$-1.172^{+0.143}_{-0.144}$ &
$\phn8.28^{+0.30}_{-0.29}$ &
$4.42^{+0.11}_{-0.11}$ & $3.80^{+0.47}_{-0.47}$ &
$3.62^{+0.22}_{-0.20}$ &
$0.122^{+0.004}_{-0.004}$ & $0.105^{+0.015}_{-0.014}$ &
$0.1$ 
\\

MS1054 & 
$0.43^{+0.02}_{-0.02}$ &
$57.7^{+8.8}_{-7.5}$ &
$0.884^{+0.155}_{-0.117}$ &
$-1.183^{+0.125}_{-0.135}$ &
$\phn9.77^{+1.10}_{-0.94}$ &
$1.06^{+0.50}_{-0.49}$ & $1.12^{+0.42}_{-0.47}$ &
$0.74^{+0.45}_{-0.38}$ &
$0.144^{+0.014}_{-0.013}$ & $0.153^{+0.034}_{-0.029}$ &
$2.8$ 
\\

MS1137 & 
$0.62^{+0.11}_{-0.08}$ &
$22.3^{+3.8}_{-3.4}$ &
$0.877^{+0.097}_{-0.078}$ &
$-0.775^{+0.094}_{-0.102}$ &
$\phn4.47^{+0.53}_{-0.44}$ &
$1.19^{+0.14}_{-0.13}$ & $2.13^{+0.26}_{-0.25}$ &
$1.03^{+0.26}_{-0.20}$ &
$0.115^{+0.014}_{-0.013}$ & $0.207^{+0.054}_{-0.045}$ &
$0.7$
\\

MC1149 & 
$0.81^{+0.04}_{-0.04}$ &
$44.7^{+2.6}_{-2.3}$ &
$0.666^{+0.020}_{-0.018}$ &
$-1.166^{+0.112}_{-0.114}$ &
$\phn9.93^{+0.80}_{-0.68}$ &
$3.09^{+0.34}_{-0.30}$ & $3.00^{+0.30}_{-0.30}$ &
$2.31^{+0.40}_{-0.34}$ &
$0.134^{+0.008}_{-0.008}$ & $0.130^{+0.022}_{-0.019}$ &
$1.9$
\\

A1413 & 
$4.31^{+0.59}_{-0.41}$ &
$37.9^{+3.7}_{-4.3}$ &
$0.536^{+0.015}_{-0.015}$ &
$-1.037^{+0.150}_{-0.150}$ &
$\phn7.51^{+0.35}_{-0.29}$ &
$2.63^{+0.11}_{-0.10}$ & $2.98^{+0.43}_{-0.42}$ &
$2.15^{+0.17}_{-0.15}$ &
$0.122^{+0.004}_{-0.004}$ & $0.139^{+0.022}_{-0.022}$ &
$0.3$
\\

CL1226 & 
$2.36^{+0.76}_{-0.48}$ &
$16.2^{+3.8}_{-3.3}$ &
$0.732^{+0.084}_{-0.060}$ &
$-1.715^{+0.202}_{-0.234}$ &
$13.47^{+2.72}_{-2.20}$ &
$3.89^{+0.51}_{-0.46}$ & $2.90^{+0.34}_{-0.30}$ &
$5.21^{+1.96}_{-1.39}$ &
$0.075^{+0.015}_{-0.014}$ & $0.056^{+0.024}_{-0.018}$ &
$0.9$
\\

MC1311 & 
$1.03^{+0.58}_{-0.46}$ &
$7.8^{+2.9}_{-1.5}$ &
$0.621^{+0.030}_{-0.022}$ &
$-1.489^{+0.309}_{-0.314}$ &
$\phn7.16^{+1.53}_{-1.15}$ &
$2.11^{+0.24}_{-0.21}$ & $2.33^{+0.38}_{-0.38}$ &
$2.17^{+0.74}_{-0.53}$ &
$0.097^{+0.019}_{-0.017}$ & $0.107^{+0.047}_{-0.035}$ &
$0.3$
\\

A1689 & 
$6.26^{+0.29}_{-0.26}$ &
$48.4^{+2.1}_{-2.0}$ &
$0.688^{+0.013}_{-0.012}$ &
$-1.651^{+0.130}_{-0.137}$ &
$10.46^{+0.46}_{-0.47}$ &
$5.06^{+0.16}_{-0.17}$ & $5.85^{+0.49}_{-0.47}$ &
$4.96^{+0.36}_{-0.36}$ &
$0.102^{+0.004}_{-0.004}$ & $0.118^{+0.014}_{-0.013}$ &
$0.3$
\\

RXJ1347 & 
$15.6^{+0.70}_{-0.60}$ &
$17.7^{+0.7}_{-0.8}$ &
$0.651^{+0.007}_{-0.007}$ &
$-2.777^{+0.294}_{-0.284}$ &
$16.48^{+0.99}_{-0.92}$ &
$8.84^{+0.37}_{-0.35}$ & $5.60^{+0.58}_{-0.60}$ &
$8.08^{+0.77}_{-0.69}$ &
$0.109^{+0.006}_{-0.005}$ & $0.069^{+0.011}_{-0.010}$ &
$0.1$
\\

MS1358 & 
$1.79^{+0.13}_{-0.11}$ &
$28.9^{+1.8}_{-1.9}$ &
$0.638^{+0.016}_{-0.016}$ &
$-0.721^{+0.096}_{-0.097}$ &
$\phn8.93^{+0.88}_{-0.74}$ &
$2.53^{+0.19}_{-0.17}$ & $2.23^{+0.29}_{-0.29}$ &
$3.13^{+0.51}_{-0.41}$ &
$0.081^{+0.006}_{-0.006}$ & $0.071^{+0.015}_{-0.014}$ &
$0.2$
\\

A1835 & 
$9.49^{+0.36}_{-0.32}$ &
$33.6^{+1.0}_{-1.0}$ &
$0.690^{+0.007}_{-0.008}$ &
$-1.636^{+0.116}_{-0.114}$ &
$11.44^{+0.66}_{-0.56}$ &
$5.78^{+0.23}_{-0.20}$ & $4.98^{+0.36}_{-0.36}$ &
$5.56^{+0.50}_{-0.42}$ &
$0.104^{+0.005}_{-0.005}$ & $0.089^{+0.011}_{-0.010}$ &
$0.1$ 
\\

MC1423 & 
$3.21^{+0.39}_{-0.34}$ &
$14.6^{+1.2}_{-1.0}$ &
$0.637^{+0.012}_{-0.011}$ &
$-1.177^{+0.210}_{-0.200}$ &
$\phn6.97^{+0.42}_{-0.40}$ &
$2.26^{+0.10}_{-0.10}$ & $2.63^{+0.44}_{-0.47}$ &
$1.94^{+0.20}_{-0.18}$ &
$0.116^{+0.006}_{-0.006}$ & $0.135^{+0.029}_{-0.028}$ &
$0.1$
\\

A1914 & 
$10.3^{+0.40}_{-0.40}$ &
$44.0^{+1.3}_{-1.3}$ &
$0.734^{+0.009}_{-0.010}$ &
$-1.565^{+0.125}_{-0.125}$ &
$\phn9.48^{+0.35}_{-0.29}$ &
$4.85^{+0.12}_{-0.10}$ & $5.11^{+0.42}_{-0.42}$ &
$4.82^{+0.30}_{-0.25}$ &
$0.100^{+0.003}_{-0.004}$ & $0.106^{+0.011}_{-0.011}$ &
$0.7$
\\

A1995 & 
$1.62^{+0.05}_{-0.05}$ &
$57.5^{+2.7}_{-2.6}$ &
$1.013^{+0.041}_{-0.038}$ &
$-0.879^{+0.053}_{-0.052}$ &
$\phn8.22^{+0.44}_{-0.45}$ &
$3.51^{+0.14}_{-0.14}$ & $4.33^{+0.28}_{-0.27}$ &
$4.74^{+0.50}_{-0.48}$ &
$0.074^{+0.005}_{-0.005}$ & $0.092^{+0.013}_{-0.011}$ &
$1.2$
\\

A2111 & 
$089^{+0.10}_{-0.08}$ &
$51.0^{+7.6}_{-6.6}$ &
$0.611^{+0.044}_{-0.036}$ &
$-0.596^{+0.117}_{-0.122}$ &
$\phn8.15^{+0.98}_{-0.83}$ &
$2.19^{+0.27}_{-0.23}$ & $2.15^{+0.42}_{-0.42}$ &
$2.49^{+0.56}_{-0.44}$ &
$0.088^{+0.008}_{-0.008}$ & $0.085^{+0.026}_{-0.021}$ &
$1.7$
\\

A2163 & 
$4.63^{+0.09}_{-0.09}$ &
$57.6^{+1.3}_{-1.2}$ &
$0.538^{+0.005}_{-0.005}$ &
$-1.903^{+0.171}_{-0.177}$ &
$14.81^{+0.39}_{-0.38}$ &
$8.08^{+0.21}_{-0.20}$ & $5.89^{+0.55}_{-0.53}$ &
$5.49^{+0.24}_{-0.23}$ &
$0.147^{+0.003}_{-0.003}$ & $0.107^{+0.011}_{-0.011}$ &
$1.2$
\\

A2204 & 
$1.37^{+0.35}_{-0.22}$ &
$33.7^{+6.6}_{-5.7}$ &
$0.614^{+0.047}_{-0.034}$ &
$-1.644^{+0.166}_{-0.166}$ &
$11.23^{+0.85}_{-0.72}$ &
$4.73^{+0.27}_{-0.24}$ & $4.30^{+0.37}_{-0.39}$ &
$4.96^{+0.80}_{-0.64}$ &
$0.096^{+0.009}_{-0.010}$ & $0.087^{+0.014}_{-0.014}$ &
$0.1$
\\

A2218 & 
$1.69^{+0.03}_{-0.02}$ &
$70.3^{+1.7}_{-1.7}$ &
$0.765^{+0.014}_{-0.013}$ &
$-0.870^{+0.079}_{-0.078}$ &
$\phn7.80^{+0.41}_{-0.37}$ &
$3.00^{+0.13}_{-0.12}$ & $4.10^{+0.38}_{-0.38}$ &
$3.33^{+0.31}_{-0.27}$ &
$0.090^{+0.004}_{-0.004}$ & $0.123^{+0.017}_{-0.015}$ &
$1.2$
\\

RXJ1716 & 
$055^{+0.22}_{-0.13}$ &
$17.6^{+5.7}_{-4.5}$ &
$0.665^{+0.097}_{-0.064}$ &
$-0.656^{+0.148}_{-0.166}$ &
$\phn6.57^{+1.08}_{-0.88}$ &
$1.23^{+0.19}_{-0.16}$ & $1.32^{+0.28}_{-0.28}$ &
$1.39^{+0.45}_{-0.33}$ &
$0.088^{+0.013}_{-0.011}$ & $0.093^{+0.034}_{-0.028}$ &
$1.0$
\\

A2259 & 
$1.48^{+0.19}_{-0.14}$ &
$59.1^{+9.1}_{-8.4}$ &
$0.663^{+0.057}_{-0.048}$ &
$-0.397^{+0.138}_{-0.139}$ &
$\phn5.81^{+0.44}_{-0.36}$ &
$1.81^{+0.14}_{-0.14}$ & $1.65^{+0.61}_{-0.58}$ &
$1.79^{+0.27}_{-0.24}$ &
$0.101^{+0.007}_{-0.007}$ & $0.092^{+0.036}_{-0.034}$ &
$0.8$ 
\\

A2261 & 
$5.92^{+0.96}_{-0.77}$ &
$28.9^{+3.5}_{-3.0}$ &
$0.624^{+0.023}_{-0.018}$ &
$-1.178^{+0.122}_{-0.121}$ &
$\phn7.45^{+0.65}_{-0.55}$ &
$3.04^{+0.20}_{-0.19}$ & $3.58^{+0.40}_{-0.38}$ &
$2.56^{+0.36}_{-0.31}$ &
$0.119^{+0.008}_{-0.008}$ & $0.139^{+0.026}_{-0.023}$ &
$0.3$
\\

MS2053 & 
$0.29^{+0.06}_{-0.05}$ &
$27.0^{+6.2}_{-4.8}$ &
$0.821^{+0.122}_{-0.085}$ &
$-0.419^{+0.069}_{-0.078}$ &
$\phn4.77^{+0.71}_{-0.58}$ &
$0.93^{+0.12}_{-0.12}$ & $1.39^{+0.23}_{-0.22}$ &
$1.23^{+0.38}_{-0.29}$ &
$0.076^{+0.012}_{-0.011}$ & $0.113^{+0.039}_{-0.030}$ &
$0.9$
\\

MC2129 & 
$1.99^{+0.25}_{-0.20}$ &
$24.4^{+2.9}_{-2.6}$ &
$0.712^{+0.040}_{-0.033}$ &
$-1.176^{+0.129}_{-0.133}$ &
$\phn8.55^{+0.97}_{-0.80}$ &
$3.26^{+0.30}_{-0.26}$ & $3.28^{+0.34}_{-0.35}$ &
$2.82^{+0.57}_{-0.45}$ &
$0.116^{+0.011}_{-0.011}$ & $0.116^{+0.027}_{-0.023}$ &
$0.8$
\\

RXJ2129 & 
$4.99^{+1.10}_{-0.83}$ &
$25.4^{+3.8}_{-3.3}$ &
$0.611^{+0.024}_{-0.020}$ &
$-0.734^{+0.113}_{-0.120}$ &
$\phn6.69^{+0.49}_{-0.47}$ &
$2.58^{+0.16}_{-0.14}$ & $2.17^{+0.33}_{-0.33}$ &
$2.08^{+0.27}_{-0.23}$ &
$0.124^{+0.008}_{-0.008}$ & $0.104^{+0.022}_{-0.019}$ &
$0.1$
\\

MC2214 & 
$1.76^{+0.15}_{-0.14}$ &
$31.5^{+3.1}_{-2.8}$ &
$0.710^{+0.038}_{-0.034}$ &
$-1.433^{+0.119}_{-0.126}$ &
$10.22^{+0.99}_{-0.88}$ &
$3.95^{+0.33}_{-0.32}$ & $4.50^{+0.36}_{-0.36}$ &
$3.85^{+0.68}_{-0.59}$ &
$0.103^{+0.009}_{-0.008}$ & $0.117^{+0.024}_{-0.020}$ &
$0.9$
\\

MC2228 & 
$2.36^{+0.17}_{-0.13}$ &
$20.2^{+2.7}_{-2.5}$ &
$0.522^{+0.015}_{-0.013}$ &
$-1.727^{+0.166}_{-0.177}$ &
$\phn8.43^{+0.78}_{-0.71}$ &
$2.81^{+0.24}_{-0.22}$ & $3.84^{+0.35}_{-0.36}$ &
$2.03^{+0.31}_{-0.27}$ &
$0.138^{+0.009}_{-0.009}$ & $0.188^{+0.035}_{-0.031}$ &
$0.9$
\\

\enddata
\tablenotetext{a}{Units are $10^{-12}$ erg cm$^{-2}$ arcmin$^{-2}$ s$^{-1}$.}
\end{deluxetable}

\begin{deluxetable}{lccccccccccccc}
\centering
\rotate
\renewcommand{\arraystretch}{0.90}
\tabletypesize{\tiny}
\tablewidth{0pt}
\tablecolumns{14}
\tablecaption{Model Parameters, Cluster Masses, and Gas Mass Fractions:
Non-isothermal Double-$\beta$ Model \label{tab:non_isoth_mod_params}}
\tablehead{
\colhead{Cluster} &
\colhead{\rth} & 
\colhead{conc} & 
\colhead{$n_{e0}^{xray}$} &
\colhead{$n_{e0}^{sze}$} &
\colhead{$\theta_{c1}$} & 
\colhead{$\beta$} & 
\colhead{$f$} & 
\colhead{$\theta_{c2}$} & 
\colhead{$M_{\mbox{\tiny gas}}^{^{\mbox{\tiny xray}}}$} &
\colhead{$M_{\mbox{\tiny gas}}^{^{\mbox{\tiny sze}}}$} &
\colhead{\Mtot} &
\colhead{$f_{\mbox{\scriptsize gas}}^{^{\mbox{\scriptsize xray}}}$} & 
\colhead{$f_{\mbox{\scriptsize gas}}^{^{\mbox{\scriptsize sze}}}$}
 \\
\colhead{} & 
\colhead{(Mpc)} & 
\colhead{} & 
\colhead{($10^{-2} cm^{-3}$)} &
\colhead{($10^{-2} cm^{-3}$)} &
\colhead{($\arcsec$)} & 
\colhead{} & 
\colhead{} & 
\colhead{($\arcsec$)} &
\colhead{($10^{13}\,\Msun$)} & 
\colhead{($10^{13}\,\Msun$)} &
\colhead{($10^{14}\,\Msun$)} & 
\colhead{} & 
\colhead{} 
} 
\startdata 
CL0016 & 
$2.12^{+0.24}_{-0.22}$ & $1.27^{+0.57}_{-0.43}$ &
$1.39^{+0.16}_{-0.12}$ & 
$1.44^{+0.20}_{-0.18}$ &
$11.0^{+3.2}_{-2.5}$ & $0.759^{+0.034}_{-0.024}$ &
$0.483^{+0.046}_{-0.054}$ & $48.1^{+3.8}_{-2.6}$ &
$4.63^{+0.48}_{-0.56}$ & $4.76^{+0.32}_{-0.36}$ &
$2.89^{+0.52}_{-0.55}$ &
$0.160^{+0.014}_{-0.011}$ & $0.164^{+0.024}_{-0.020}$
\\ 
 
A68 & 
$2.02^{+0.60}_{-0.32}$ & $2.96^{+2.09}_{-1.67}$ &
$0.82^{+0.03}_{-0.03}$ & 
$0.77^{+0.18}_{-0.15}$ &
\ldots & $0.707^{+0.031}_{-0.031}$ &
\ldots & $48.7^{+3.3}_{-3.3}$ &
$3.84^{+0.50}_{-0.50}$ & $3.63^{+0.59}_{-0.55}$ &
$3.57^{+0.98}_{-0.87}$ &
$0.107^{+0.016}_{-0.013}$ & $0.102^{+0.034}_{-0.025}$
\\ 
 
A267 & 
$1.55^{+0.25}_{-0.20}$ & $4.19^{+2.41}_{-1.48}$ &
$1.04^{+0.04}_{-0.04}$ & 
$1.22^{+0.18}_{-0.16}$ &
\ldots & $0.694^{+0.029}_{-0.029}$ &
\ldots & $40.6^{+2.8}_{-2.7}$ &
$2.35^{+0.19}_{-0.19}$ & $2.76^{+0.35}_{-0.34}$ &
$1.95^{+0.36}_{-0.32}$ &
$0.120^{+0.013}_{-0.011}$ & $0.141^{+0.027}_{-0.022}$
\\ 
 
A370 & 
$1.51^{+0.10}_{-0.10}$ & $5.32^{+1.98}_{-1.40}$ &
$0.54^{+0.01}_{-0.01}$ & 
$0.57^{+0.07}_{-0.06}$ &
\ldots & $0.737^{+0.031}_{-0.027}$ &
\ldots & $55.5^{+2.8}_{-2.4}$ &
$3.02^{+0.19}_{-0.19}$ & $3.20^{+0.30}_{-0.31}$ &
$2.52^{+0.26}_{-0.26}$ &
$0.120^{+0.005}_{-0.005}$ & $0.127^{+0.018}_{-0.015}$
\\ 
 
MS0451 & 
$1.97^{+0.53}_{-0.33}$ & $2.18^{+1.66}_{-1.07}$ &
$1.35^{+0.03}_{-0.03}$ & 
$1.33^{+0.17}_{-0.14}$ &
\ldots & $0.774^{+0.021}_{-0.019}$ &
\ldots & $34.0^{+1.2}_{-1.2}$ &
$5.25^{+0.37}_{-0.40}$ & $5.20^{+0.43}_{-0.49}$ &
$3.56^{+0.54}_{-0.51}$ &
$0.147^{+0.012}_{-0.010}$ & $0.146^{+0.024}_{-0.019}$
\\ 

MC0647 & 
$2.02^{+0.28}_{-0.22}$ & $4.71^{+1.78}_{-1.35}$ &
$1.70^{+0.06}_{-0.06}$ & 
$1.10^{+0.14}_{-0.13}$ &
\ldots & $0.660^{+0.017}_{-0.016}$ &
\ldots & $20.5^{+1.0}_{-1.0}$ &
$4.90^{+0.39}_{-0.36}$ & $3.16^{+0.41}_{-0.38}$ &
$6.19^{+1.12}_{-0.95}$ &
$0.079^{+0.008}_{-0.007}$ & $0.051^{+0.009}_{-0.008}$
\\ 
 
A586 & 
$1.60^{+0.19}_{-0.15}$ & $6.99^{+2.40}_{-2.04}$ &
$1.71^{+0.05}_{-0.05}$ & 
$1.50^{+0.23}_{-0.21}$ &
\ldots & $0.627^{+0.015}_{-0.014}$ &
\ldots & $31.9^{+1.4}_{-1.4}$ &
$2.46^{+0.18}_{-0.17}$ & $2.15^{+0.35}_{-0.31}$ &
$2.52^{+0.46}_{-0.38}$ &
$0.098^{+0.010}_{-0.009}$ & $0.085^{+0.016}_{-0.013}$
\\ 
 
MC0744 & 
$1.19^{+0.24}_{-0.13}$ & $5.40^{+3.81}_{-2.47}$ &
$11.9^{+2.1}_{-1.7}$ & 
$12.0^{+2.7}_{-2.2}$ &
$3.02^{+0.83}_{-0.67}$ & $0.584^{+0.072}_{-0.049}$ &
$0.930^{+0.009}_{-0.012}$ & $22.2^{+7.1}_{-5.6}$ &
$2.88^{+0.24}_{-0.21}$ & $2.87^{+0.30}_{-0.31}$ &
$1.82^{+0.31}_{-0.26}$ &
$0.158^{+0.013}_{-0.012}$ & $0.157^{+0.029}_{-0.022}$
\\ 
 
A611 & 
$1.49^{+0.12}_{-0.10}$ & $5.08^{+1.18}_{-1.01}$ &
$3.28^{+1.00}_{-0.61}$ & 
$3.12^{+0.94}_{-0.66}$ &
$4.59^{+2.89}_{-1.52}$ & $0.602^{+0.015}_{-0.011}$ &
$0.570^{+0.083}_{-0.097}$ & $23.3^{+2.5}_{-1.6}$ &
$2.55^{+0.12}_{-0.11}$ & $2.40^{+0.29}_{-0.28}$ &
$2.10^{+0.23}_{-0.19}$ &
$0.122^{+0.007}_{-0.007}$ & $0.114^{+0.014}_{-0.014}$
\\ 
 
A665 & 
$2.11^{+0.24}_{-0.20}$ & $2.35^{+0.76}_{-0.64}$ &
$0.82^{+0.06}_{-0.05}$ & 
$0.91^{+0.11}_{-0.11}$ &
$4.16^{+1.81}_{-1.23}$ & $0.661^{+0.014}_{-0.014}$ &
$0.077^{+0.073}_{-0.056}$ & $65.2^{+2.0}_{-2.0}$ &
$2.79^{+0.12}_{-0.12}$ & $3.09^{+0.30}_{-0.31}$ &
$2.35^{+0.20}_{-0.19}$ &
$0.119^{+0.005}_{-0.005}$ & $0.131^{+0.015}_{-0.015}$
\\ 
 
A697 & 
$2.12^{+0.28}_{-0.24}$ & $2.75^{+1.38}_{-0.95}$ &
$0.98^{+0.02}_{-0.02}$ & 
$0.92^{+0.18}_{-0.15}$ &
\ldots & $0.593^{+0.012}_{-0.011}$ &
\ldots & $42.2^{+1.5}_{-1.5}$ &
$4.68^{+0.35}_{-0.38}$ & $4.38^{+0.63}_{-0.65}$ &
$3.57^{+0.52}_{-0.51}$ &
$0.131^{+0.010}_{-0.008}$ & $0.122^{+0.028}_{-0.021}$
\\ 
 
A773 & 
$1.56^{+0.19}_{-0.13}$ & $3.78^{+1.41}_{-1.18}$ &
$0.93^{+0.02}_{-0.02}$ & 
$1.31^{+0.18}_{-0.15}$ &
\ldots & $0.586^{+0.016}_{-0.016}$ &
\ldots & $41.7^{+1.9}_{-1.9}$ &
$3.06^{+0.18}_{-0.18}$ & $4.34^{+0.50}_{-0.46}$ &
$2.13^{+0.24}_{-0.22}$ &
$0.144^{+0.008}_{-0.007}$ & $0.204^{+0.029}_{-0.025}$
\\ 
 
Zw3146 & 
$2.03^{+0.07}_{-0.06}$ & $3.52^{+0.20}_{-0.19}$ &
$15.7^{+0.5}_{-0.5}$ & 
$14.2^{+1.7}_{-1.6}$ &
$4.62^{+0.18}_{-0.19}$ & $0.668^{+0.006}_{-0.006}$ &
$0.882^{+0.004}_{-0.004}$ & $26.0^{+0.9}_{-0.9}$ &
$4.71^{+0.10}_{-0.10}$ & $4.25^{+0.42}_{-0.43}$ &
$3.85^{+0.19}_{-0.18}$ &
$0.122^{+0.004}_{-0.004}$ & $0.110^{+0.012}_{-0.011}$
\\ 
 
MS1054 & 
$1.94^{+0.37}_{-0.33}$ & $0.90^{+0.63}_{-0.30}$ &
$0.55^{+0.01}_{-0.01}$ & 
$0.53^{+0.08}_{-0.07}$ &
\ldots & $1.743^{+0.188}_{-0.236}$ &
\ldots & $86.1^{+6.1}_{-8.3}$ &
$1.06^{+0.25}_{-0.25}$ & $1.03^{+0.16}_{-0.16}$ &
$1.00^{+0.34}_{-0.29}$ &
$0.106^{+0.009}_{-0.010}$ & $0.102^{+0.020}_{-0.018}$
\\ 
 
MS1137 & 
$0.96^{+0.15}_{-0.10}$ & $8.12^{+4.95}_{-3.34}$ &
$1.70^{+0.09}_{-0.09}$ & 
$2.25^{+0.32}_{-0.32}$ &
\ldots & $0.680^{+0.041}_{-0.039}$ &
\ldots & $14.6^{+1.3}_{-1.2}$ &
$1.27^{+0.11}_{-0.10}$ & $1.69^{+0.17}_{-0.17}$ &
$1.17^{+0.27}_{-0.22}$ &
$0.109^{+0.014}_{-0.012}$ & $0.144^{+0.032}_{-0.027}$
\\ 
 
MC1149 & 
$1.96^{+0.14}_{-0.12}$ & $2.46^{+0.87}_{-0.67}$ &
$0.68^{+0.02}_{-0.02}$ & 
$0.53^{+0.06}_{-0.06}$ &
\ldots & $0.625^{+0.019}_{-0.017}$ &
\ldots & $39.3^{+2.1}_{-2.0}$ &
$3.03^{+0.24}_{-0.28}$ & $2.36^{+0.21}_{-0.22}$ &
$2.82^{+0.36}_{-0.40}$ &
$0.107^{+0.007}_{-0.005}$ & $0.084^{+0.013}_{-0.011}$
\\ 
 
A1413 & 
$1.54^{+0.12}_{-0.13}$ & $6.48^{+1.87}_{-1.80}$ &
$4.78^{+0.61}_{-0.47}$ & 
$5.40^{+1.09}_{-1.03}$ &
$5.86^{+1.14}_{-1.20}$ & $0.519^{+0.010}_{-0.009}$ &
$0.752^{+0.021}_{-0.021}$ & $36.6^{+2.9}_{-2.8}$ &
$2.85^{+0.10}_{-0.18}$ & $3.18^{+0.51}_{-0.47}$ &
$2.21^{+0.15}_{-0.27}$ &
$0.129^{+0.009}_{-0.004}$ & $0.146^{+0.020}_{-0.018}$
\\ 
 
CL1226 & 
$1.53^{+0.34}_{-0.17}$ & $5.76^{+4.63}_{-3.51}$ &
$2.49^{+0.16}_{-0.14}$ & 
$1.88^{+0.42}_{-0.33}$ &
\ldots & $0.704^{+0.044}_{-0.042}$ &
\ldots & $15.5^{+1.6}_{-1.5}$ &
$4.17^{+0.44}_{-0.52}$ & $3.15^{+0.33}_{-0.32}$ &
$4.69^{+1.28}_{-1.28}$ &
$0.089^{+0.018}_{-0.012}$ & $0.067^{+0.028}_{-0.017}$
\\ 
 
MC1311 & 
$1.25^{+0.52}_{-0.13}$ & $9.44^{+9.76}_{-7.35}$ &
$4.23^{+0.39}_{-0.29}$ & 
$4.70^{+1.20}_{-0.95}$ &
\ldots & $0.615^{+0.025}_{-0.022}$ &
\ldots & $\phn9.2^{+7.2}_{-7.1}$ &
$2.22^{+0.17}_{-0.19}$ & $2.46^{+0.39}_{-0.37}$ &
$2.00^{+0.41}_{-0.42}$ &
$0.111^{+0.018}_{-0.012}$ & $0.123^{+0.038}_{-0.028}$
\\ 
 
A1689 & 
$1.88^{+0.07}_{-0.07}$ & $6.56^{+1.07}_{-0.73}$ &
$4.13^{+0.11}_{-0.10}$ & 
$4.35^{+0.36}_{-0.37}$ &
$21.8^{+1.5}_{-1.1}$ & $0.889^{+0.053}_{-0.041}$ &
$0.867^{+0.004}_{-0.004}$ & $104^{+8}_{-6}$ &
$5.56^{+0.12}_{-0.12}$ & $5.85^{+0.46}_{-0.49}$ &
$4.48^{+0.26}_{-0.22}$ &
$0.124^{+0.004}_{-0.004}$ & $0.130^{+0.010}_{-0.009}$
\\ 
 
R1347 & 
$2.10^{+0.07}_{-0.12}$ & $6.41^{+0.59}_{-0.42}$ &
$25.3^{+0.5}_{-0.6}$ & 
$16.9^{+1.9}_{-1.8}$ &
$3.88^{+0.12}_{-0.13}$ & $0.629^{+0.008}_{-0.008}$ &
$0.942^{+0.003}_{-0.004}$ & $22.9^{+1.4}_{-1.3}$ &
$9.55^{+0.26}_{-0.47}$ & $6.40^{+0.60}_{-0.58}$ &
$7.96^{+0.50}_{-0.87}$ &
$0.120^{+0.008}_{-0.004}$ & $0.081^{+0.012}_{-0.009}$
\\ 
 
MS1358 & 
$1.83^{+0.15}_{-0.15}$ & $3.93^{+0.69}_{-0.53}$ &
$10.2^{+0.7}_{-0.5}$ & 
$8.60^{+1.26}_{-1.18}$ &
$3.30^{+0.23}_{-0.24}$ & $0.676^{+0.016}_{-0.017}$ &
$0.934^{+0.003}_{-0.003}$ & $37.4^{+1.8}_{-2.0}$ &
$2.82^{+0.18}_{-0.18}$ & $2.39^{+0.30}_{-0.28}$ &
$3.14^{+0.44}_{-0.41}$ &
$0.090^{+0.007}_{-0.006}$ & $0.078^{+0.013}_{-0.012}$
\\ 
 
A1835 & 
$1.95^{+0.46}_{-0.40}$ & $4.23^{+0.13}_{-0.09}$ &
$12.6^{+0.2}_{-0.2}$ & 
$12.5^{+0.6}_{-0.8}$ &
$9.34^{+0.21}_{-0.22}$ & $0.797^{+0.009}_{-0.013}$ &
$0.940^{+0.001}_{-0.001}$ & $63.6^{+1.3}_{-1.8}$ &
$6.49^{+0.30}_{-0.33}$ & $6.41^{+0.43}_{-0.41}$ &
$5.81^{+0.88}_{-0.89}$ &
$0.112^{+0.018}_{-0.019}$ & $0.108^{+0.015}_{-0.016}$
\\ 
 
MC1423 & 
$1.30^{+0.06}_{-0.04}$ & $6.43^{+0.45}_{-0.47}$ &
$16.7^{+0.4}_{-0.3}$ & 
$16.7^{+2.7}_{-2.5}$ &
$4.08^{+0.15}_{-0.14}$ & $0.701^{+0.024}_{-0.019}$ &
$0.976^{+0.002}_{-0.002}$ & $36.0^{+1.4}_{-1.5}$&
$2.56^{+0.10}_{-0.08}$ & $2.56^{+0.36}_{-0.37}$ &
$2.08^{+0.18}_{-0.15}$ &
$0.123^{+0.005}_{-0.006}$ & $0.122^{+0.022}_{-0.019}$
\\ 
 
A1914 & 
$1.89^{+0.12}_{-0.09}$ & $8.85^{+1.73}_{-1.57}$ &
$1.49^{+0.03}_{-0.02}$ & 
$1.27^{+0.11}_{-0.11}$ &
$5.57^{+6.60}_{-2.90}$ & $0.890^{+0.012}_{-0.011}$ &
$0.009^{+0.021}_{-0.008}$ & $67.7^{+1.0}_{-1.0}$ &
$5.02^{+0.16}_{-0.13}$ & $4.29^{+0.42}_{-0.38}$ &
$4.82^{+0.46}_{-0.34}$ &
$0.104^{+0.005}_{-0.006}$ & $0.089^{+0.008}_{-0.008}$
\\ 
 
A1995 & 
$2.62^{+0.75}_{-0.39}$ & $1.70^{+0.75}_{-0.71}$ &
$0.90^{+0.01}_{-0.01}$ & 
$1.09^{+0.08}_{-0.08}$ &
\ldots & $0.916^{+0.024}_{-0.023}$ &
\ldots & $50.3^{+1.5}_{-1.4}$ &
$3.79^{+0.20}_{-0.20}$ & $4.56^{+0.35}_{-0.37}$ &
$4.42^{+0.62}_{-0.54}$ &
$0.086^{+0.007}_{-0.007}$ & $0.103^{+0.013}_{-0.012}$
\\ 
 
A2111 & 
$2.04^{+0.59}_{-0.44}$ & $2.21^{+2.36}_{-1.22}$ &
$0.55^{+0.03}_{-0.03}$ & 
$0.57^{+0.14}_{-0.12}$ &
\ldots & $0.596^{+0.032}_{-0.029}$ &
\ldots & $49.9^{+4.5}_{-4.3}$ &
$2.31^{+0.35}_{-0.37}$ & $2.39^{+0.45}_{-0.45}$ &
$2.34^{+0.64}_{-0.59}$ &
$0.099^{+0.013}_{-0.010}$ & $0.102^{+0.032}_{-0.026}$
\\ 
 
A2163 & 
$2.89^{+0.32}_{-0.23}$ & $1.57^{+0.56}_{-0.45}$ &
$0.94^{+0.01}_{-0.01}$ & 
$0.90^{+0.09}_{-0.11}$ &
$4.61^{+1.81}_{-1.15}$ & $0.559^{+0.006}_{-0.005}$ &
$0.013^{+0.014}_{-0.006}$ & $67.0^{+1.2}_{-0.9}$ &
$8.62^{+0.49}_{-0.54}$ & $8.21^{+0.66}_{-0.72}$ &
$5.40^{+0.53}_{-0.55}$ &
$0.160^{+0.007}_{-0.006}$ & $0.152^{+0.019}_{-0.021}$
\\ 
 
A2204 & 
$2.43^{+0.21}_{-0.12}$ & $4.25^{+0.33}_{-0.44}$ &
$21.1^{+0.4}_{-0.4}$ & 
$21.2^{+1.9}_{-1.7}$ &
$7.93^{+0.34}_{-0.30}$ & $0.741^{+0.028}_{-0.027}$ &
$0.957^{+0.003}_{-0.003}$ & $68.8^{+2.1}_{-1.9}$ &
$5.00^{+0.29}_{-0.22}$ & $5.03^{+0.47}_{-0.43}$ &
$5.79^{+1.00}_{-0.55}$ &
$0.086^{+0.007}_{-0.008}$ & $0.086^{+0.010}_{-0.011}$
\\ 
 
A2218 & 
$1.83^{+0.20}_{-0.15}$ & $4.27^{+1.34}_{-1.14}$ &
$0.72^{+0.01}_{-0.01}$ & 
$0.81^{+0.10}_{-0.09}$ &
\ldots & $0.746^{+0.015}_{-0.015}$ &
\ldots & $69.1^{+1.8}_{-1.8}$ &
$3.17^{+0.14}_{-0.14}$ & $3.56^{+0.48}_{-0.42}$ &
$3.07^{+0.31}_{-0.28}$ &
$0.103^{+0.005}_{-0.005}$ & $0.116^{+0.013}_{-0.012}$
\\ 
 
R1716 & 
$1.11^{+0.38}_{-0.16}$ & $5.65^{+6.76}_{-4.36}$ &
$1.53^{+0.20}_{-0.16}$ & 
$1.35^{+0.42}_{-0.37}$ &
\ldots & $0.540^{+0.052}_{-0.043}$ &
\ldots & $11.2^{+2.1}_{-1.7}$ &
$1.33^{+0.22}_{-0.29}$ & $1.16^{+0.26}_{-0.27}$ &
$1.39^{+0.49}_{-0.54}$ &
$0.095^{+0.026}_{-0.014}$ & $0.083^{+0.040}_{-0.027}$
\\ 
 
A2259 & 
$1.47^{+0.23}_{-0.17}$ & $3.74^{+1.76}_{-1.37}$ &
$0.94^{+0.04}_{-0.04}$ & 
$1.39^{+0.40}_{-0.34}$ &
\ldots & $0.557^{+0.021}_{-0.019}$ &
\ldots & $40.5^{+2.8}_{-2.7}$ &
$1.95^{+0.13}_{-0.13}$ & $1.99^{+0.56}_{-0.55}$ &
$1.54^{+0.20}_{-0.19}$ &
$0.126^{+0.009}_{-0.008}$ & $0.131^{+0.037}_{-0.038}$
\\ 
 
A2261 & 
$1.54^{+0.15}_{-0.14}$ & $6.86^{+2.16}_{-1.53}$ &
$3.93^{+0.39}_{-0.29}$ &
$4.14^{+0.64}_{-0.59}$ &
$11.7^{+1.9}_{-2.1}$ & $0.659^{+0.043}_{-0.032}$ &
$0.803^{+0.034}_{-0.039}$ & $44.6^{+10.2}_{-6.5}$ &
$3.34^{+0.22}_{-0.23}$ & $3.50^{+0.44}_{-0.38}$ &
$2.55^{+0.40}_{-0.38}$ &
$0.131^{+0.013}_{-0.011}$ & $0.138^{+0.022}_{-0.019}$
\\ 
 
MS2053 & 
$0.92^{+0.16}_{-0.16}$ & $6.58^{+6.41}_{-4.06}$ &
$1.15^{+0.14}_{-0.11}$ & 
$1.77^{+0.60}_{-0.45}$ &
\ldots & $0.552^{+0.051}_{-0.055}$ &
\ldots & $12.5^{+1.8}_{-2.1}$ &
$0.99^{+0.18}_{-0.21}$ & $1.53^{+0.25}_{-0.23}$ &
$0.84^{+0.36}_{-0.33}$ &
$0.117^{+0.035}_{-0.021}$ & $0.183^{+0.082}_{-0.056}$
\\ 
 
MC2129 & 
$1.28^{+0.17}_{-0.12}$ & $6.29^{+0.38}_{-0.26}$ &
$1.75^{+0.07}_{-0.07}$ & 
$1.69^{+0.27}_{-0.23}$ &
\ldots & $0.602^{+0.022}_{-0.020}$ &
\ldots & $18.6^{+1.2}_{-1.1}$ &
$3.53^{+0.23}_{-0.23}$ & $3.41^{+0.40}_{-0.40}$ &
$2.35^{+0.32}_{-0.30}$ &
$0.150^{+0.012}_{-0.010}$ & $0.145^{+0.027}_{-0.022}$
\\ 
 
R2129 & 
$1.68^{+0.14}_{-0.15}$ & $4.17^{+0.88}_{-0.75}$ &
$14.1^{+1.6}_{-1.2}$ & 
$11.3^{+1.9}_{-2.0}$ &
$3.44^{+0.44}_{-0.44}$ & $0.584^{+0.016}_{-0.015}$ &
$0.910^{+0.009}_{-0.010}$ & $25.7^{+2.9}_{-2.6}$ &
$2.80^{+0.16}_{-0.19}$ & $2.23^{+0.31}_{-0.34}$ &
$2.29^{+0.30}_{-0.32}$ &
$0.122^{+0.011}_{-0.008}$ & $0.098^{+0.019}_{-0.018}$
\\ 
 
MC2214 & 
$1.64^{+0.23}_{-0.20}$ & $4.95^{+2.62}_{-1.69}$ &
$1.46^{+0.05}_{-0.05}$ & 
$1.58^{+0.20}_{-0.17}$ &
\ldots & $0.606^{+0.019}_{-0.019}$ &
\ldots & $22.3^{+1.3}_{-1.3}$ &
$4.15^{+0.37}_{-0.44}$ & $4.49^{+0.37}_{-0.38}$ &
$3.48^{+0.67}_{-0.70}$ &
$0.119^{+0.015}_{-0.011}$ & $0.129^{+0.032}_{-0.019}$
\\ 
 
MC2228 & 
$1.94^{+0.29}_{-0.22}$ & $1.92^{+1.05}_{-0.83}$ &
$1.31^{+0.07}_{-0.06}$ & 
$1.58^{+0.28}_{-0.22}$ &
\ldots & $0.518^{+0.013}_{-0.013}$ &
\ldots & $21.4^{+1.5}_{-1.4}$ &
$2.99^{+0.37}_{-0.45}$ & $3.63^{+0.33}_{-0.34}$ &
$2.19^{+0.49}_{-0.54}$ &
$0.136^{+0.018}_{-0.012}$ & $0.165^{+0.040}_{-0.027}$
\\ 
 
\enddata 
\end{deluxetable}

\begin{deluxetable}{lccccc}
\tablewidth{0pt}
\tablecaption{Model Parameters, Cluster Masses and Gas Mass Fractions:
Isothermal $\beta$ Model Constraints Derived from SZE Imaging Data\label{tab:sze_masses}}
\tablehead{
\colhead{Cluster} &
\colhead{$\theta_c$} &
\colhead{\dTo} &
\colhead{\Mgas(\rtfh)} & 
\colhead{\Mtot(\rtfh)} &
\colhead{\fg(\rtfh)} 
\\
\colhead{} & 
\colhead{($\arcsec$)} &
\colhead{(mK)} &
\colhead{($10^{13}\,\Msun$)} & 
\colhead{($10^{14}\,\Msun$)} & 
\colhead{} 
}
\tablecolumns{6}
\startdata
CL0016 &
$\phn52.0^{+\phn9.0}_{-\phn8.1}$ & $-1.449^{+0.088}_{-0.094}$ &
$4.65^{+0.42}_{-0.39}$ & $2.83^{+0.32}_{-0.32}$ &
$0.165^{+0.033}_{-0.027}$ 
\\

A267 &
$\phn38.6^{+15.8}_{-12.8}$ & $-0.735^{+0.084}_{-0.099}$ &
$2.74^{+0.68}_{-0.57}$ & $1.99^{+0.25}_{-0.23}$ &
$0.137^{+0.054}_{-0.038}$ 
\\                     

A370 &
$\phn21.5^{+\phn9.3}_{-\phn7.4}$ & $-1.072^{+0.144}_{-0.244}$ &
$2.58^{+0.47}_{-0.40}$ & $3.18^{+0.22}_{-0.22}$ &
$0.081^{+0.021}_{-0.016}$ 
\\                     

MS0451 &
$\phn31.4^{+\phn7.1}_{-\phn6.4}$ & $-1.610^{+0.100}_{-0.112}$ &
$5.03^{+0.83}_{-0.71}$ & $3.46^{+0.32}_{-0.31}$ &
$0.145^{+0.039}_{-0.030}$ 
\\                     

MC0647 &
$\phn19.1^{+\phn8.1}_{-\phn6.6}$ & $-1.390^{+0.151}_{-0.242}$ &
$2.81^{+0.72}_{-0.57}$ & $6.21^{+0.86}_{-0.74}$ &
$0.045^{+0.018}_{-0.012}$ 
\\                     

A\phn665 &
$\phn56.8^{+23.5}_{-17.5}$ & $-0.801^{+0.083}_{-0.105}$ &
$2.43^{+0.71}_{-0.53}$ & $2.77^{+0.21}_{-0.29}$ &
$0.087^{+0.038}_{-0.023}$ 
\\

A773 &
$\phn30.9^{+11.7}_{-\phn9.8}$ & $-0.985^{+0.109}_{-0.128}$ &
$2.56^{+0.67}_{-0.54}$ & $3.12^{+0.23}_{-0.22}$ &
$0.082^{+0.027}_{-0.020}$ 
\\                     

Zw3146 &
$\phn29.4^{+11.9}_{-\phn9.7}$ & $-1.223^{+0.177}_{-0.242}$ &
$3.95^{+0.82}_{-0.73}$ & $3.43^{+0.17}_{-0.19}$ &
$0.115^{+0.030}_{-0.024}$ 
\\                     

MS1054 &
$\phn54.5^{+11.4}_{-10.4}$ & $-1.483^{+0.191}_{-0.253}$ &
$1.17^{+0.16}_{-0.15}$ & $0.62^{+0.21}_{-0.16}$ &
$0.187^{+0.081}_{-0.054}$ 
\\

MS1137 &
$\phn12.2^{+\phn8.4}_{-\phn4.4}$ & $-1.021^{+0.301}_{-0.441}$ &
$2.09^{+0.38}_{-0.36}$ & $0.96^{+0.15}_{-0.14}$ &
$0.221^{+0.064}_{-0.051}$ 
\\                     

MC1149 &
$\phn53.2^{+19.1}_{-16.6}$ & $-1.181^{+0.127}_{-0.160}$ &
$3.00^{+0.42}_{-0.37}$ & $2.17^{+0.55}_{-0.53}$ &
$0.139^{+0.059}_{-0.038}$ 
\\

CL1226 &
$\phn11.1^{+\phn8.2}_{-\phn4.5}$ & $-2.171^{+0.677}_{-1.164}$ &
$2.82^{+0.63}_{-0.51}$ & $5.08^{+1.05}_{-0.88}$ &
$0.056^{+0.026}_{-0.017}$ 
\\

A1689 &
$\phn25.2^{+\phn6.9}_{-\phn5.9}$ & $-1.749^{+0.171}_{-0.213}$ &
$3.57^{+0.69}_{-0.57}$ & $5.21^{+0.23}_{-0.23}$ &
$0.069^{+0.015}_{-0.012}$ 
\\

R1347 &
$\phn28.7^{+10.5}_{-\phn8.1}$ & $-2.235^{+0.316}_{-0.429}$ &
$5.88^{+0.92}_{-0.85}$ & $8.35^{+0.56}_{-0.56}$ &
$0.070^{+0.015}_{-0.013}$ 
\\

A1835 &
$\phn50.1^{+\phn9.2}_{-\phn8.2}$ & $-1.590^{+0.112}_{-0.113}$ &
$6.20^{+0.93}_{-0.82}$ & $5.38^{+0.34}_{-0.31}$ &
$0.115^{+0.024}_{-0.020}$ 
\\

A1914 &
$\phn29.4^{+11.3}_{-\phn8.9}$ & $-1.776^{+0.200}_{-0.317}$ &
$4.33^{+0.96}_{-0.75}$ & $4.69^{+0.18}_{-0.16}$ &
$0.092^{+0.023}_{-0.017}$ 
\\                     

A1995 &
$\phn34.6^{+\phn5.5}_{-\phn5.4}$ & $-1.051^{+0.062}_{-0.063}$ &
$4.49^{+0.63}_{-0.57}$ & $3.63^{+0.22}_{-0.21}$ &
$0.124^{+0.024}_{-0.020}$ 
\\

A2163 &
$102.7^{+17.0}_{-15.1}$ & $-1.898^{+0.276}_{-0.336}$ &
$7.57^{+1.61}_{-1.38}$ & $6.17^{+0.38}_{-0.44}$ &
$0.123^{+0.036}_{-0.028}$ 
\\

A2204 &
$100.2^{+20.6}_{-18.5}$ & $-1.938^{+0.296}_{-0.382}$ &
$8.96^{+2.57}_{-2.13}$ & $4.95^{+0.46}_{-0.41}$ &
$0.180^{+0.069}_{-0.051}$ 
\\

A2218 &
$\phn79.0^{+13.6}_{-12.4}$ & $-1.059^{+0.143}_{-0.176}$ &
$5.34^{+1.23}_{-1.03}$ & $2.95^{+0.21}_{-0.20}$ &
$0.181^{+0.055}_{-0.042}$ 
\\

A2261 &
$\phn20.9^{+\phn9.8}_{-\phn7.5}$ & $-1.179^{+0.148}_{-0.239}$ &
$2.46^{+0.75}_{-0.54}$ & $2.91^{+0.27}_{-0.23}$ &
$0.084^{+0.031}_{-0.021}$ 
\\

MC2214 &
$\phn27.3^{+\phn8.3}_{-\phn7.1}$ & $-1.518^{+0.160}_{-0.227}$ &
$4.37^{+0.60}_{-0.55}$ & $3.87^{+0.41}_{-0.38}$ &
$0.113^{+0.028}_{-0.022}$ 
\\

MC2228 &
$\phn47.3^{+18.7}_{-15.5}$ & $-1.315^{+0.137}_{-0.164}$ &
$3.91^{+0.65}_{-0.56}$ & $2.15^{+0.40}_{-0.43}$ &
$0.182^{+0.076}_{-0.049}$ 
\\

\enddata
\end{deluxetable}

\begin{deluxetable}{lcc}
\tablecaption{Mean Gas Mass Fractions \label{tab:fgas}}
\tablecolumns{3}\
\tablewidth{0pt}
\tablehead{
\colhead{Method used in spatial analysis} &
\colhead{$f_{\mbox{\scriptsize gas}}^{\mbox{\scriptsize X-ray}}$} &
\colhead{$f_{\mbox{\scriptsize gas}}^{\mbox{\scriptsize SZE}}$}
}
\startdata
Joint fit to X-ray+SZE data, 100~kpc-cut isothermal $\beta$ model &
$0.110^{+0.003}_{-0.003}\,^{+0.006}_{-0.018}$ & $0.116^{+0.005}_{-0.005}\,^{+0.009}_{-0.026}$ \\

Joint fit to X-ray+SZE data, non-isothermal double-$\beta$ model &
$0.119^{+0.003}_{-0.003}\,^{+0.007}_{-0.014}$ & $0.121^{+0.005}_{-0.005}\,^{+0.009}_{-0.016}$ \\

Spatial fit to SZE data only, isothermal $\beta$ model &
\nodata & $0.120^{+0.009}_{-0.009}\,^{+0.009}_{-0.027}$ \\
\enddata
\end{deluxetable}

\begin{deluxetable}{lccc}
\tablecaption{Comparison of Cool-Core and Non Cool-Core Results \label{tab:cores_nocores}}
\tablecolumns{4}\
\tablewidth{0pt}
\tablehead{
\colhead{Analysis Method} &
\colhead{Dataset} &
\colhead{Cool-Core} &
\colhead{Non Cool-Core}
}
\startdata
Joint fit X-ray+SZE 100~kpc-cut isothermal $\beta$ model & X-ray &
$0.110\pm0.004$ & $0.109\pm0.004$ \\
 & SZE &
$0.107\pm0.007$ & $0.120\pm0.006$ \\

 & & & \\

Joint fit X-ray+SZE non-isothermal double-$\beta$ model & X-ray &
$0.118\pm0.005$ & $0.120\pm0.005$ \\

 & SZE &
$0.113\pm0.007$ & $0.125\pm0.007$ \\

 & & & \\

SZE-data only, 100~kpc-cut X-ray temperature & SZE &
$0.098\pm0.015$ & $0.129\pm0.011$ \\

 & & & \\

Joint fit X-ray+SZE no-cut isothermal $\beta$ model & X-ray &
$0.139\pm0.009$ & $0.106\pm0.004$ \\

 & SZE &
$0.158\pm0.014$ & $0.118\pm0.006$ \\

 & & & \\

SZE-data only, no-cut X-ray temperature$^{\rm a}$ & SZE &
$0.148\pm0.037$ & $0.123\pm0.009$ \\
\enddata
\tablenotetext{a}{
The large statistical uncertainties for this
subsample reflect the small sample sizes: six clusters in the cool-core case and
seventeen in the non cool-core case.}
\end{deluxetable}

\begin{deluxetable}{lcc}
\tablecaption{Constraints on \mbox{$\eta_{\mbox{\scriptsize gas}}$} \label{tab:fg_bc_results}}
\tablecolumns{3}\
\tablewidth{0pt}

\tablehead{
\colhead{Method used in spatial analysis} &
\colhead{\mbox{$\eta_{\mbox{\scriptsize gas}}^{\mbox{\scriptsize X-ray}}$}} &
\colhead{\mbox{$\eta_{\mbox{\scriptsize gas}}^{\mbox{\scriptsize SZE}}$}}
}

\startdata

Joint fit to X-ray+SZE data, 100~kpc-cut isothermal $\beta$ model &
$0.65^{+0.03}_{-0.03}\,^{+0.04}_{-0.10}$ & 
$0.65^{+0.02}_{-0.03}\,^{+0.05}_{-0.15}$ \\

Joint fit to X-ray+SZE data, non-isothermal double-$\beta$ model &
$0.71^{+0.03}_{-0.03}\,^{+0.04}_{-0.09}$ & 
$0.67^{+0.03}_{-0.03}\,^{+0.05}_{-0.09}$ \\

Spatial fit to SZE data only, isothermal $\beta$ model &
\nodata & $0.64^{+0.05}_{-0.03}\,^{+0.06}_{-0.14}$ \\

\enddata
\end{deluxetable}

\clearpage

\newpage
\section*{Appendix A: Surface brightness profiles \label{appendixA}}

\begin{figure}[hp]
\centerline{
  \includegraphics[scale=0.28,angle=270]{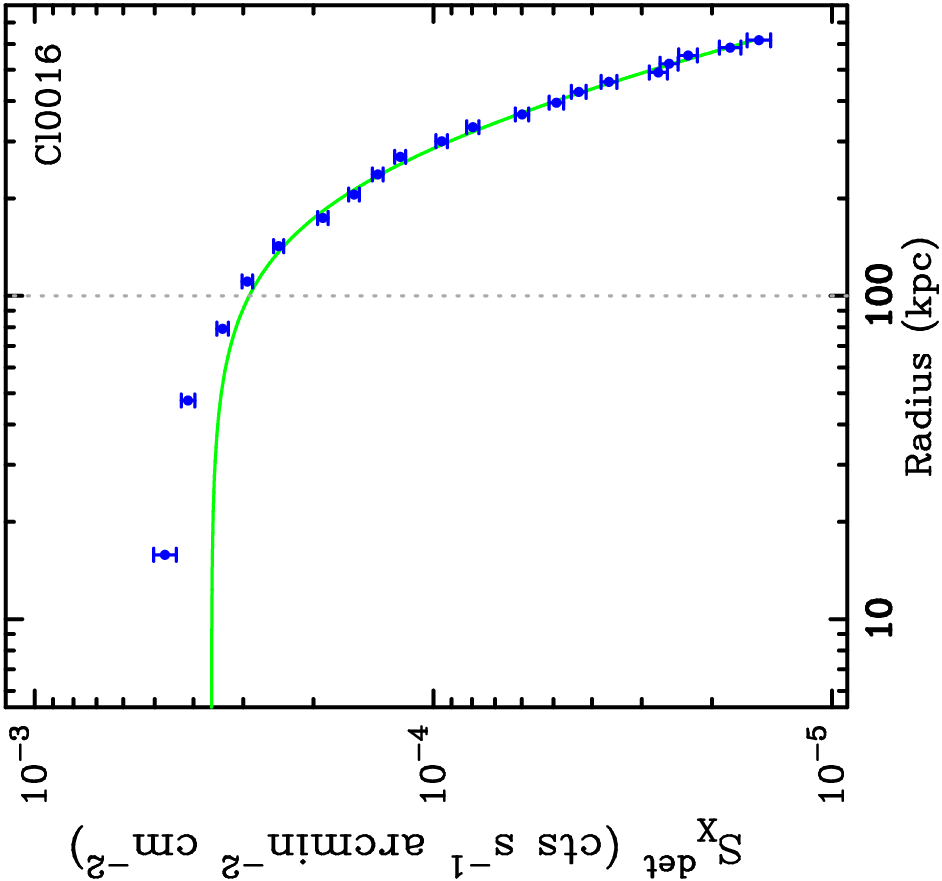}
  \includegraphics[scale=0.28,angle=270]{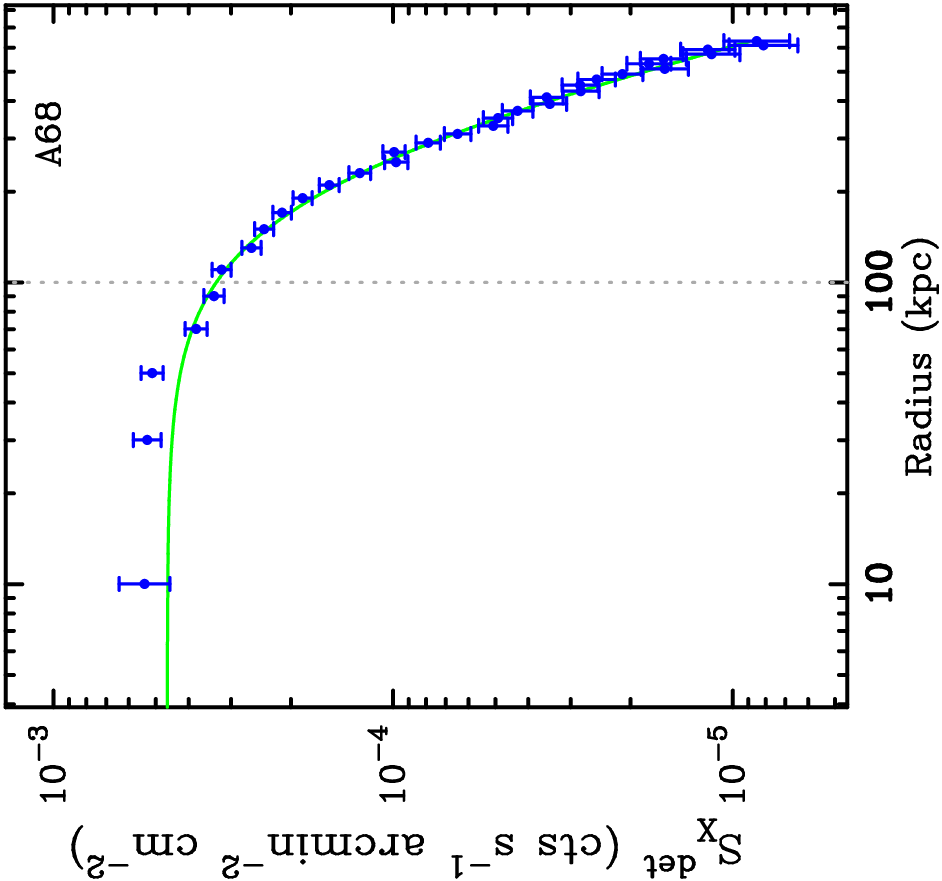}
  \includegraphics[scale=0.28,angle=270]{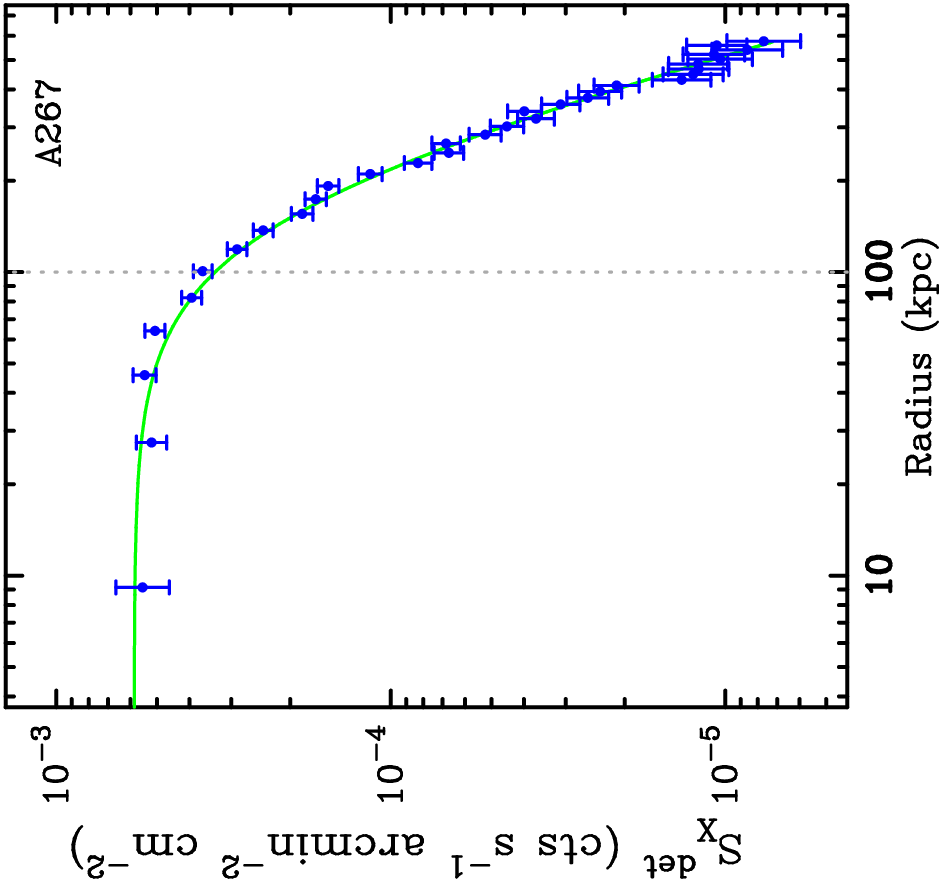}
}
\vspace{0.15in}
\centerline{
  \includegraphics[scale=0.28,angle=270]{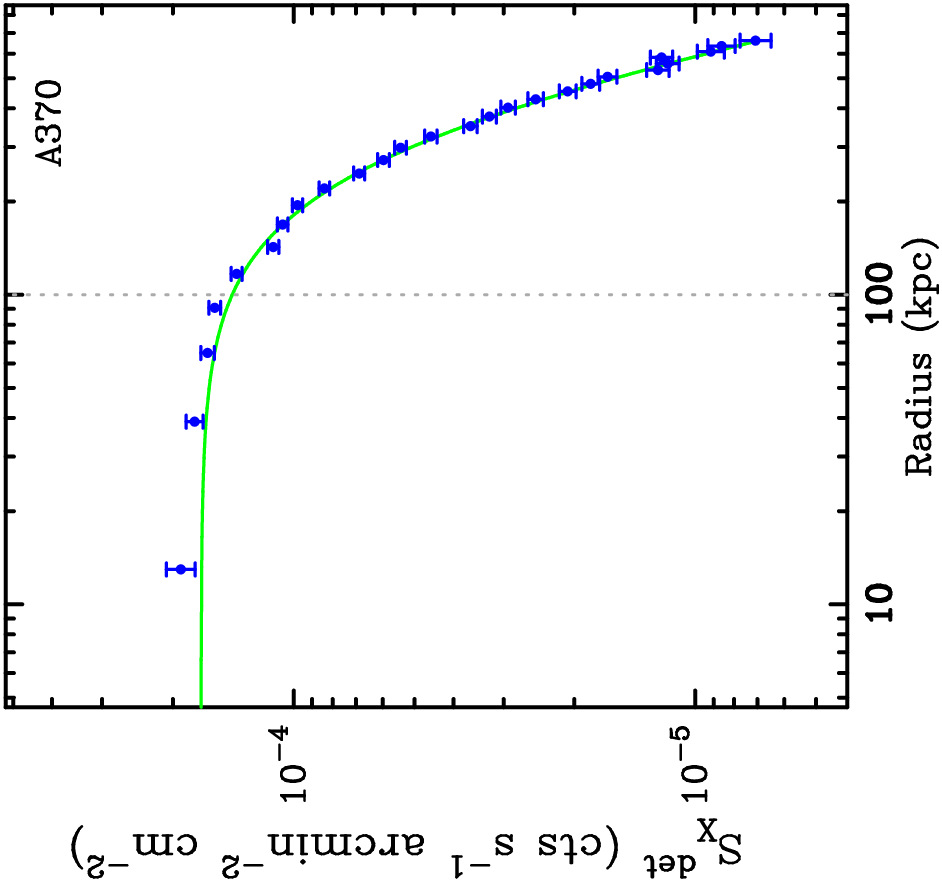}
  \includegraphics[scale=0.28,angle=270]{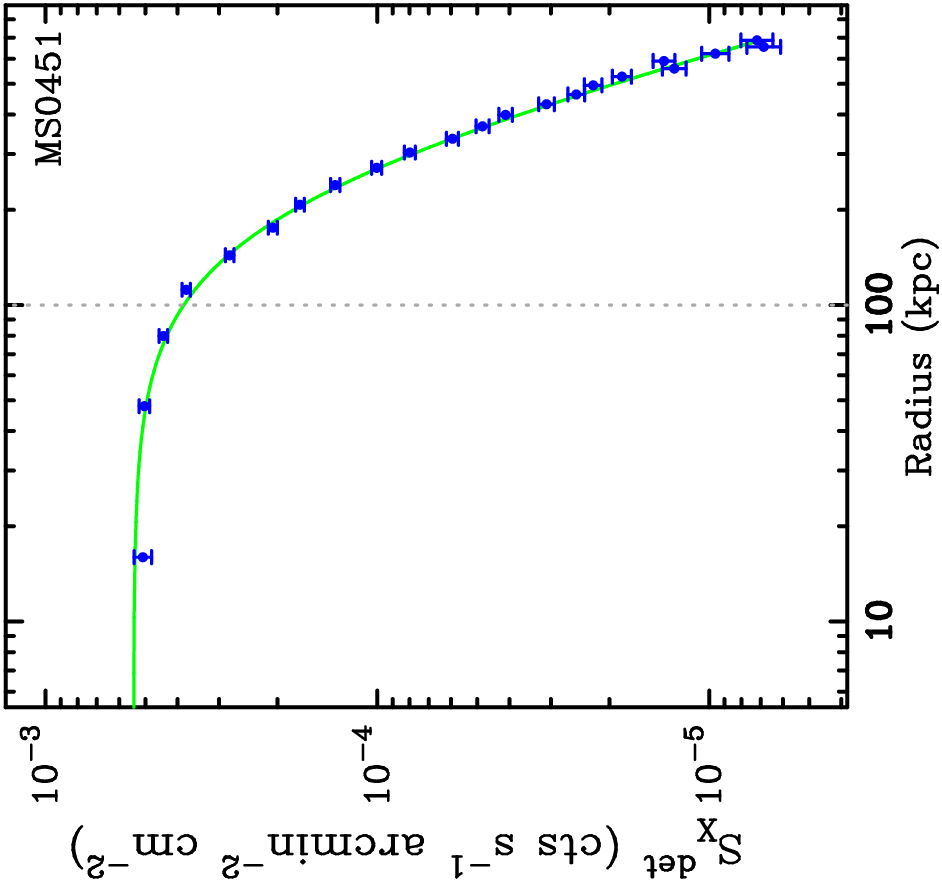}
  \includegraphics[scale=0.28,angle=270]{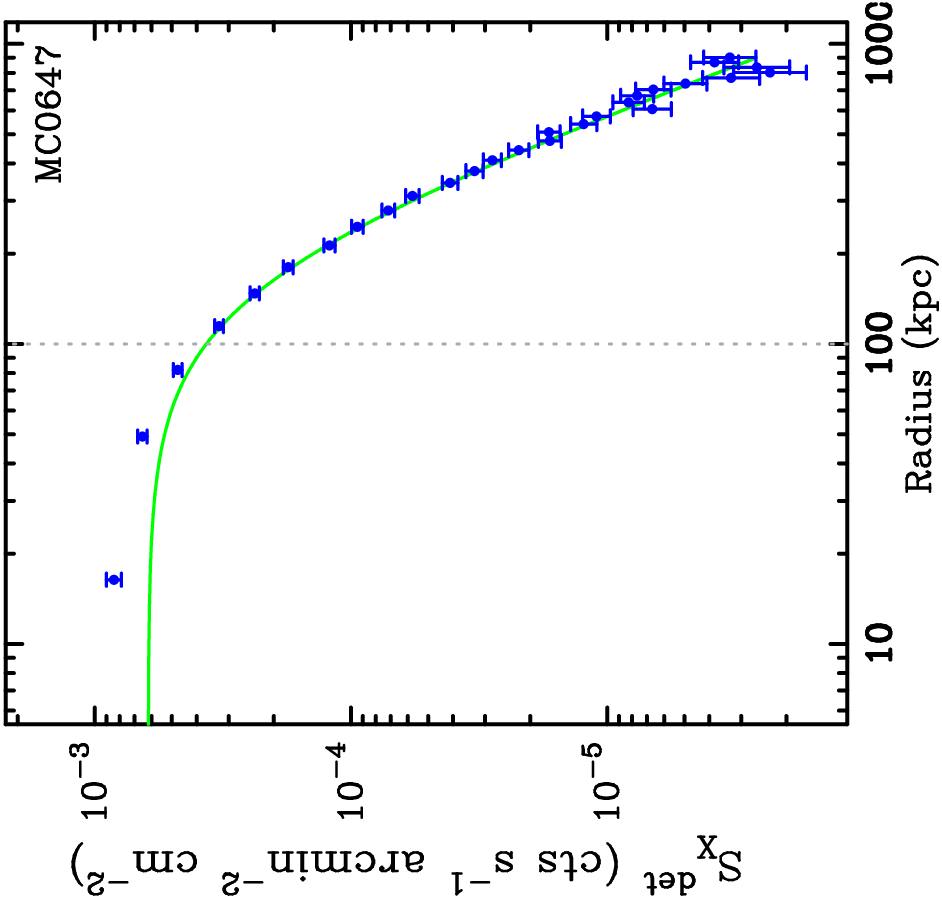}
}
\vspace{0.15in}
\centerline{
  \includegraphics[scale=0.28,angle=270]{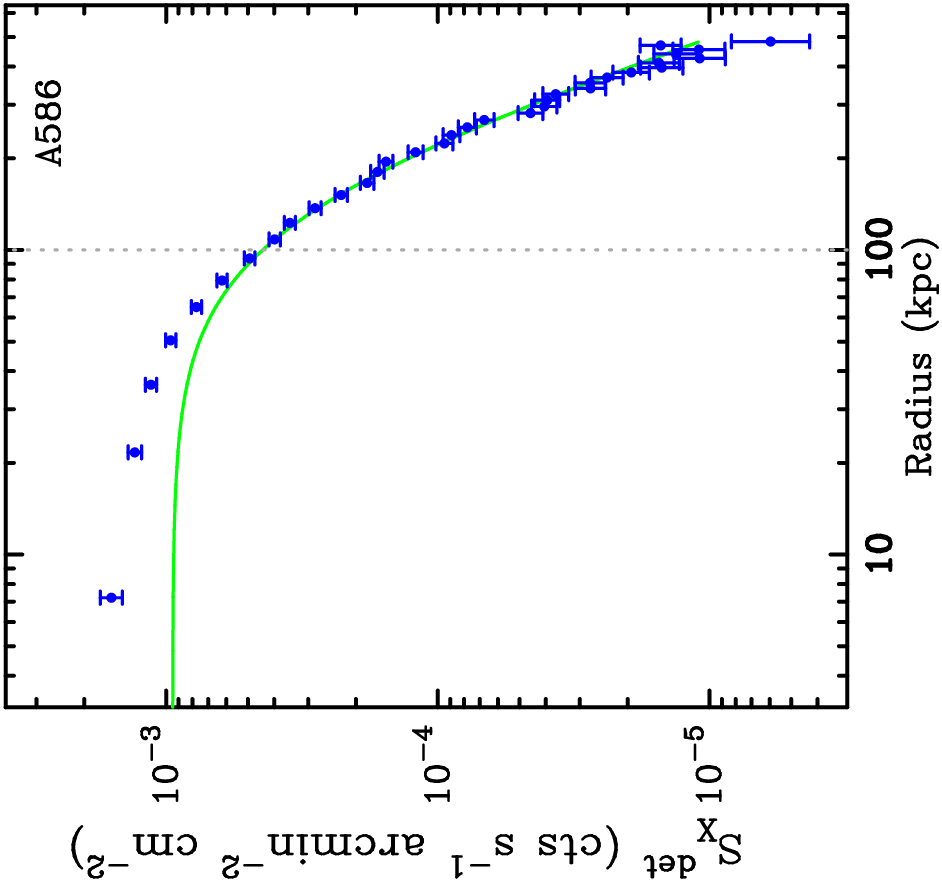}
  \includegraphics[scale=0.28,angle=270]{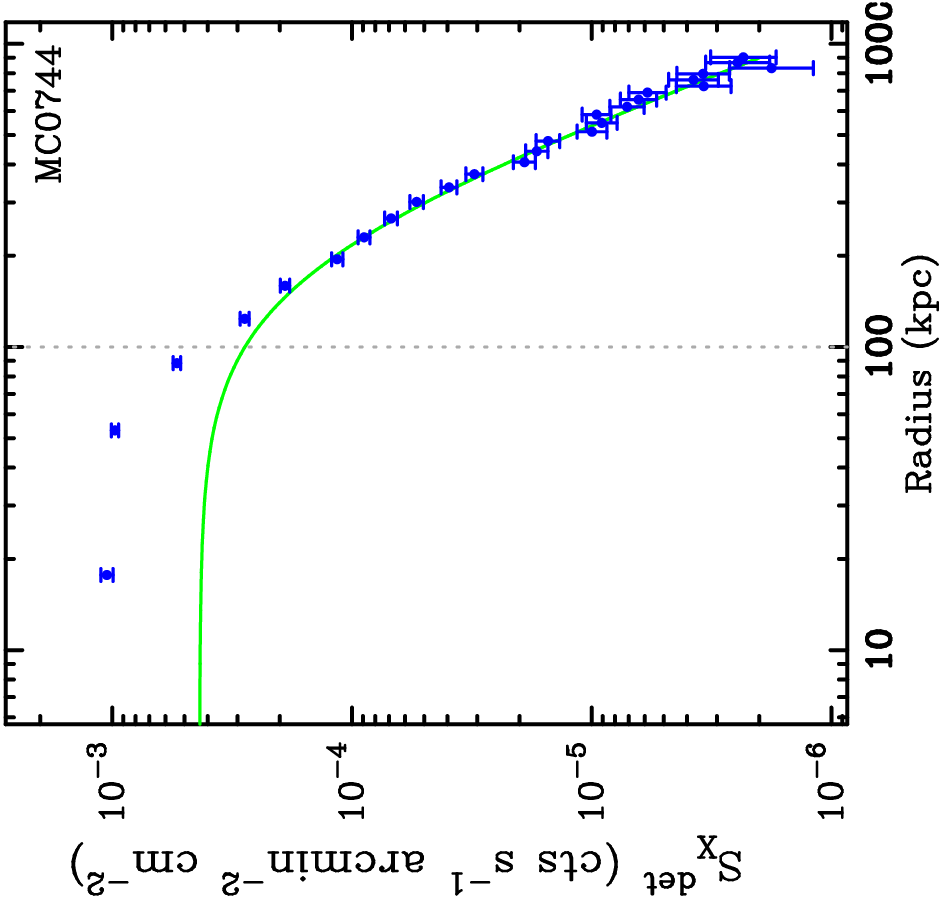}
  \includegraphics[scale=0.28,angle=270]{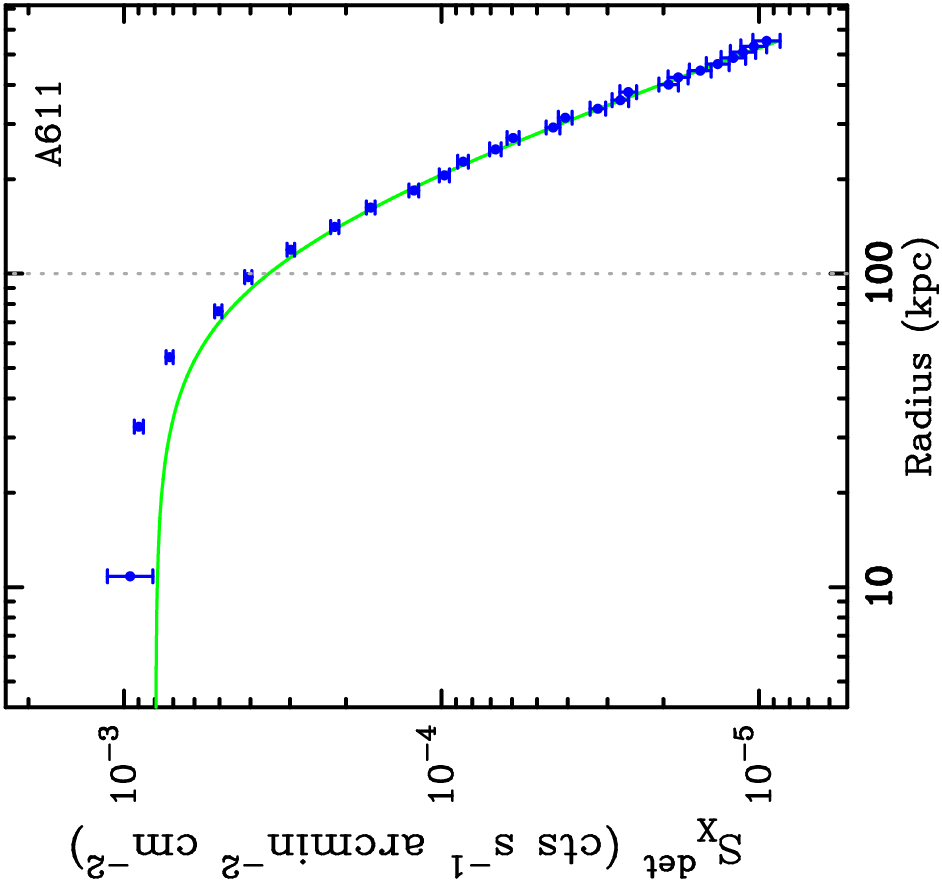}
}
\caption[X-ray surface brightness radial profiles with best fit
models]{X-ray surface brightness radial profiles (points) with best
fit 100~kpc-cut models (solid lines).  Radii are calculated using a
reasonable input cosmology of \Om$=0.3$, \Ol$=0.7$, and $h=0.7$.  The
vertical dotted line is drawn at 100~kpc; data within this radius are
not used in the fit.}
\label{fig:radprof}
\end{figure}

\begin{figure}
\centerline{
  \includegraphics[scale=0.3,angle=270]{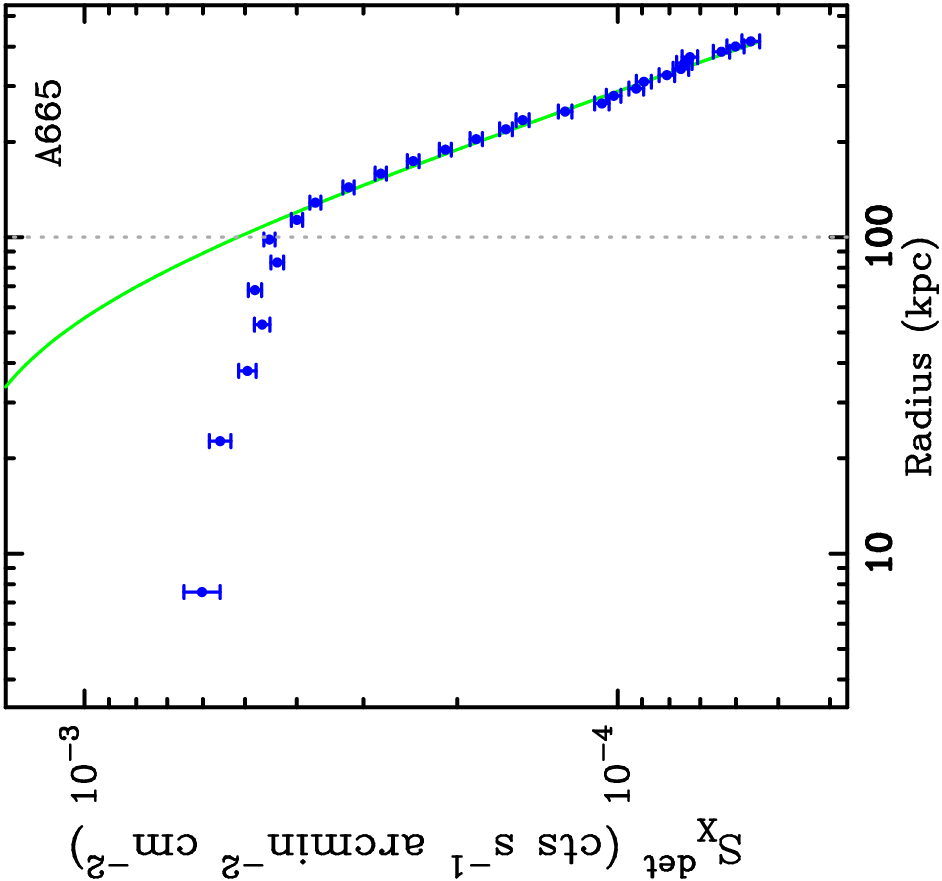}
  \includegraphics[scale=0.3,angle=270]{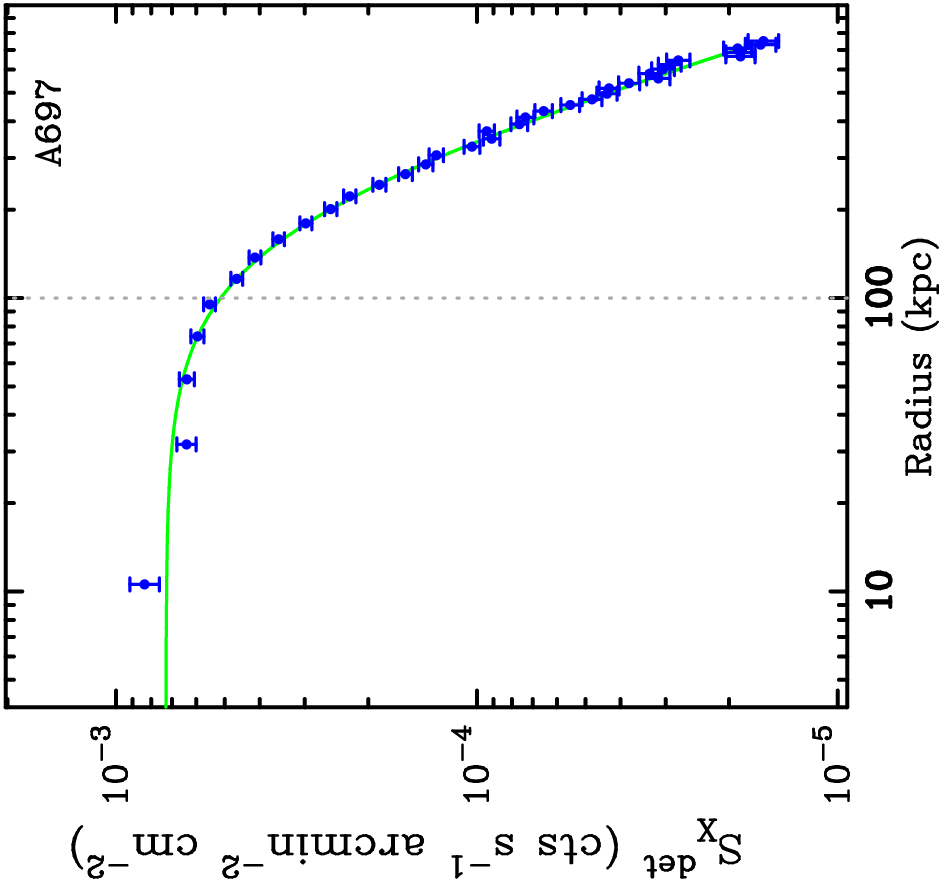}
  \includegraphics[scale=0.3,angle=270]{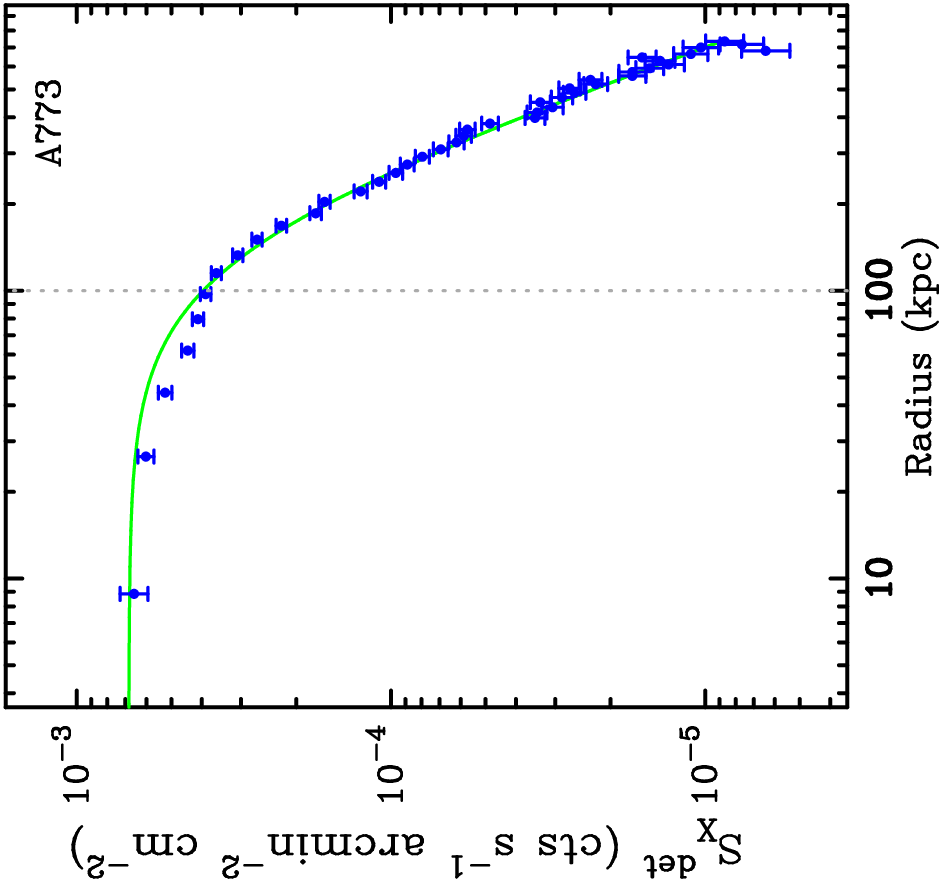}
}
\vspace{0.15in}
\centerline{
  \includegraphics[scale=0.3,angle=270]{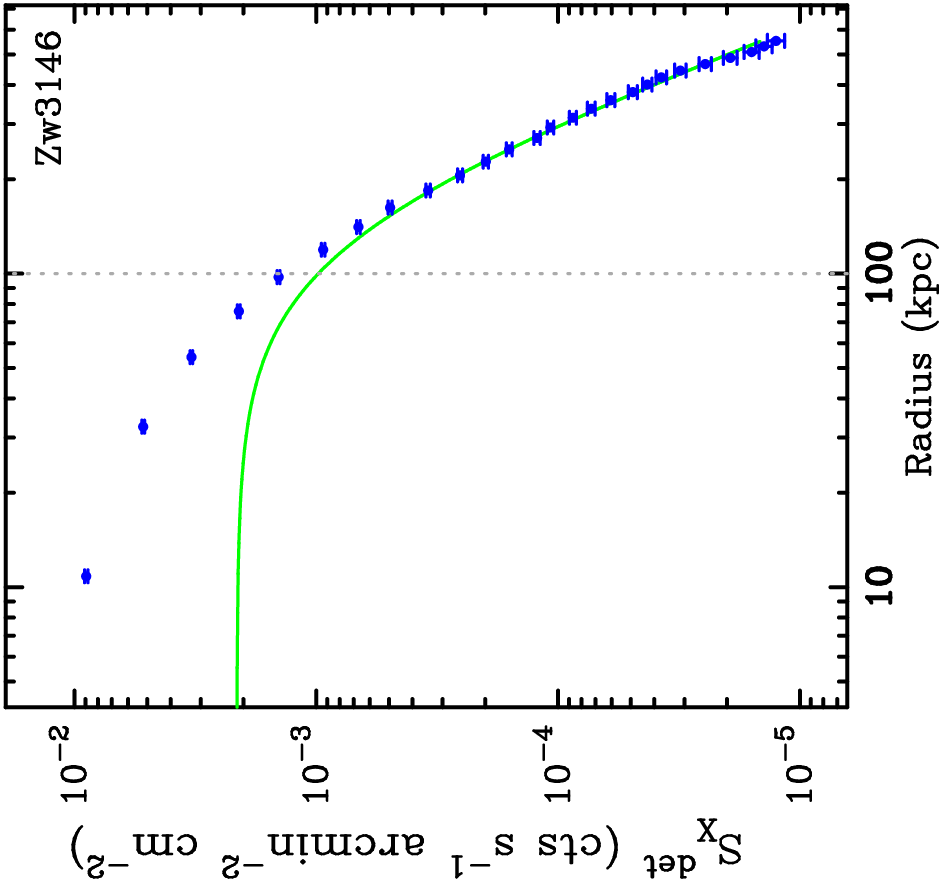}
  \includegraphics[scale=0.3,angle=270]{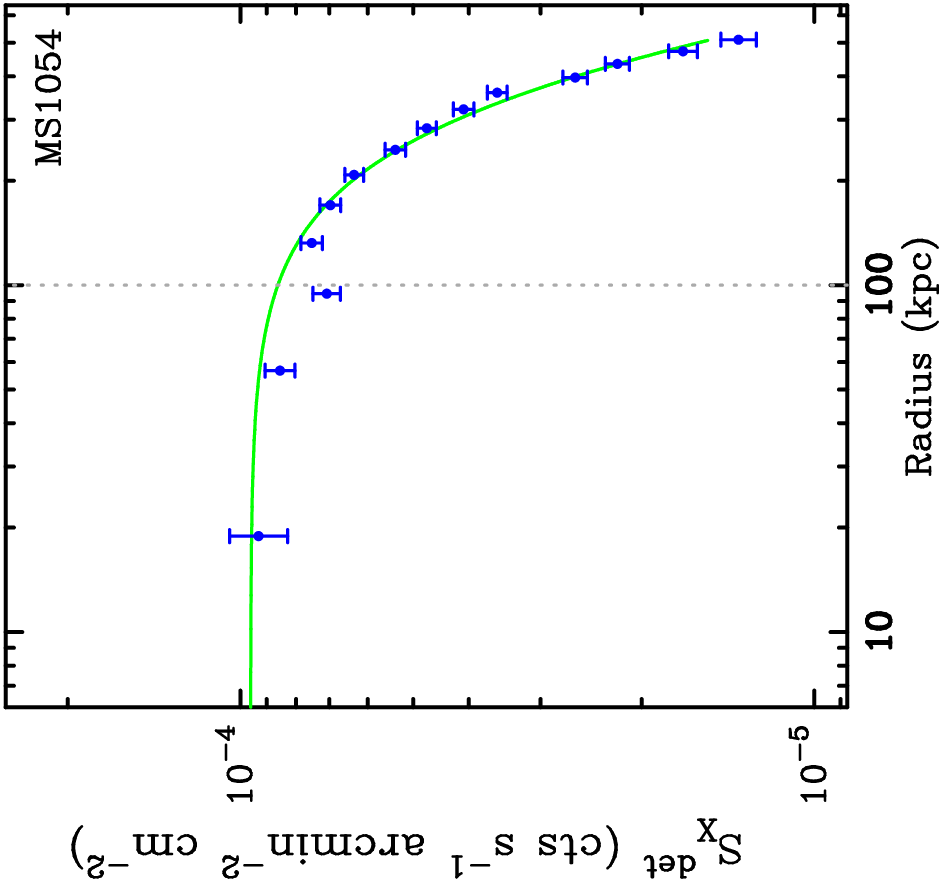}
  \includegraphics[scale=0.3,angle=270]{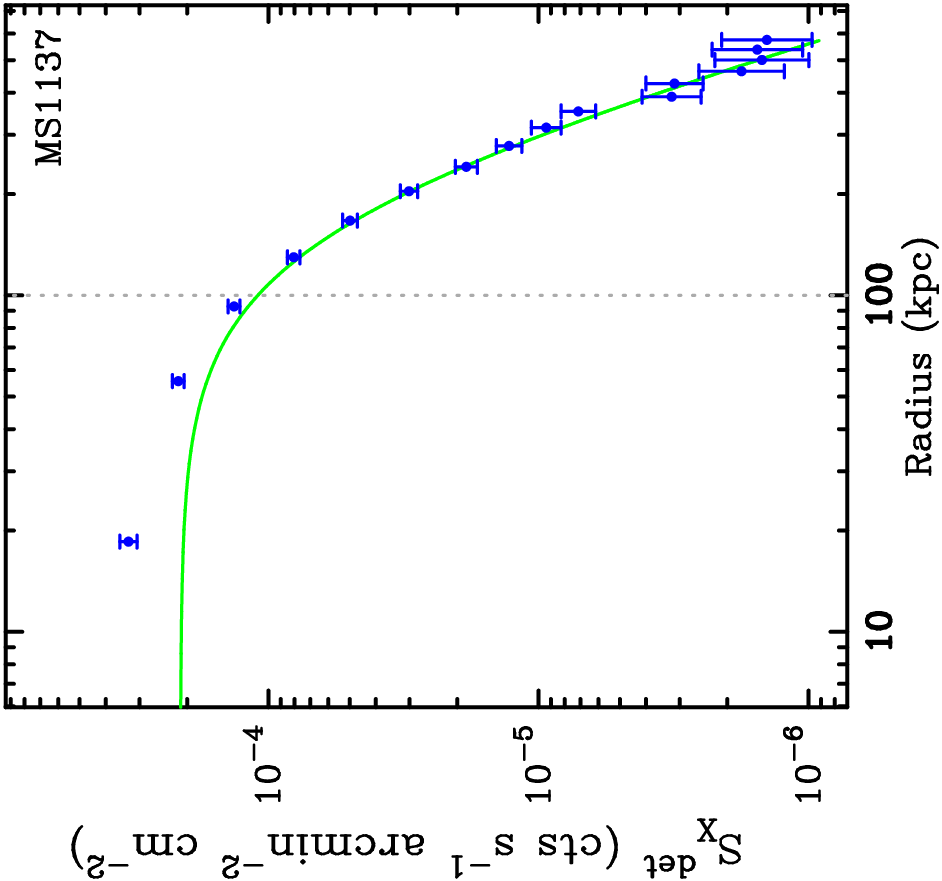}
}
\vspace{0.15in}
\centerline{
  \includegraphics[scale=0.3,angle=270]{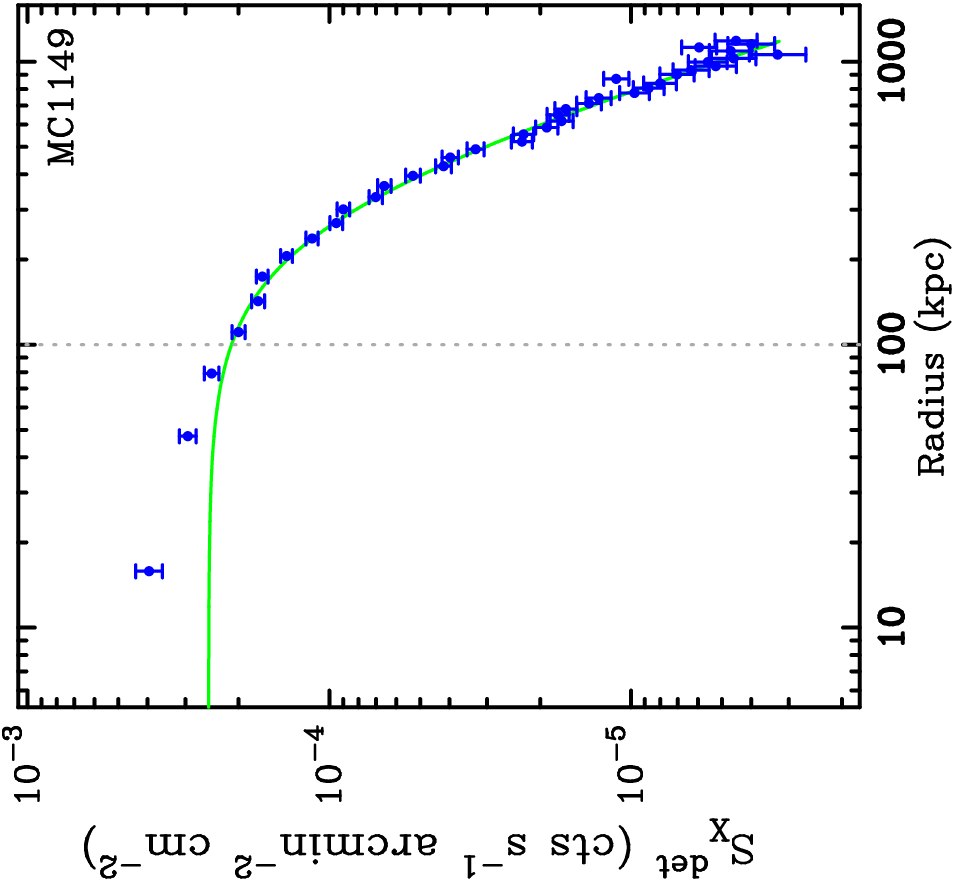}
  \includegraphics[scale=0.3,angle=270]{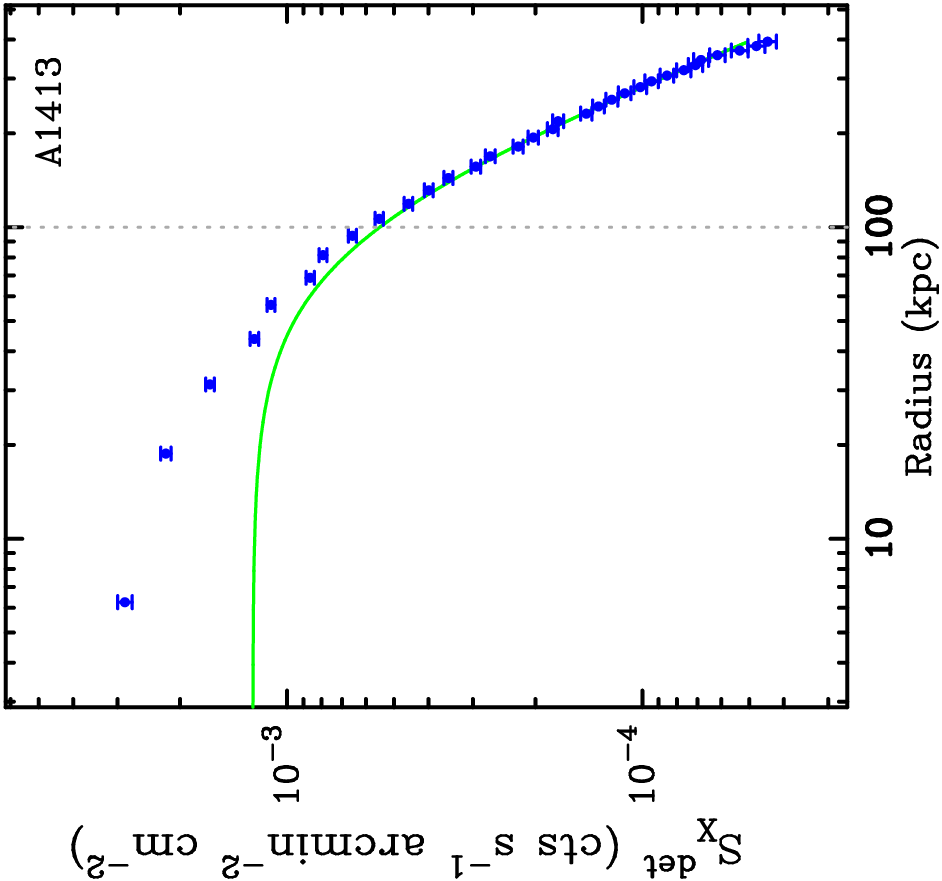}
  \includegraphics[scale=0.3,angle=270]{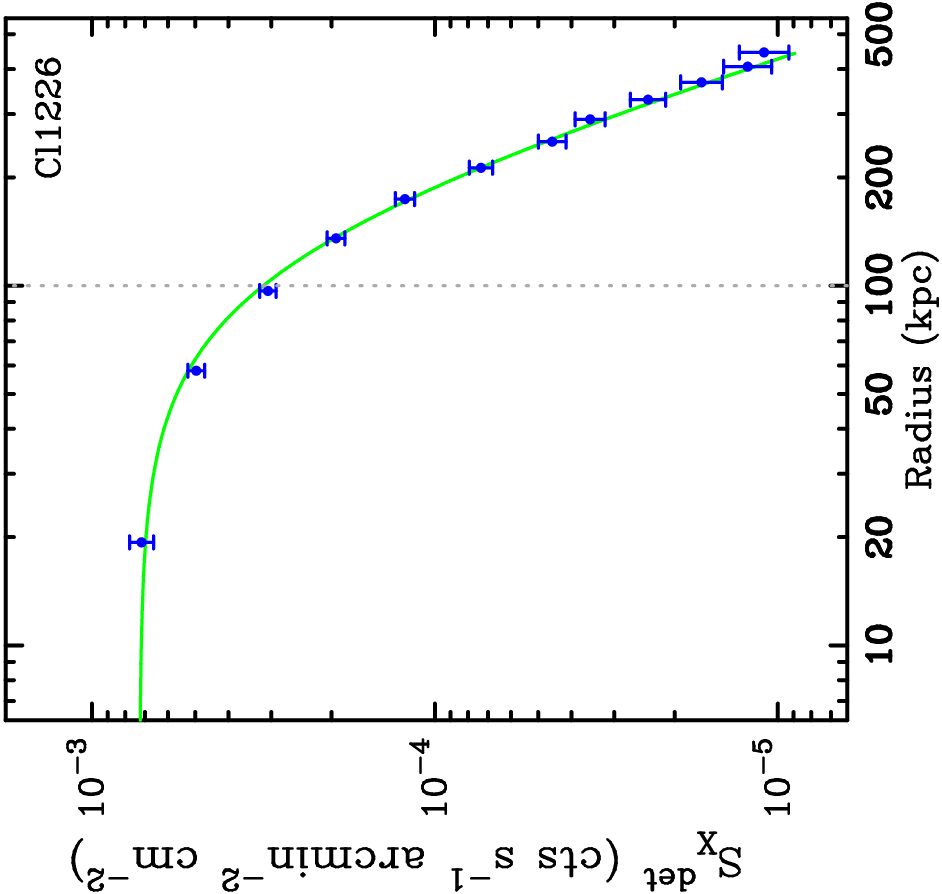}
}
\contcaption{Cont.}
\end{figure}
                                                                                                                                           
\begin{figure}
\centerline{
  \includegraphics[scale=0.3,angle=270]{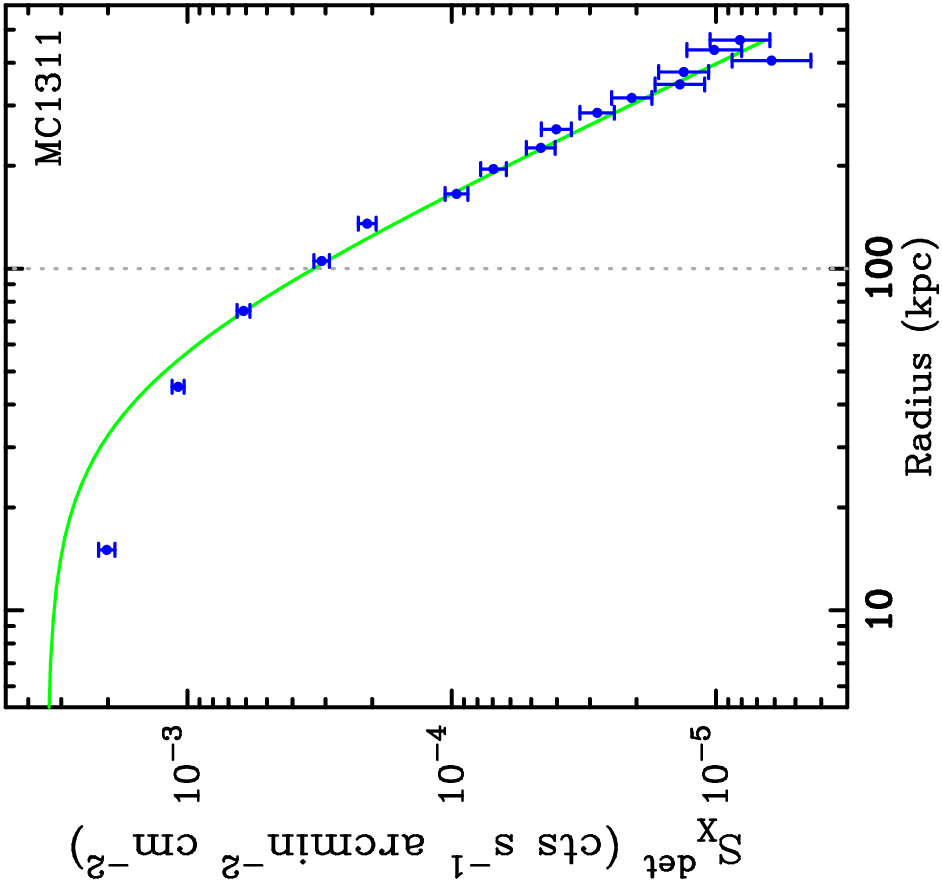}
  \includegraphics[scale=0.3,angle=270]{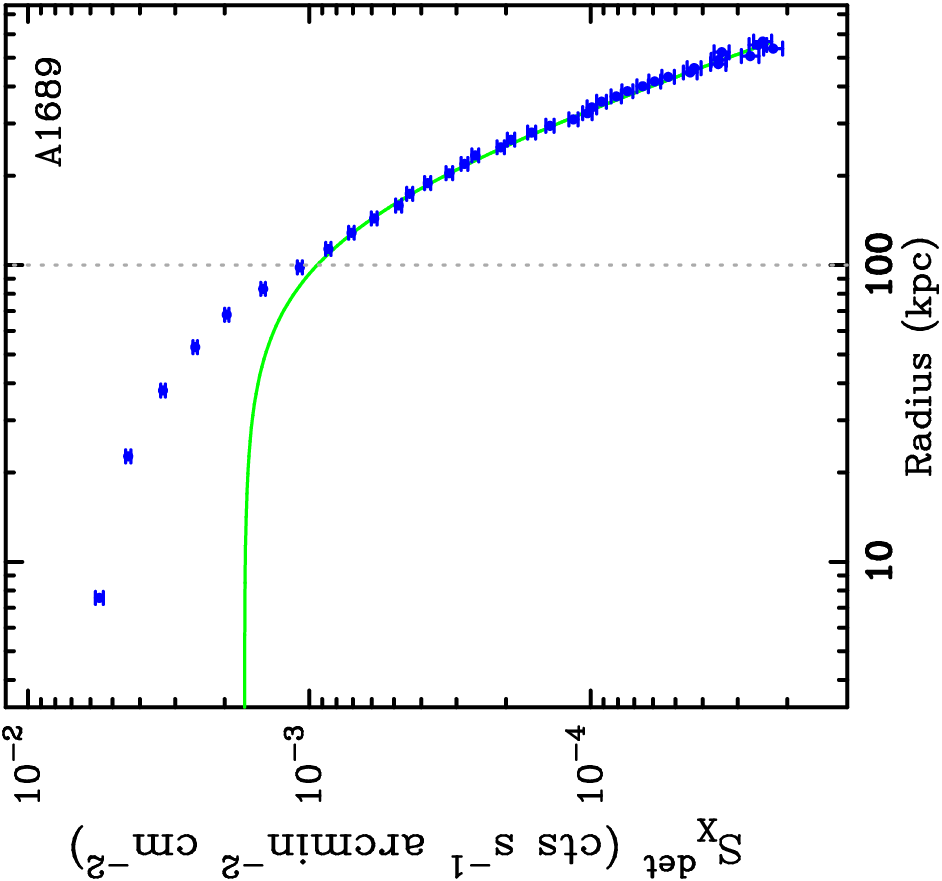}
  \includegraphics[scale=0.3,angle=270]{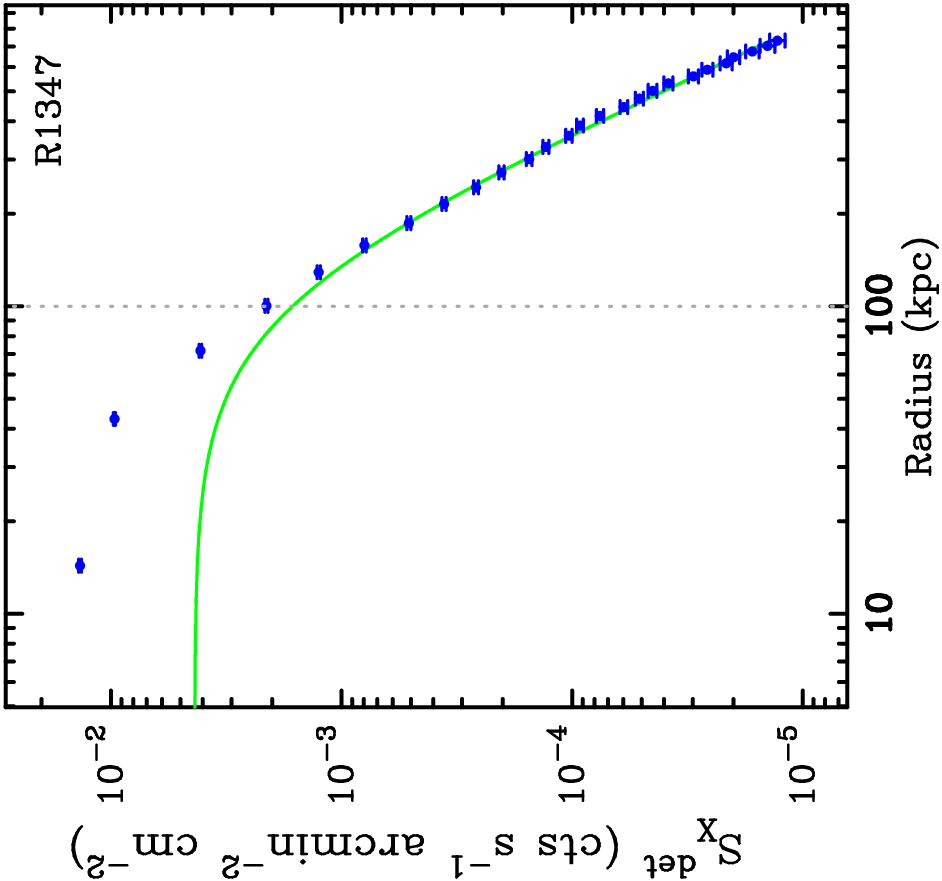}
}
\vspace{0.15in}
\centerline{
  \includegraphics[scale=0.3,angle=270]{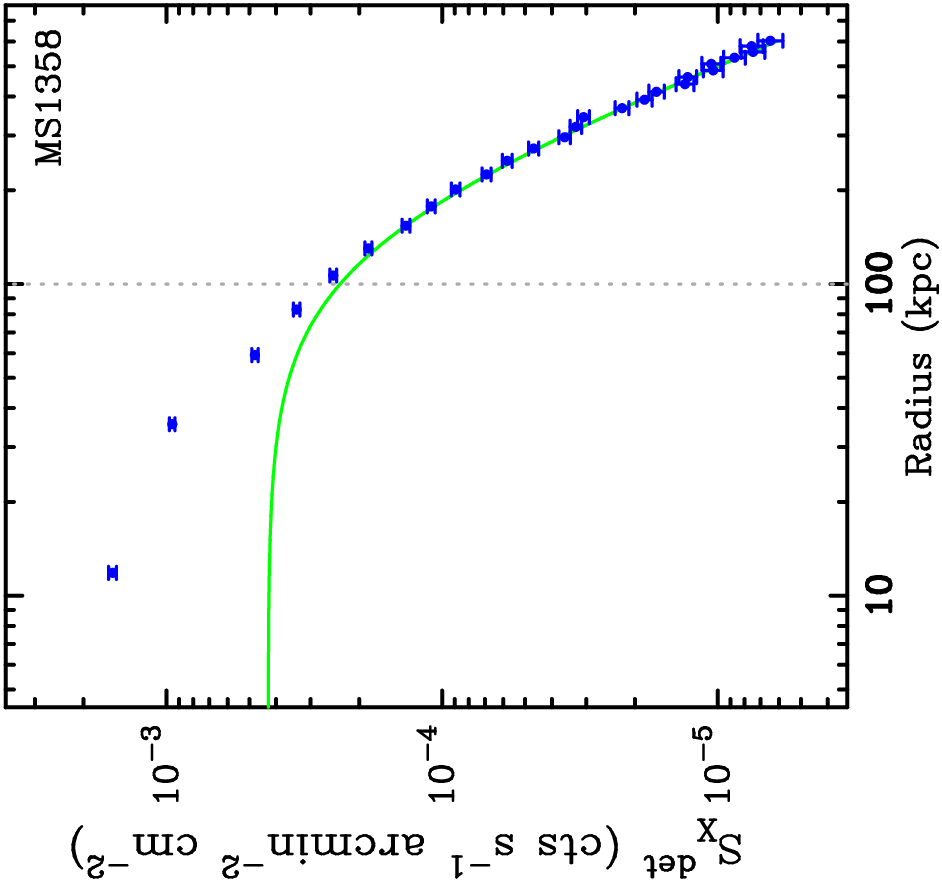}
  \includegraphics[scale=0.3,angle=270]{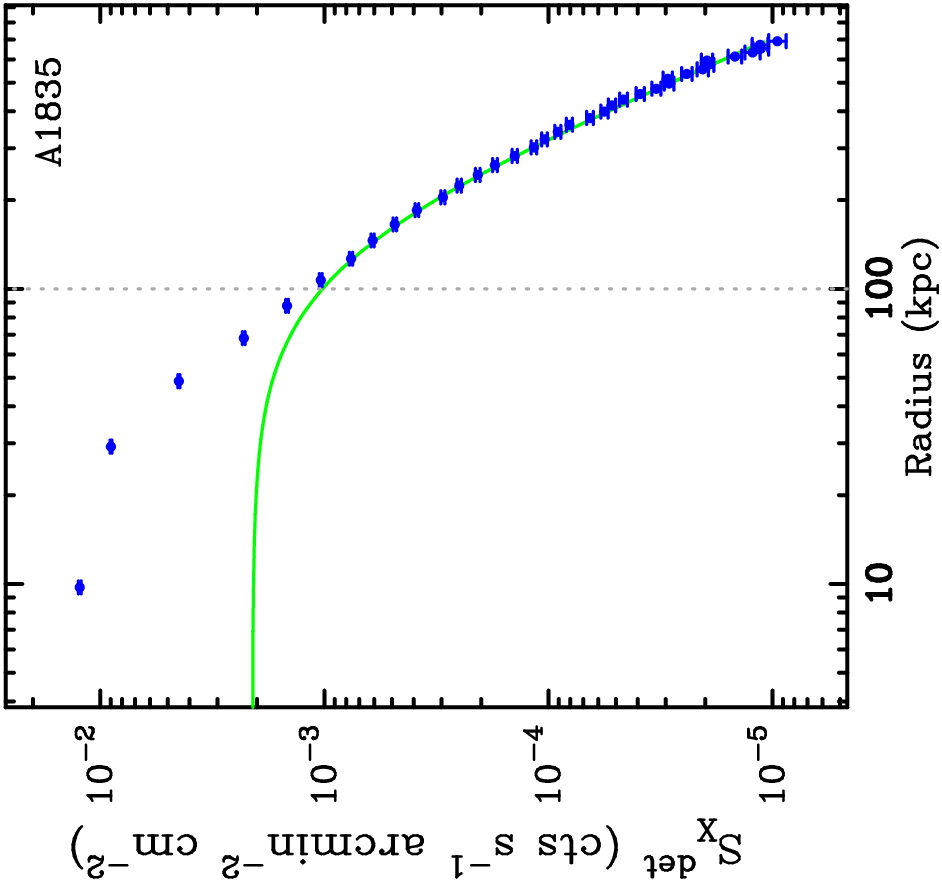}
  \includegraphics[scale=0.3,angle=270]{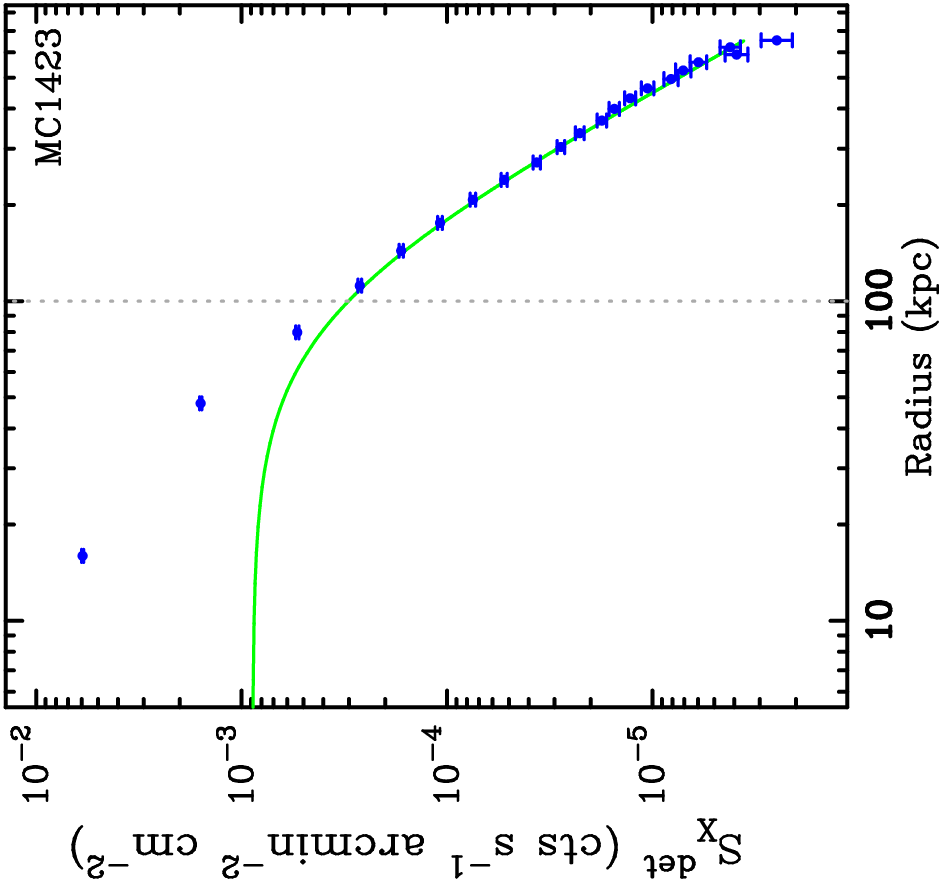}
}
\vspace{0.15in}
\centerline{
  \includegraphics[scale=0.3,angle=270]{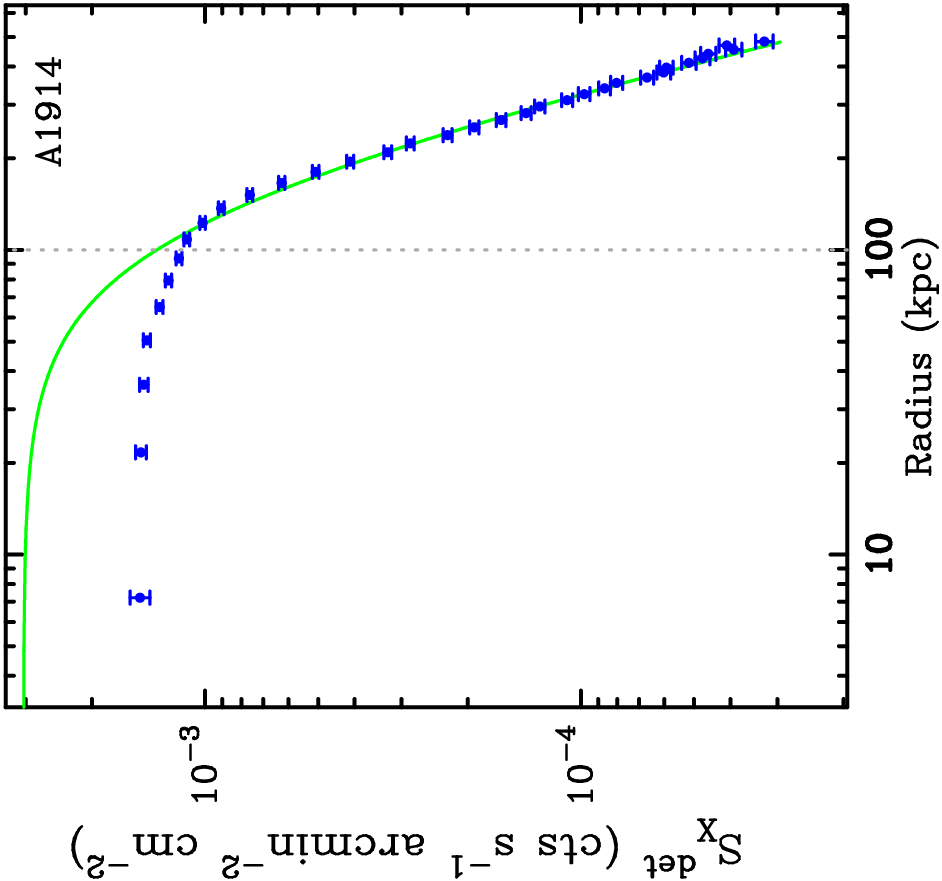}
  \includegraphics[scale=0.3,angle=270]{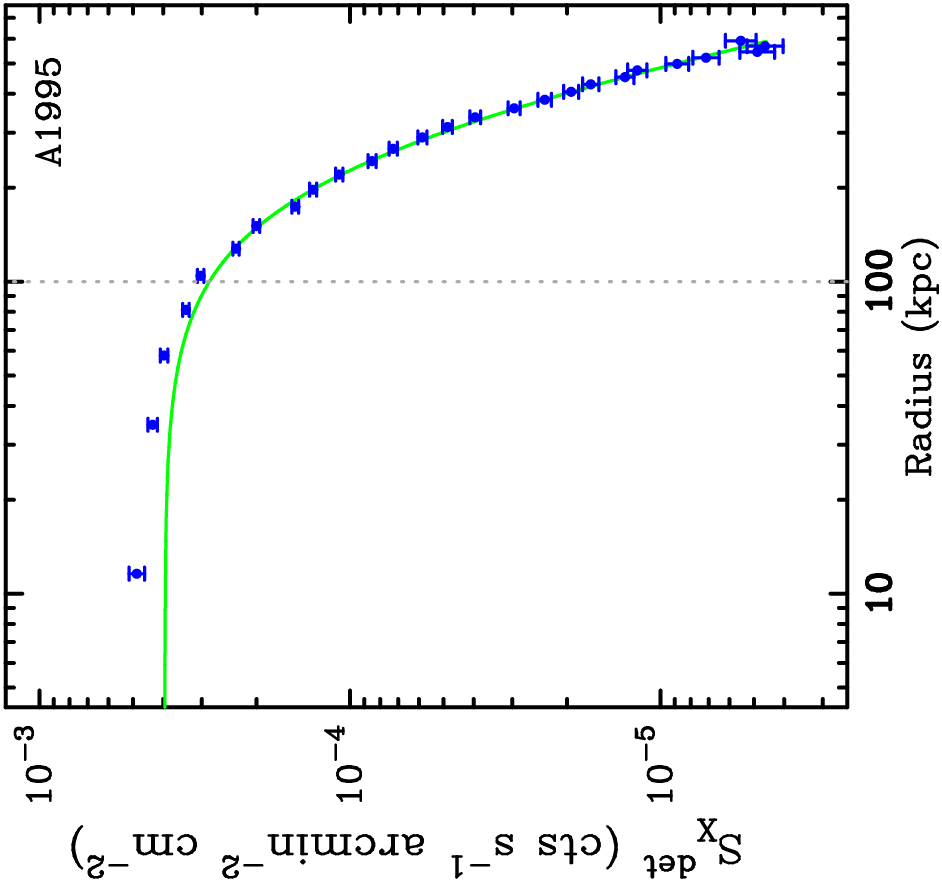}
  \includegraphics[scale=0.3,angle=270]{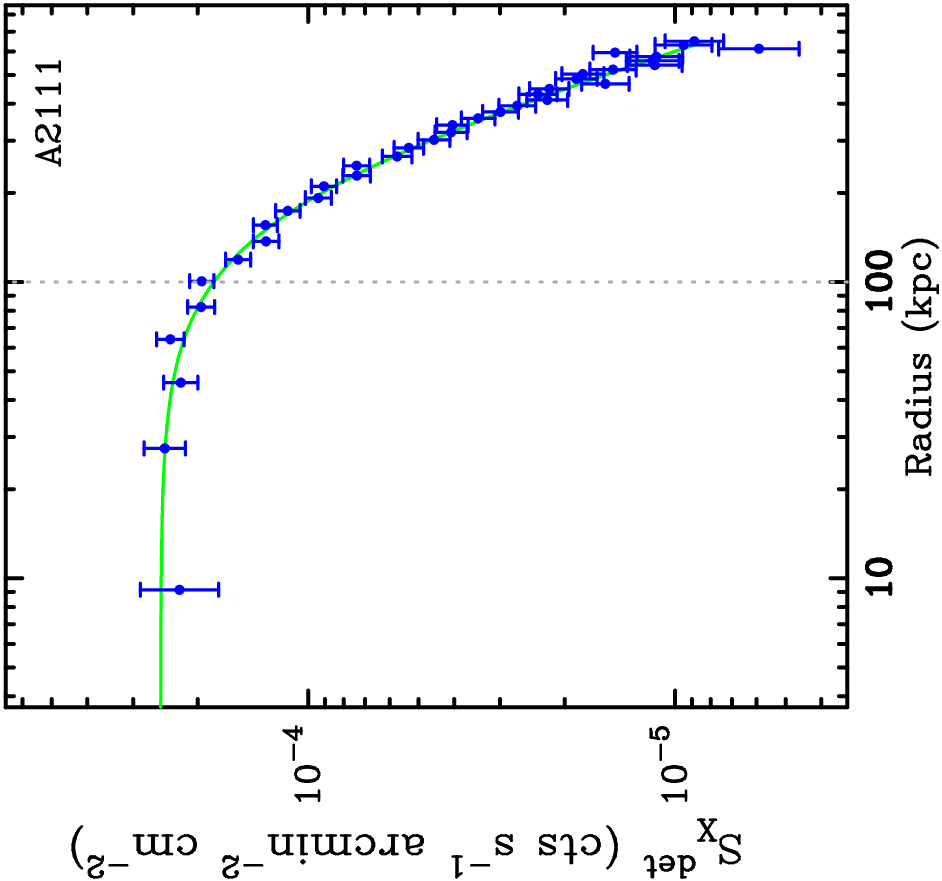}
}
\contcaption{Cont.}
\end{figure}

\begin{figure}
\centerline{
  \includegraphics[scale=0.3,angle=270]{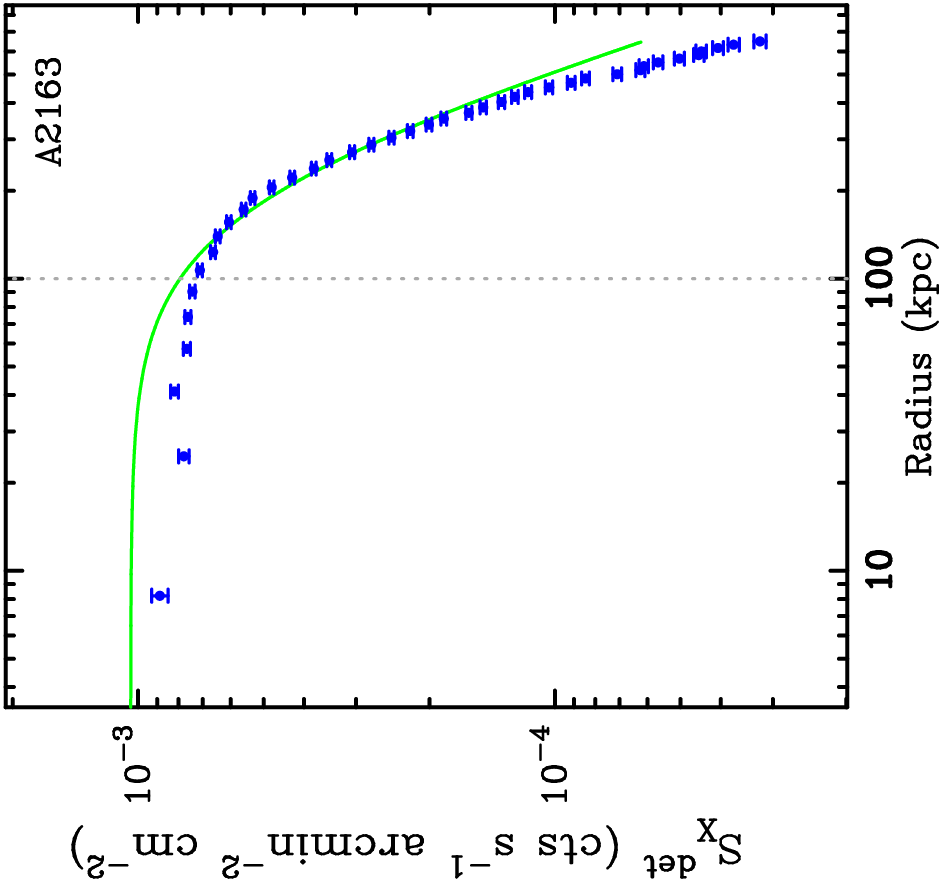}
  \includegraphics[scale=0.3,angle=270]{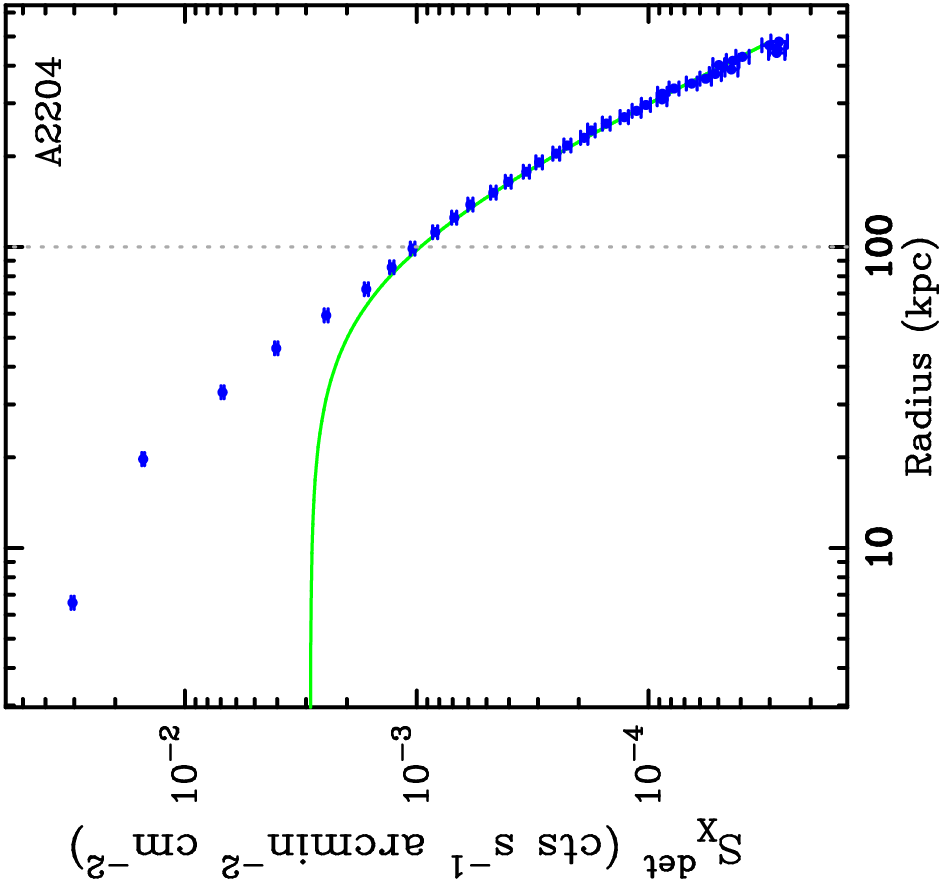}
  \includegraphics[scale=0.3,angle=270]{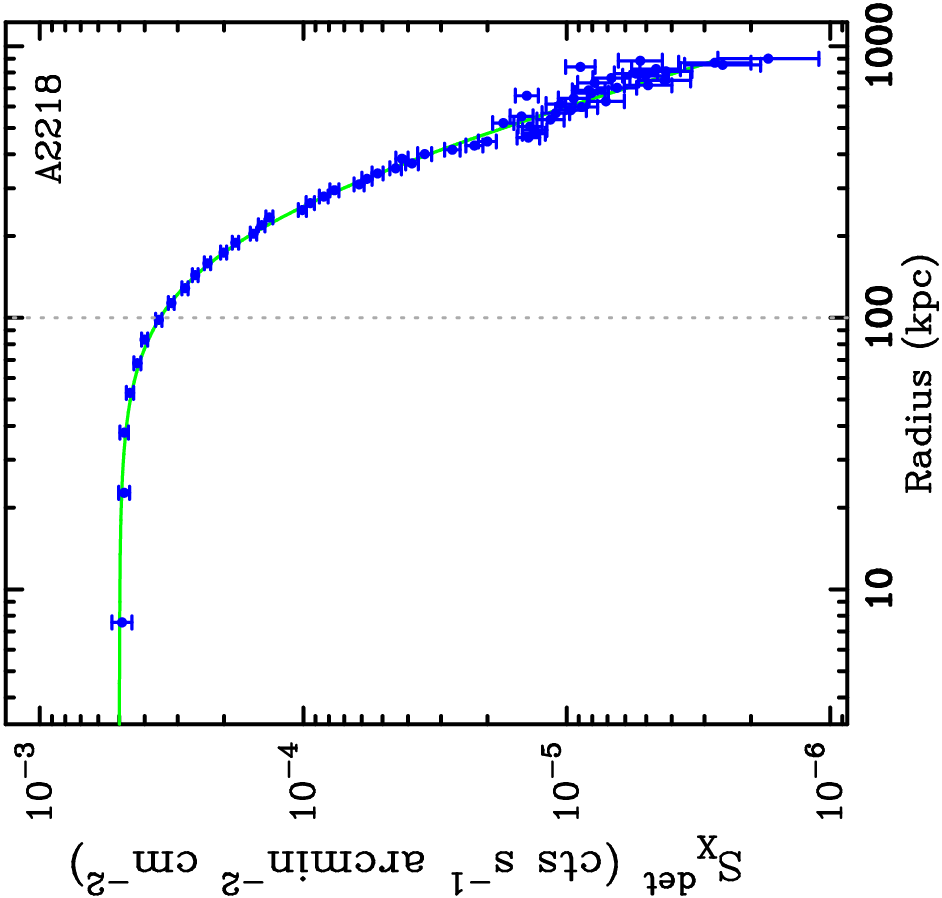}
}
\vspace{0.15in}
\centerline{
  \includegraphics[scale=0.3,angle=270]{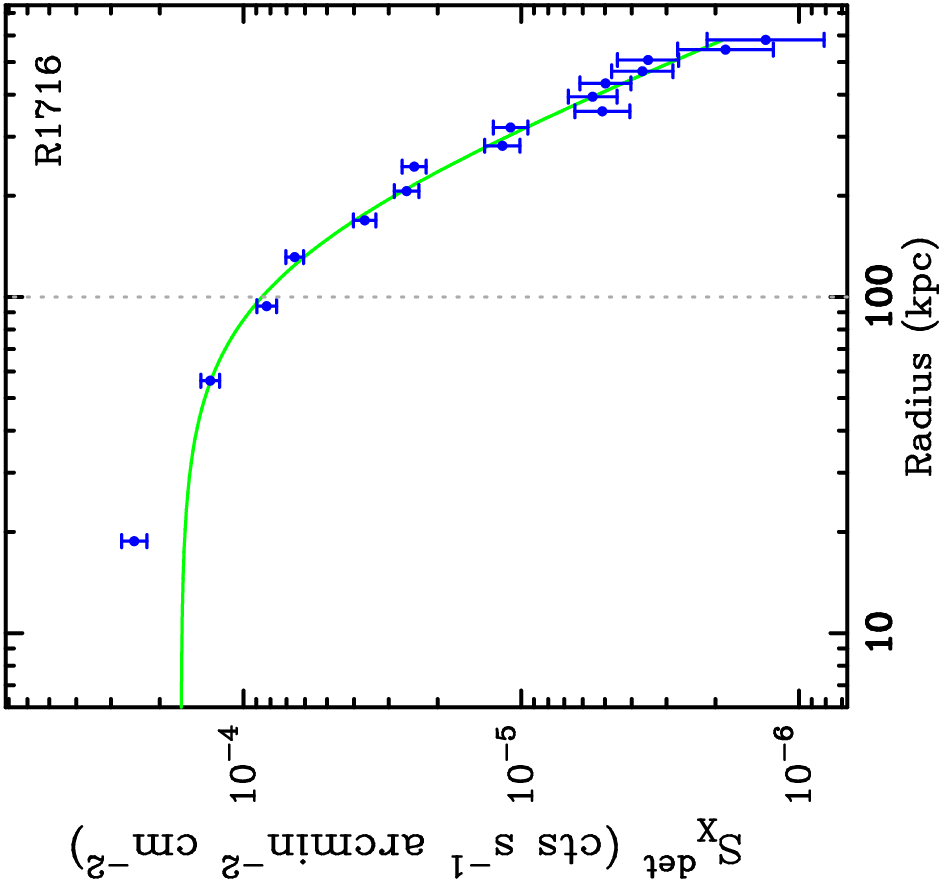}
  \includegraphics[scale=0.3,angle=270]{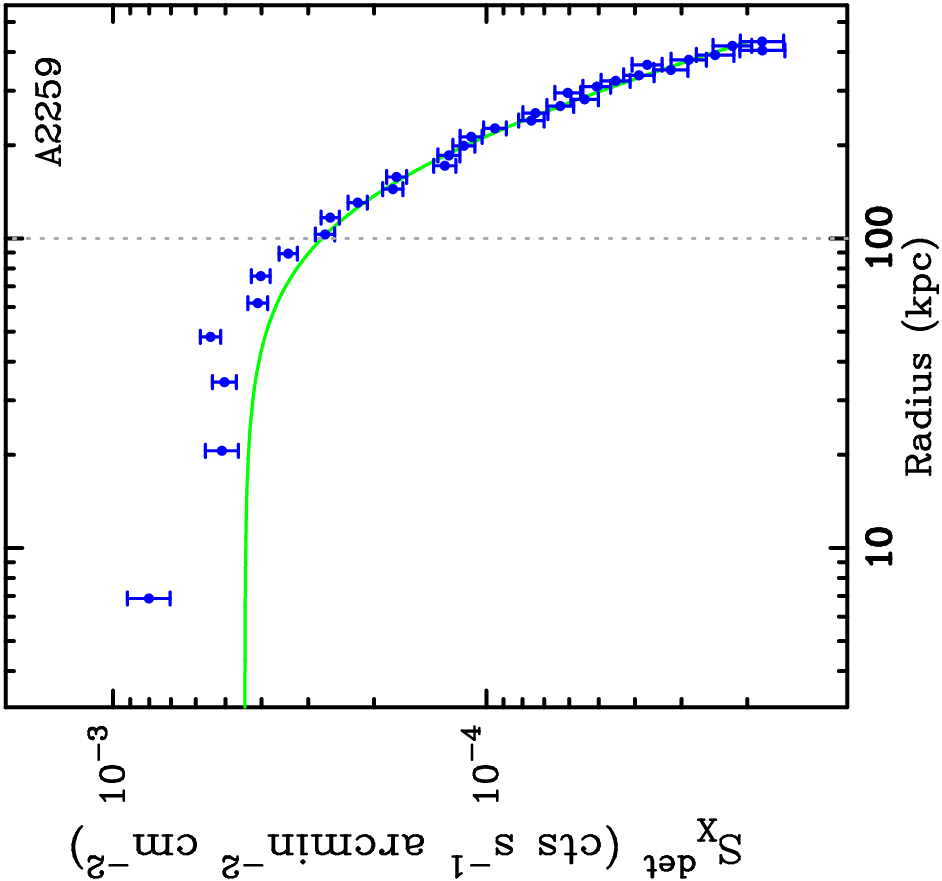}
  \includegraphics[scale=0.3,angle=270]{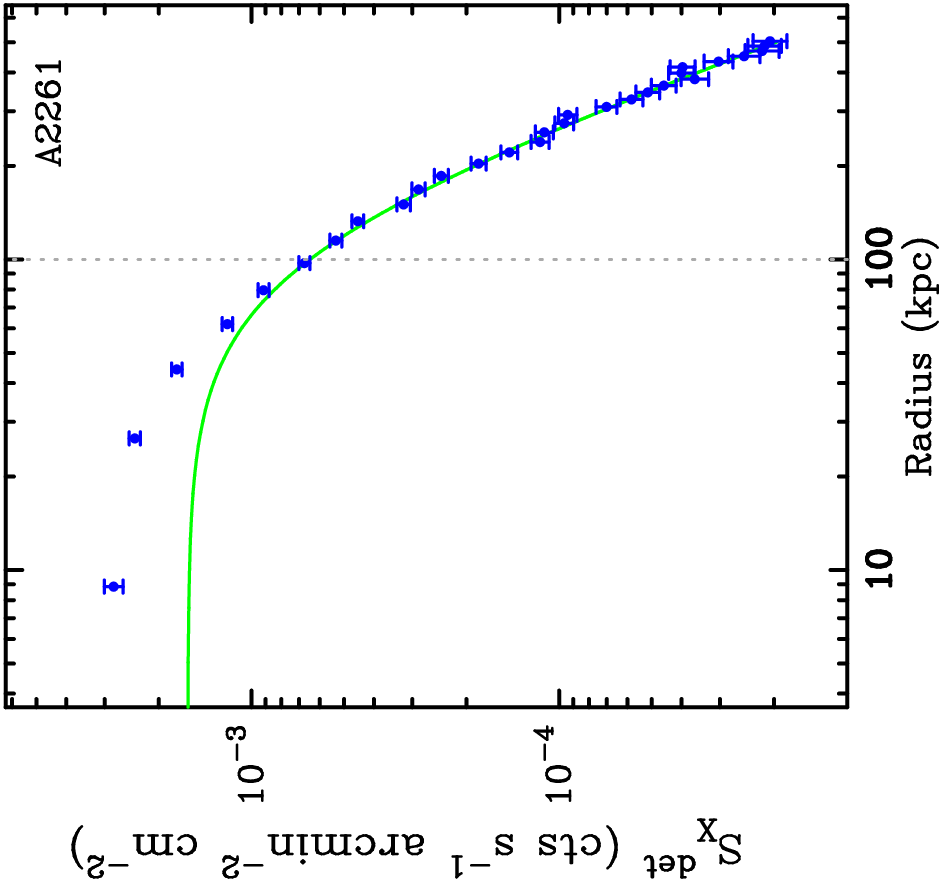}
}
\vspace{0.15in}
\hspace{0.5in}
  \includegraphics[scale=0.3,angle=270]{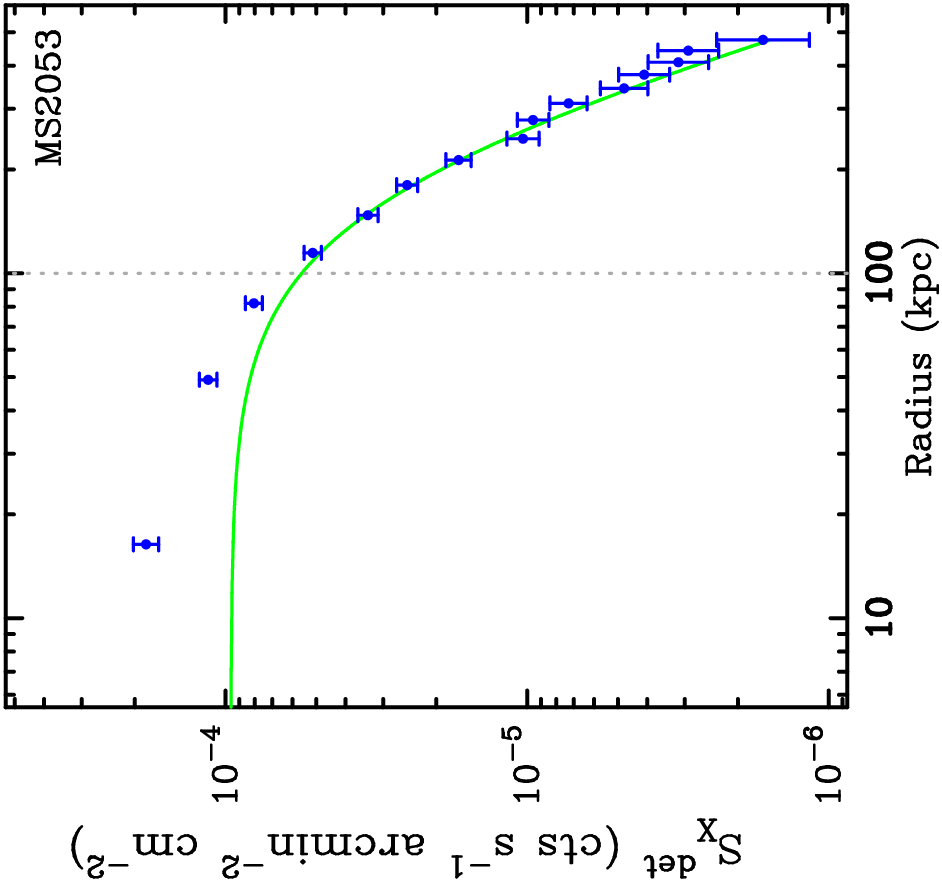}
  \includegraphics[scale=0.3,angle=270]{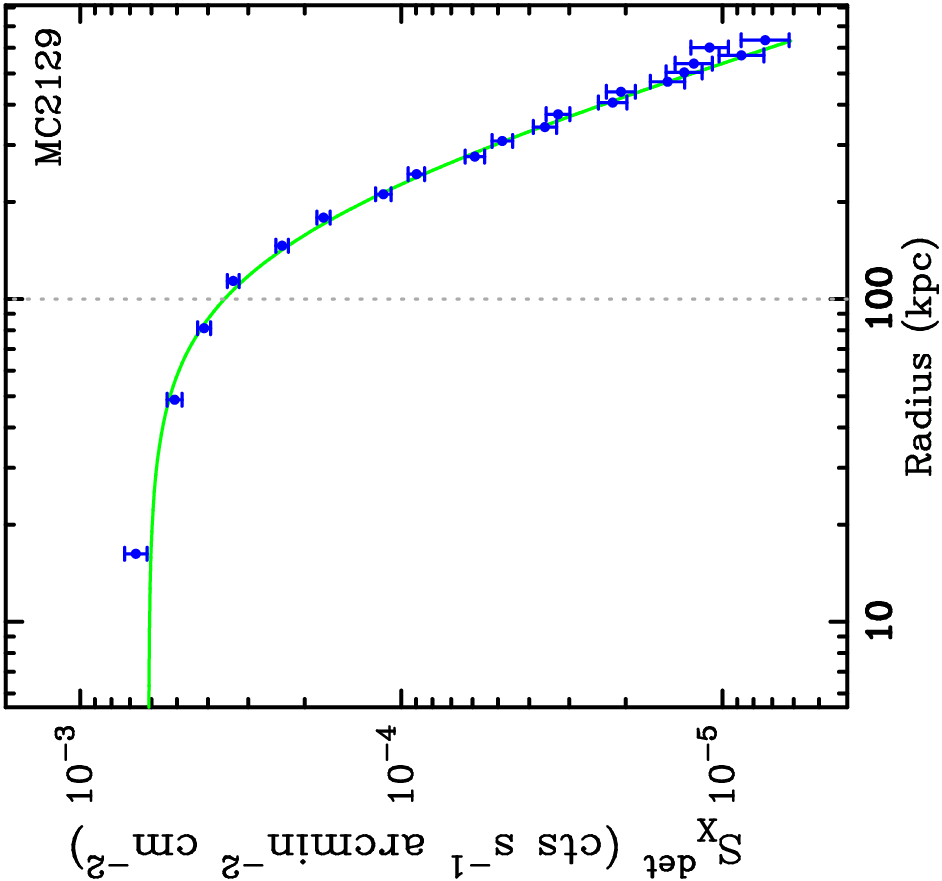}
  \includegraphics[scale=0.3,angle=270]{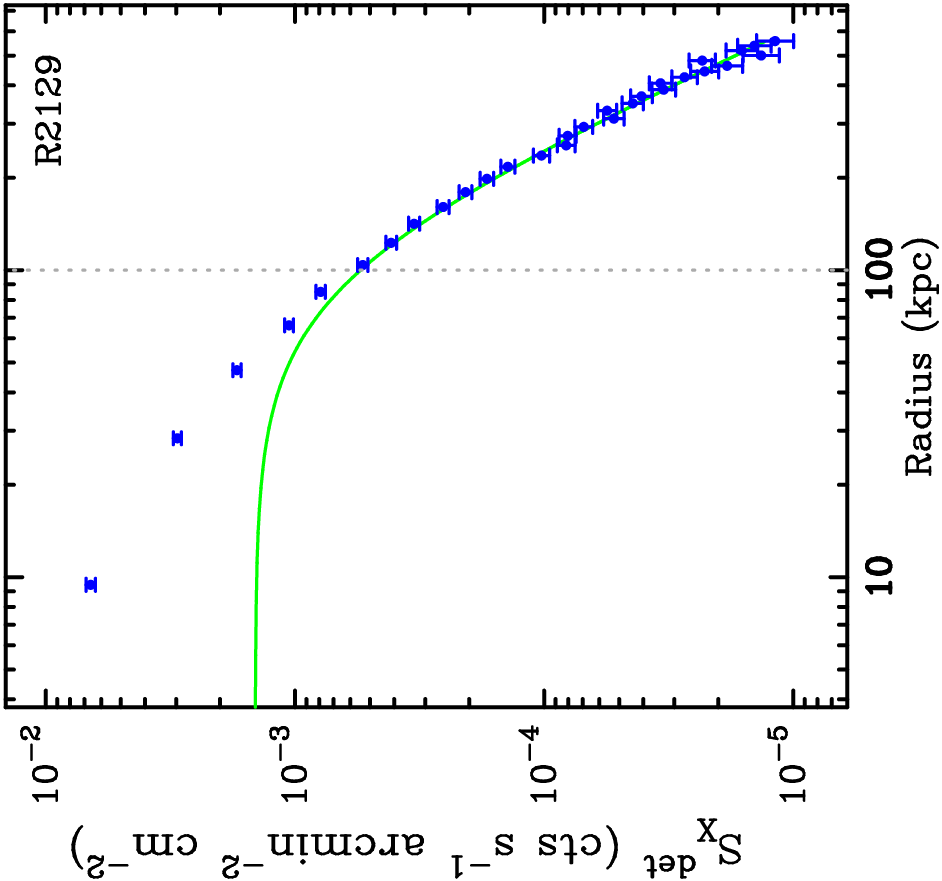}
\contcaption{Cont.}
\end{figure}

\newpage

\begin{figure}
\centerline{
  \includegraphics[scale=0.3,angle=270]{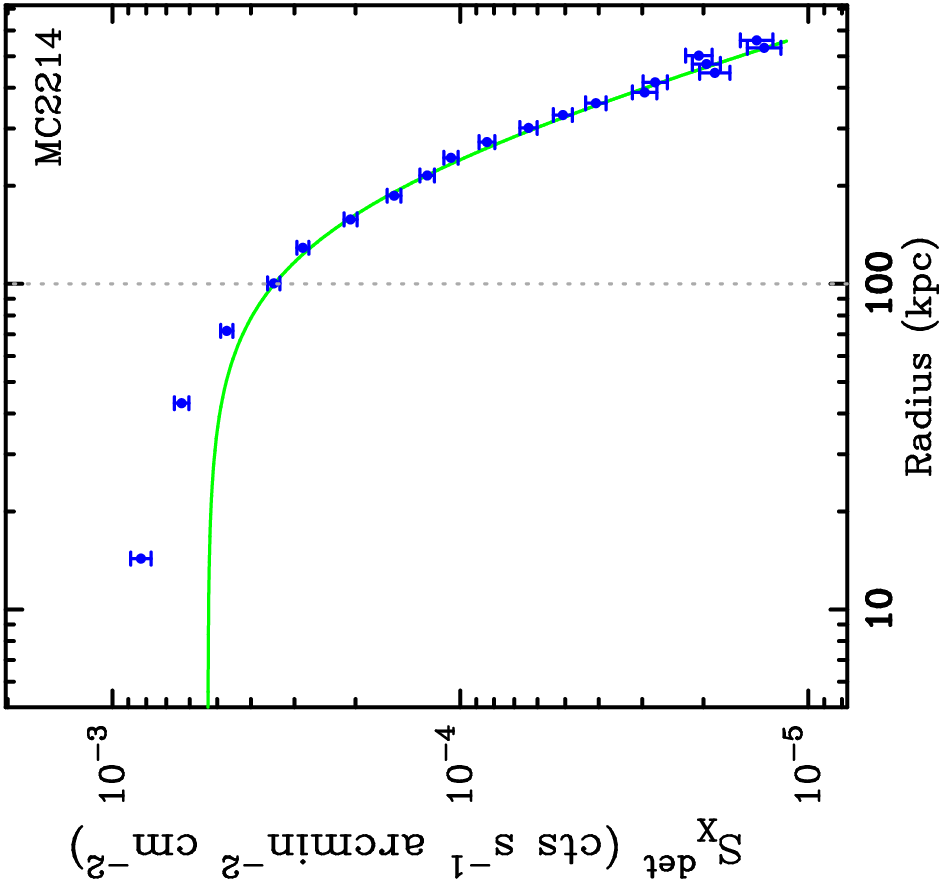}
  \includegraphics[scale=0.3,angle=270]{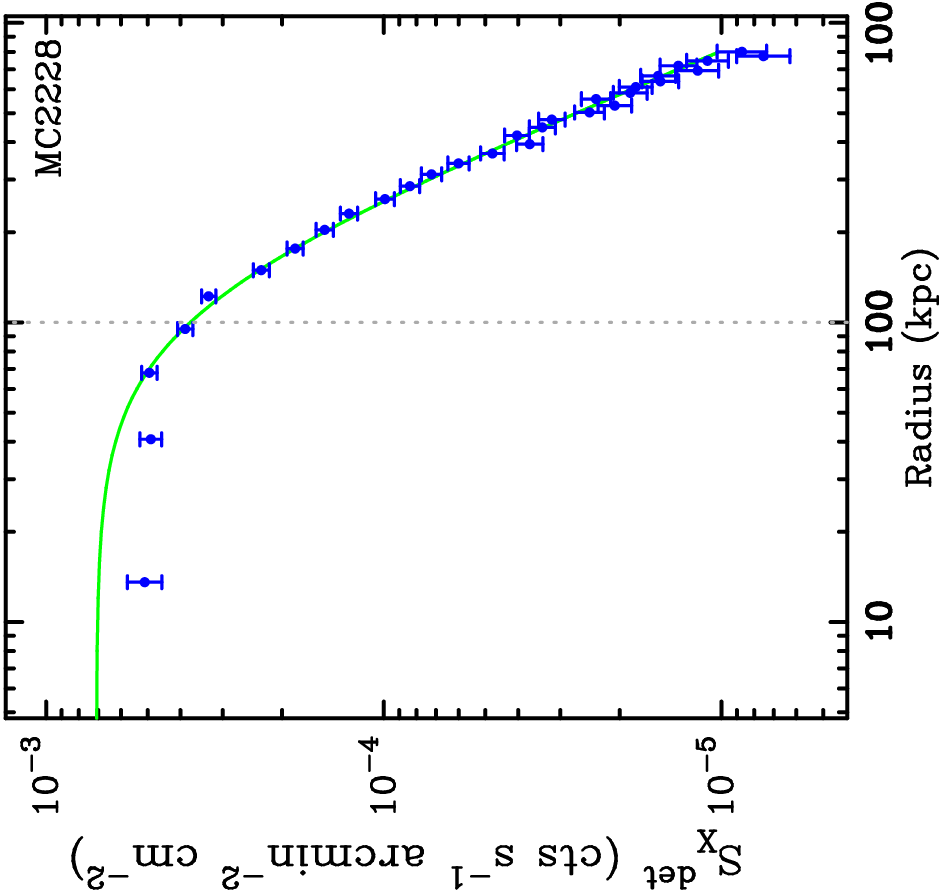}
}
\contcaption{Cont.}
\end{figure}

\end{document}